\documentclass[11pt, a4paper]{article}
\pdfoutput=1
\usepackage{jheppub}
\usepackage{braket}
\usepackage{mathrsfs}
\usepackage{comment}
\usepackage{tikz}

\usepackage{amsfonts}
\usepackage{amscd}
\usepackage{amssymb}
\usepackage{amsmath,bbm}
\usepackage{graphicx}
\usepackage{epsfig}
\usepackage{latexsym}
\usepackage{mathtools}
\usepackage{hyperref}
\usepackage{tikz-cd}
\usepackage[vcentermath]{youngtab}
    
\def \be  {\begin{equation}}
\def \ee  {\end{equation}}
\def \bea {\begin{equation}\begin{aligned}}
\def \eea {\end{aligned}\end{equation}}
\def \ba  {\begin{eqnarray}}
\def \ea  {\end{eqnarray}}
\def \bb  {\begin{thebibliography}}
\def \eb  {\end{thebibliography}}
\def \lab #1 {\label{#1}}
\def\Re{{\rm Re}} \def\Im{{\rm Im}}

\newcommand\ep{\epsilon}

\newcommand\cE{\mathcal{E}}

\newcommand\cL{\mathcal{L}}

\newcommand\cN{\mathcal{N}}
\newcommand\cO{\mathcal{O}}

\newcommand\cV{\mathcal{V}}

\newcommand\cX{\mathcal{X}}

\newcommand\cZ{\mathcal{Z}}
\newcommand\al{\alpha}

\newcommand\bC{\mathbb{C }}
\newcommand\bP{\mathbb{P}}

\newcommand\bR{\mathbb{R}}
\newcommand\bT{\mathbb{T}}
\newcommand\bZ{\mathbb{Z}}

\newcommand\Hom{\mathrm{Hom}}

\definecolor{cardinal}{rgb}{0.6,0,0}
\definecolor{darkgreen}{rgb}{0,0.5,0}
\definecolor{golden}{rgb}{0.92, 0.7, 0}
\definecolor{midnight}{rgb}{0, 0, 0.5}
\definecolor{darkblue}{rgb}{0.2, 0, 0.8}

\usepackage[framemethod=TikZ]{mdframed}
\mdfsetup{nobreak=true}

\makeatletter
\gdef\@fpheader{}
\makeatother

\begin{document}

\vspace*{5mm}

\title{Berry Connections for 2d $(2,2)$ Theories, Monopole Spectral Data \& (Generalised) Cohomology Theories}

\author[1,2,3]{Andrea E. V. Ferrari}
\author[3,4,5]{Daniel Zhang}

\affiliation[1]{Department of Mathematical Sciences, Durham University, Stockton Road, Durham, U.K.}
\affiliation[2]{School of Mathematics, The University of Edinburgh, Mayfield Road, Edinburgh, U.K.}
\affiliation[3]{Mathematical Institute, University of Oxford, Woodstock Road, Oxford, U.K.}
\affiliation[4]{St John’s College, University of Oxford, St Giles’, Oxford, U.K.}
\affiliation[5]{Theoretical Sciences Visiting Program, Okinawa Institute of Science and Technology Graduate University, Onna, Japan}

\emailAdd{andrea.e.v.ferrari@gmail.com}
\emailAdd{daniel.zhang@sjc.ox.ac.uk}

\abstract{
We study Berry connections for supersymmetric ground states of 2d $\mathcal{N}=(2,2)$ GLSMs quantised on a circle, which are generalised periodic monopoles. Periodic monopole solutions may be encoded into difference modules, as shown by Mochizuki, or into an alternative algebraic construction given in terms of vector bundles endowed with filtrations. By studying the ground states in terms of a one--parameter family of supercharges, we relate these two different kinds of spectral data to the physics of the GLSMs. From the difference modules we derive novel difference equations for brane amplitudes, which in the conformal limit yield novel difference equations for hemisphere or vortex partition functions. When the GLSM flows to a nonlinear sigma model with K\"ahler target $X$, we show that the two kinds of spectral data are related to different (generalised) cohomology theories: the difference modules are related to the equivariant quantum cohomology of $X$, whereas the vector bundles with filtrations are related to its equivariant K--theory.
}

\maketitle

\setcounter{page}{1}


\section{Introduction}

Supersymmetric gauge theories in low dimensions have been an inexhaustible source of deep mathematical constructions and problems. This is undoubtedly the case for 2d $\mathcal{N}=(2,2)$ GLSMs, which represent, amongst other things, a physical arena for the study of homological mirror symmetry. In this paper we revisit some physical phenomena related to the supersymmetric ground states of 2d $(2,2)$ GLSMs quantised on a circle, either in a cylindrical or cigar geometry, expanding upon them and demonstrating that this particular source still has much to give. 

The starting, fundamental observation is that moduli spaces of solutions to supersymmetric Berry connections over a twisted mass deformation and associated holonomy for an abelian flavor symmetry correspond to moduli spaces of periodic monopoles. This allows us to relate supersymmetric ground states in the cohomology of a one--parameter family of supercharges to mathematical constructions that have recently received significant attention, namely the difference modules representing monopole solutions of Mochizuki \cite{mochizuki2017periodic, mochizuki2019doubly} as well as a Riemann--Hilbert correspondence between these and holomorphic vector bundles endowed with filtrations \cite{Kontsevich_Soibelman_1}. These can be thought of as encoding different kinds of spectral data for the monopole. As a result of the relation between Berry connections and difference modules, we derive novel difference equations satisfied by brane amplitudes and hemisphere and vortex partition functions. Moreover, in the case of a GLSM that flows to an NLSM with (GKM) target $X$, we relate the different algebraic representations entering the Riemann--Hilbert correspondence to a quantisation of a certain action on the equivariant quantum cohomology of $X$ and the equivariant K--theory of $X$, respectively. To the best of our knowledge, this connection has also not appeared so far in the literature.

\subsection{Overview}

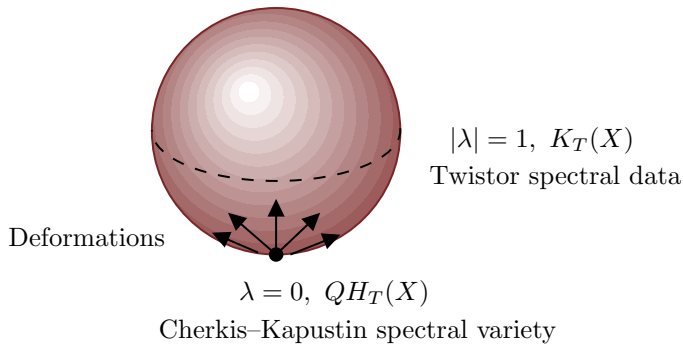
\begin{figure}
\centering
  
\tikzset {_dp4fhqamn/.code = {\pgfsetadditionalshadetransform{ \pgftransformshift{\pgfpoint{89.1 bp } { -128.7 bp }  }  \pgftransformscale{1.32 }  }}}
\pgfdeclareradialshading{_qbgonrwsr}{\pgfpoint{-72bp}{104bp}}{rgb(0bp)=(1,1,1);
rgb(0bp)=(1,1,1);
rgb(25bp)=(0.48,0.15,0.15);
rgb(400bp)=(0.48,0.15,0.15)}
\tikzset{every picture/.style={line width=0.75pt}} 

\begin{tikzpicture}[x=0.75pt,y=0.75pt,yscale=-1,xscale=1]

\path  [shading=_qbgonrwsr,_dp4fhqamn] (143,74) .. controls (143,39.76) and (170.76,12) .. (205,12) .. controls (239.24,12) and (267,39.76) .. (267,74) .. controls (267,108.24) and (239.24,136) .. (205,136) .. controls (170.76,136) and (143,108.24) .. (143,74) -- cycle ; 
 \draw  [color={rgb, 255:red, 122; green, 38; blue, 38 }  ,draw opacity=1 ] (143,74) .. controls (143,39.76) and (170.76,12) .. (205,12) .. controls (239.24,12) and (267,39.76) .. (267,74) .. controls (267,108.24) and (239.24,136) .. (205,136) .. controls (170.76,136) and (143,108.24) .. (143,74) -- cycle ; 

\draw  [color={rgb, 255:red, 0; green, 0; blue, 0 }  ,draw opacity=1 ][fill={rgb, 255:red, 0; green, 0; blue, 0 }  ,fill opacity=1 ] (201.88,136) .. controls (201.88,134.27) and (203.27,132.88) .. (205,132.88) .. controls (206.73,132.88) and (208.13,134.27) .. (208.13,136) .. controls (208.13,137.73) and (206.73,139.13) .. (205,139.13) .. controls (203.27,139.13) and (201.88,137.73) .. (201.88,136) -- cycle ;
\draw  [draw opacity=0][dash pattern={on 4.5pt off 4.5pt}] (266.95,72.48) .. controls (266.98,72.82) and (267,73.16) .. (267,73.5) .. controls (267,87.58) and (239.24,99) .. (205,99) .. controls (170.76,99) and (143,87.58) .. (143,73.5) -- (205,73.5) -- cycle ; \draw  [dash pattern={on 4.5pt off 4.5pt}] (266.95,72.48) .. controls (266.98,72.82) and (267,73.16) .. (267,73.5) .. controls (267,87.58) and (239.24,99) .. (205,99) .. controls (170.76,99) and (143,87.58) .. (143,73.5) ;  
\draw    (205,136) -- (225.55,116.38) ;
\draw [shift={(227,115)}, rotate = 136.33] [fill={rgb, 255:red, 0; green, 0; blue, 0 }  ][line width=0.08]  [draw opacity=0] (8.93,-4.29) -- (0,0) -- (8.93,4.29) -- cycle    ;
\draw    (205,136) -- (182.51,116.32) ;
\draw [shift={(181,115)}, rotate = 41.19] [fill={rgb, 255:red, 0; green, 0; blue, 0 }  ][line width=0.08]  [draw opacity=0] (8.93,-4.29) -- (0,0) -- (8.93,4.29) -- cycle    ;
\draw    (205,137) -- (205,110) ;
\draw [shift={(205,108)}, rotate = 90] [fill={rgb, 255:red, 0; green, 0; blue, 0 }  ][line width=0.08]  [draw opacity=0] (8.93,-4.29) -- (0,0) -- (8.93,4.29) -- cycle    ;
\draw    (212.13,136) -- (235.27,126.74) ;
\draw [shift={(237.13,126)}, rotate = 158.2] [fill={rgb, 255:red, 0; green, 0; blue, 0 }  ][line width=0.08]  [draw opacity=0] (8.93,-4.29) -- (0,0) -- (8.93,4.29) -- cycle    ;

\draw    (196,135) -- (174.71,125.79) ;
\draw [shift={(172.88,125)}, rotate = 23.39] [fill={rgb, 255:red, 0; green, 0; blue, 0 }  ][line width=0.08]  [draw opacity=0] (8.93,-4.29) -- (0,0) -- (8.93,4.29) -- cycle    ;

\draw (185,147.4) node [anchor=north west][inner sep=0.75pt]    {\small $\lambda =0,\ QH_{T}( X)$};
\draw (290,70.4) node [anchor=north west][inner sep=0.75pt]    {\small $|\lambda |=1,\ K_{T}( X)$};
\draw (70,121) node [anchor=north west][inner sep=0.75pt]   [align=left] {\small Deformations};
\draw (145,167) node [anchor=north west][inner sep=0.75pt]   [align=left] {\small Cherkis--Kapustin spectral variety};
\draw (280,90) node [anchor=north west][inner sep=0.75pt]   [align=left] {\small Twistor spectral data};

\end{tikzpicture}

\caption{The twistor sphere of supercharges and related constructions for a GLSM with Higgs branch $X$, or equivalently an NLSM with smooth target $X$. At $\lambda=0$, operators in $Q_\lambda$-cohomology reproduce the equivariant quantum cohomology ring of the target $X$. The Cherkis--Kapustin spectral variety of the Berry connection encodes properties of this ring. Away from it, the spectral variety is quantised. At $|\lambda|=1$ an alternative twistorial description of the spectral data emerges that is related to the equivariant K-theory of $X$.}
\label{fig:intro-twistor-sphere}
\end{figure}

We now provide a narrative overview of our results. In Section~\ref{sec:setup} we describe the setup for this work. We consider a 2d $\mathcal{N}=(2,2)$ GLSM with an abelian flavour symmetry $T$ on $\bR\times S^1$ and cigar geometries, highlighting aspects of the Hilbert space of the theory on the spatial $S^1$. We first identify a $\mathbb{P}^1$ (twistor) family of $\mathcal{N}=2$ Supersymmetric Quantum Mechanics (SQMs) along the non--compact direction $\bR$, labelled by a local coordinate  $\lambda $ on $\mathbb{P}^1$. The SQMs contain a distinguished supercharge $Q_\lambda$, and our conventions are such that
\begin{equation}
    Q_{\lambda}|_{\lambda=0} = Q_A, \quad Q_{\lambda}|_{\lambda=1} = Q.
\end{equation}
Here $Q_A$ is the A--model supercharge whose cohomology is the twisted chiral ring and therefore, when the GLSM flows to an NLSM with target $X$, it is known to reproduce the equivariant quantum cohomology of the target $QH_T^\bullet (X)$. $Q$ is a dimensional reduction of the 3d $\mathcal{N}=2$ supercharge whose cohomology is related to the equivariant elliptic cohomology \cite{Bullimore:2021rnr, Dedushenko:2021mds}; in this 2d setting we obtain equivariant K--theory $K_T(X)$. See Figure~\ref{fig:intro-twistor-sphere} for a pictorial representation of the twistor sphere.

For simplicity, in this overview let us consider a GLSM with a $T=U(1)$ flavour symmetry. This is the main focus of the mathematical literature \cite{mochizuki2017periodic, mochizuki2019doubly}. However, our results can naturally be generalised to an abelian flavour symmetry $T=U(1)^n$, $n>1$, and much of our discussions in the main body will be devoted to the study of these higher--rank cases.

We can deform the theory by introducing a holonomy $t$ for $T$ on the circle as well as a complex mass $w$, so that the pair $(t,w)$ takes values in $M=S^1\times \mathbb{R}^2$. It will be useful for us to understand $t$ as an $\mathbb{R}$-valued variable in the first place and then quotient by a $\mathbb{Z}$-action. The space of supersymmetric ground states of a theory with $N$ vacua forms a rank-$N$ vector bundle $E \rightarrow M$ endowed with a Berry connection. The Berry connection is determined via (by now standard) considerations in $tt^*$ geometry, and it can be shown to correspond to a periodic monopole solution on $M$, solving the Bogomolny equations $F =\star D \phi $, where $\phi$ as an adjoint Higgs field. Under some physical assumptions, the monopole we obtain is of generalised Cherkis--Kapustin type \cite{Cherkis:2000cj, Cherkis:2000ft}, that is (approximately speaking) it has only Dirac singularities (corresponding to parameters where the theory is no longer gapped) and asymptotically approaches in $w$ a direct sum of Dirac monopoles. In particular, the asymptotics of the Berry connection can be related to physical data, in the form of the effective twisted superpotential:
\begin{equation}
\begin{array}{cc}
     A_{t}^{\alpha} + i \phi^{\alpha}\sim -2 i \partial_{w} W_{\text{eff}}^{(\alpha)}  &  \\
     A_w \sim O(\frac{1}{w
}) & 
\end{array}
    \quad \text{as} \quad { |w| \rightarrow \infty.}
\end{equation}
where $\alpha=1,\ldots,N$ label isolated massive vacua, and $W^{(\alpha)}_{\text{eff}}$ is the effective twisted superpotential evaluated on these vacua.

Our twistor family of supercharges can naturally be thought of as a family of mini--complex structures on the space $M$ (thus justifying the name). These correspond to collections of open charts $\mathbb{R}\times \mathbb{C}$, glued in a way that it makes sense to speak of functions locally constant along $\mathbb{R}$ and holomorphic along $\mathbb{C}$.\footnote{It may help the reader to keep in mind that a mini--complex structure can be understood in terms of a dimensional reduction of complex structures from $\mathbb{R}^4$. Our twistor sphere is a direct descendant of the sphere of complex structures of this hyperk\"ahler manifold.} Adapted to a certain mini--complex structure on $M$ we have mini--complex coordinates $(t_1(\lambda),\beta_1(\lambda) )$ such that $(t_1(\lambda=0),\beta_1(\lambda = 0))=(t,w)$. To emphasise this structure, we sometimes write $M^\lambda$. Moreover, the vector bundle $E \rightarrow M^\lambda$ is mini--holomorphic, in a suitable sense. In particular, when restricted to $t_1$ it is naturally holomorphic with Dolbeault operator $\partial_{E,\beta_1}$. The Bogomolny (monopole) equations imply a complex equation
\begin{equation}\label{eq:intro-Bog}
    [\partial_{E,t_1},\partial_{E,\beta_1}] = 0
\end{equation}
that holds at all values of $\lambda$, where $\partial_{E,t_1} = D_{t_1} - i \phi$ is the covariant derivative operator in the $t_1$ direction complexified by the Higgs field. We review these facts in greater detail at the beginning of Section~\ref{sec:spectral_data_1}. Following Mochizuki~\cite{mochizuki2017periodic}, we can encode the monopole solutions in terms of Hitchin--Kobayashi correspondences that arise from the above complex equation, and which encode the other, real equations into stability conditions.

\paragraph{Product case, Cherkis–Kapustin spectral curve \& quantum cohomology. } 

Consider the case $\lambda=0$ case first (the so--called product case), which we investigate in Section~\ref{sec:product_case}. In this case, as mini--complex manifolds, we have
\begin{equation}
    M^{\lambda=0} \cong S^1\times \mathbb{C}.
\end{equation}
The vector bundle of supersymmetric ground states restricted at $t=0$, $\cE^{0}:=E|_{t=0}$, is holomorphic with respect to $\partial_{E,w}$ and we can loosely denote the space of meromorphic sections by $V$. Along the locus $D \in \mathbb{C}_w$ where monopole singularities are located, the sections may acquire some poles, so one na\"ive definition would be\footnote{Here and below, $\mathbb{C}(w)$ denotes the ring of rational functions in $w$}
\begin{equation}
    V := H^{0}(\mathbb{C}_w,\cE^0(\star D))\otimes_{\mathbb{C}[w]} \mathbb{C}(w),
\end{equation}
where we are allowing for these poles. In fact, $V$ is a more refined version of this, as we will review. These technicalities aside, we can parallel transport the holomorphic sections around a full circle by means of the operator $\partial_{E,t}$. By~\eqref{eq:intro-Bog} we then get an automorphism
\begin{equation}
    F(w) : V \rightarrow V,
\end{equation}
which endows $V$ with the structure of a $0$-difference $\mathbb{C}(w)$-module. Mochizuki showed in~\cite{mochizuki2017periodic}, following work of Charbonneau--Hurtubise \cite{charbonneau2011singular}, that by keeping track of some additional structure of the module (related to the aforementioned stability, as well as to the behaviour of the monopole at infinity and at the singularities) one gets a 1:1 correspondence between monopoles of GCK type and these modules. Moreover, one can consider the curve spanned by the eigenvalues of this automorphism, which essentially corresponds to the Cherkis--Kapustin spectral curve $\mathcal{L}$ for the monopole
\begin{equation}\label{eq:intro_cherkis_kapustin}
    \mathcal{L} = \{(p,w)~|~\det(p\mathbf{1}-F(w)) = 0\}.
\end{equation}

We demonstrate that the pair $(V,F)$ has a neat physical interpretation. The module $V$ can be generated by states obtained via the A--twisted topological path integral on a cigar, by inserting operators in $Q_A$-cohomology at the tip. The automorphism arises from the insertion of an operator $p$ that acts on this module. It can be interpreted as a defect operator inserting flux for $T$, whose eigenvalues can be efficiently computed. We find that they correspond to $e^{-2i \partial_w W_{\text{eff}}^{(\alpha)}}$. In other words, they can be obtained from the equations
\begin{equation}\label{eq:mom-space-qc}
    e^{-2i\frac{\partial W_{\text{eff}}(\sigma, w)}{\partial_{w} }} = p,\qquad
    e^{\frac{\partial W_{\text{eff}}(\sigma, w)}{\partial_{\sigma_a} }} = 1, \quad\, a=1\ldots r.
\end{equation}
In the above $\sigma_a$ are scalars parametrising the Cartan of the gauge group, lying in abelian vector multiplets in the IR. For non-abelian gauge groups, one most quotient the set of solutions by the Weyl group. Once eliminated from the above equations, one recovers the spectral curve \eqref{eq:intro_cherkis_kapustin}.

In the case of a GLSM that flows to a target $X$, the $Q_A$-cohomology ring is known to correspond to the quantum equivariant cohomology $QH_T^{\bullet}(X)$ \cite{Vafa:1991uz}. Thus, the module $V$ may be identified with $QH_T^{\bullet}(X)$. The set of equations on the right of~\eqref{eq:mom-space-qc}, are known to precisely describe this ring~\cite{Nekrasov:2009uh, Nekrasov:2009ui}. From the above observations, the module $(V,F)$ describes the action of the algebra of functions on $\mathbb{C}\times \mathbb{C}^*$ generated by $(w,p)$ on $QH_T^{\bullet}(X)$. This defines a sheaf on $\mathbb{C}\times \mathbb{C}^*$ with support on the solutions in $(p,w)$ to the whole set of equations~\eqref{eq:mom-space-qc}.\footnote{(This set of equations can nai\"vely be viewed as a `momentum space' representation of the quantum equivariant cohomology relations) } In upcoming work, we will interpret this action by coupling the 2d theory to a bulk--boundary 3d theory, relating it to work of Teleman \cite{Teleman:2018wac}.

\paragraph{Branes, difference modules \& curve quantisation.}

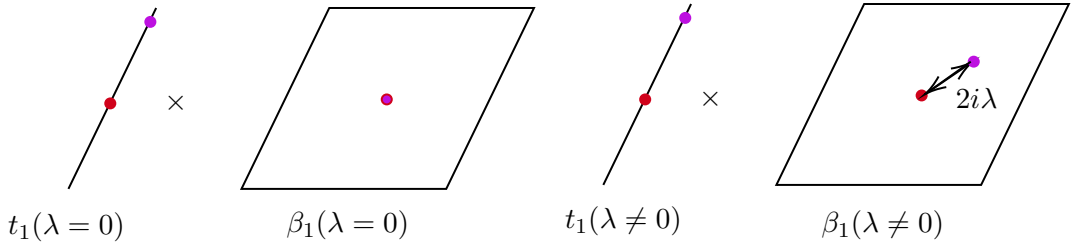
\begin{figure}
    \centering

\tikzset{every picture/.style={line width=0.75pt}} 

\tikzset{every picture/.style={line width=0.75pt}} 

\begin{tikzpicture}[x=0.75pt,y=0.75pt,yscale=-1,xscale=1]

\draw    (489,59.5) -- (511.35,44.13) ;
\draw [shift={(513,43)}, rotate = 145.49] [color={rgb, 255:red, 0; green, 0; blue, 0 }  ][line width=0.75]    (10.93,-3.29) .. controls (6.95,-1.4) and (3.31,-0.3) .. (0,0) .. controls (3.31,0.3) and (6.95,1.4) .. (10.93,3.29)   ;
\draw    (372.4,14) -- (328.6,105) ;
\draw   (458.8,14) -- (561,14) -- (517.2,105) -- (415,105) -- cycle ;
\draw  [color={rgb, 255:red, 208; green, 2; blue, 27 }  ,draw opacity=1 ][fill={rgb, 255:red, 208; green, 2; blue, 27 }  ,fill opacity=1 ] (485,60) .. controls (485,58.62) and (486.12,57.5) .. (487.5,57.5) .. controls (488.88,57.5) and (490,58.62) .. (490,60) .. controls (490,61.38) and (488.88,62.5) .. (487.5,62.5) .. controls (486.12,62.5) and (485,61.38) .. (485,60) -- cycle ;
\draw  [color={rgb, 255:red, 208; green, 2; blue, 27 }  ,draw opacity=1 ][fill={rgb, 255:red, 208; green, 2; blue, 27 }  ,fill opacity=1 ] (347,62) .. controls (347,60.62) and (348.12,59.5) .. (349.5,59.5) .. controls (350.88,59.5) and (352,60.62) .. (352,62) .. controls (352,63.38) and (350.88,64.5) .. (349.5,64.5) .. controls (348.12,64.5) and (347,63.38) .. (347,62) -- cycle ;
\draw  [color={rgb, 255:red, 189; green, 16; blue, 224 }  ,draw opacity=1 ][fill={rgb, 255:red, 189; green, 16; blue, 224 }  ,fill opacity=1 ] (511,43) .. controls (511,41.62) and (512.12,40.5) .. (513.5,40.5) .. controls (514.88,40.5) and (516,41.62) .. (516,43) .. controls (516,44.38) and (514.88,45.5) .. (513.5,45.5) .. controls (512.12,45.5) and (511,44.38) .. (511,43) -- cycle ;
\draw  [color={rgb, 255:red, 189; green, 16; blue, 224 }  ,draw opacity=1 ][fill={rgb, 255:red, 189; green, 16; blue, 224 }  ,fill opacity=1 ] (367,21) .. controls (367,19.62) and (368.12,18.5) .. (369.5,18.5) .. controls (370.88,18.5) and (372,19.62) .. (372,21) .. controls (372,22.38) and (370.88,23.5) .. (369.5,23.5) .. controls (368.12,23.5) and (367,22.38) .. (367,21) -- cycle ;
\draw    (105.4,16) -- (61.6,107) ;
\draw   (191.8,16) -- (294,16) -- (250.2,107) -- (148,107) -- cycle ;
\draw  [color={rgb, 255:red, 208; green, 2; blue, 27 }  ,draw opacity=1 ][fill={rgb, 255:red, 189; green, 16; blue, 224 }  ,fill opacity=1 ] (218,62) .. controls (218,60.62) and (219.12,59.5) .. (220.5,59.5) .. controls (221.88,59.5) and (223,60.62) .. (223,62) .. controls (223,63.38) and (221.88,64.5) .. (220.5,64.5) .. controls (219.12,64.5) and (218,63.38) .. (218,62) -- cycle ;
\draw  [color={rgb, 255:red, 208; green, 2; blue, 27 }  ,draw opacity=1 ][fill={rgb, 255:red, 208; green, 2; blue, 27 }  ,fill opacity=1 ] (80,64) .. controls (80,62.62) and (81.12,61.5) .. (82.5,61.5) .. controls (83.88,61.5) and (85,62.62) .. (85,64) .. controls (85,65.38) and (83.88,66.5) .. (82.5,66.5) .. controls (81.12,66.5) and (80,65.38) .. (80,64) -- cycle ;
\draw  [color={rgb, 255:red, 189; green, 16; blue, 224 }  ,draw opacity=1 ][fill={rgb, 255:red, 189; green, 16; blue, 224 }  ,fill opacity=1 ] (100,23) .. controls (100,21.62) and (101.12,20.5) .. (102.5,20.5) .. controls (103.88,20.5) and (105,21.62) .. (105,23) .. controls (105,24.38) and (103.88,25.5) .. (102.5,25.5) .. controls (101.12,25.5) and (100,24.38) .. (100,23) -- cycle ;
\draw    (512,43) -- (490.63,58.33) ;
\draw [shift={(489,59.5)}, rotate = 324.34] [color={rgb, 255:red, 0; green, 0; blue, 0 }  ][line width=0.75]    (10.93,-3.29) .. controls (6.95,-1.4) and (3.31,-0.3) .. (0,0) .. controls (3.31,0.3) and (6.95,1.4) .. (10.93,3.29)   ;

\draw (375,55.4) node [anchor=north west][inner sep=0.75pt]    {$\times $};
\draw (308,114.4) node [anchor=north west][inner sep=0.75pt]    {$t_{1}( \lambda \neq 0)$};
\draw (436,116.4) node [anchor=north west][inner sep=0.75pt]    {$\beta _{1}( \lambda \neq 0)$};
\draw (108,57.4) node [anchor=north west][inner sep=0.75pt]    {$\times $};
\draw (30,117.4) node [anchor=north west][inner sep=0.75pt]    {$t_{1}( \lambda =0)$};
\draw (169,116.4) node [anchor=north west][inner sep=0.75pt]    {$\beta _{1}( \lambda =0)$};
\draw (503,54.65) node [anchor=north west][inner sep=0.75pt]    {$2i\lambda $};

\end{tikzpicture}

    \caption{The mini--holomorphic coordinates $(t_1,\beta_1)$ at different $\lambda$. The purple and red points are identified in the underlying smooth manifold $M \cong S^1\times \mathbb{R}^2$. In the product case ($\lambda=0$, left), moving along the real coordinate brings one back to the same point in $M$. In the non--product case ($\lambda \neq 0$, right), an additional shift by $2i\lambda$ is necessary. }
    \label{fig:intro-mini-coord}
\end{figure}

In Section~\ref{sec:difference_modules}, we consider the case $\lambda \neq 0$. We obtain a similar structure as before, namely the bundle $\mathcal{E}^{t_1}$ still has a holomorphic structure, but now parallel transport with respect to $t_1$ needs to be composed with a $2i\lambda$ shift in order to obtain an automorphism of its sheaf of meromorphic sections. In turn, we no longer get a $0$-difference module, but a $2i\lambda$-difference module instead. See Figure~\ref{fig:intro-mini-coord}.

In this paper, we propose that supersymmetric ground states viewed as elements in $Q_\lambda$ cohomology are holomorphic with respect to $\partial_{E,\bar{\beta}_1}$, and are natural candidates for a physical representation of the elements of Mochizuki's modules encoding the Berry connection. We denote a suitable basis for the holomorphic sections as $\{\ket{a^\lambda}\}$. In the $\lambda \rightarrow 0 $ limit, such a basis can be generated by chiral ring insertions $\cO_a$ at the tip of an adjoining cigar. Moreover, we consider what we call brane states $\ket{D}$, generated by appropriate D-branes preserving $Q_{\lambda}$, as well as their projections onto the ground state sector $\Pi[D]$. The latter can be expanded in a selected basis
\begin{equation}
    \Pi [D] := \sum_{a^\lambda} \Pi [D,a^{\lambda}] \eta^{ab }\ket{b^\lambda}.
\end{equation}
In the above, $\Pi [D,a^\lambda] = \braket{a^\lambda|D}$ are known as \textit{brane amplitudes}, and $\eta^{ab}$ is the inverse of the non--degenerate pairing $\eta_{ab}=\braket{a^\lambda|b^\lambda}$. Notice that overlaps $\braket{a|D}$ can be generated by a path integral on a cigar geometry with $\cO_a$ inserted at the tip, and a B-brane D at the boundary $S^1$  (see Figure~\ref{fig:brane_amplitude}). 

The detailed properties of D-brane states are difficult to understand, however, we can show on general grounds that they must be annihilated by both $\partial_{E,t_1}$ and $\partial_{E,\bar{\beta_1}}$. This is because of the well--known fact that the D-branes are flat sections of the Lax connection. From this, we derive remarkable difference equations satisfied by brane amplitudes $\langle b | D_\alpha \rangle$. These take the form
\begin{equation}\label{eq:intro_difference_equation}
    (\Phi^*_1)^{-1} \langle a^\lambda | D \rangle = G_{a}^{\,\,b}(\beta_1) \langle b^\lambda | D\rangle
\end{equation}
where $G_a^b$ is a matrix of holomorphic functions of $\beta_1$, and $\Phi^*_1$ is the automorphism given by a $2i\lambda$-shift $\beta_1 \mapsto \beta_1 + 2i\lambda$. To our knowledge, these difference equations are novel.

There is a distinguished basis of flat sections of the $tt^*$ Lax connection given by a class of branes known as \textit{thimbles}. For GLSMs which flow in the IR to NLSMs, they are supported on the holomorphic Lagrangian submanifolds of the Higgs branch $X$ corresponding to attracting submanifolds of fixed points of $T$. Such boundary conditions were analysed explicitly for massive $(2,2)$ theories in \cite{Hori:2000ck, Gaiotto:2015aoa} and for 3d $\mathcal{N}=4$ theories in \cite{Bullimore:2016nji, Bullimore:2020jdq, Bullimore:2021rnr, Crew:2023tky}. The asymptotic behaviour of these brane amplitudes is known:
\begin{equation}\label{eq:intro_thimble_brane_asymptotic}
    \braket{b^\lambda|D_\alpha} \sim e^{\frac{W_{\text{eff}}^{(\alpha)}}{\lambda}} \cO_b|_{\alpha},\qquad\text{as } \lambda \rightarrow 0.
\end{equation}

Utilising the asymptotic behaviour of this basis, we show that the difference equations \eqref{eq:intro_difference_equation} must reduce in the $\lambda \rightarrow 0$ limit to the defining equation of the Cherkis--Kapustin spectral curve of the monopole. In particular: 
\begin{equation}\label{eq:intro_deformation}
   \lim_{\lambda \rightarrow 0}  \cL(w, G) =0,
\end{equation}
where $\cL$ is the Cherkis--Kapustin spectral curve \eqref{eq:intro_cherkis_kapustin}. That is, the difference equations quantise the Cherkis-Kapustin spectral curve. In the case of a GLSM that flows to an NLSM with target $X$, the difference equations are therefore related to a quantisation of the aforementioned action on $QH_T^\bullet (X)$. The situation is represented in Figure~\ref{fig:intro-twistor-sphere}.

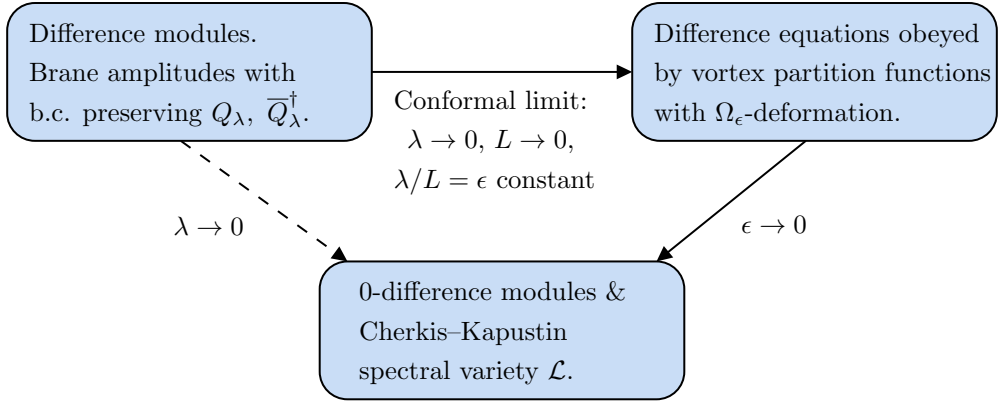
\begin{figure}
\centering

\tikzset{every picture/.style={line width=0.75pt}} 

\begin{tikzpicture}[x=0.65pt,y=0.65pt,yscale=-1,xscale=1]

\draw  [fill={rgb, 255:red, 74; green, 144; blue, 226 }  ,fill opacity=0.3 ] (40,56) .. controls (40,47.16) and (47.16,40) .. (56,40) -- (234,40) .. controls (242.84,40) and (250,47.16) .. (250,56) -- (250,104) .. controls (250,112.84) and (242.84,120) .. (234,120) -- (56,120) .. controls (47.16,120) and (40,112.84) .. (40,104) -- cycle ;
\draw    (250,80) -- (397,80) ;
\draw [shift={(400,80)}, rotate = 180] [fill={rgb, 255:red, 0; green, 0; blue, 0 }  ][line width=0.08]  [draw opacity=0] (8.93,-4.29) -- (0,0) -- (8.93,4.29) -- cycle    ;
\draw    (500,120) -- (416.33,188.11) ;
\draw [shift={(414,190)}, rotate = 320.86] [fill={rgb, 255:red, 0; green, 0; blue, 0 }  ][line width=0.08]  [draw opacity=0] (8.93,-4.29) -- (0,0) -- (8.93,4.29) -- cycle    ;
\draw  [dash pattern={on 4.5pt off 4.5pt}]  (140,120) -- (233.58,188.23) ;
\draw [shift={(236,190)}, rotate = 216.1] [fill={rgb, 255:red, 0; green, 0; blue, 0 }  ][line width=0.08]  [draw opacity=0] (8.93,-4.29) -- (0,0) -- (8.93,4.29) -- cycle    ;
\draw  [fill={rgb, 255:red, 74; green, 144; blue, 226 }  ,fill opacity=0.3 ] (400,56) .. controls (400,47.16) and (407.16,40) .. (416,40) -- (594,40) .. controls (602.84,40) and (610,47.16) .. (610,56) -- (610,104) .. controls (610,112.84) and (602.84,120) .. (594,120) -- (416,120) .. controls (407.16,120) and (400,112.84) .. (400,104) -- cycle ;
\draw  [fill={rgb, 255:red, 74; green, 144; blue, 226 }  ,fill opacity=0.3 ] (220,206) .. controls (220,197.16) and (227.16,190) .. (236,190) -- (414,190) .. controls (422.84,190) and (430,197.16) .. (430,206) -- (430,254) .. controls (430,262.84) and (422.84,270) .. (414,270) -- (236,270) .. controls (227.16,270) and (220,262.84) .. (220,254) -- cycle ;

\draw (51,50) node [anchor=north west][inner sep=0.75pt]   [align=left] {\small Difference modules.\\ \small Brane amplitudes with \\ \small b.c. preserving $\displaystyle Q_{\lambda } ,\ \overline{Q}_{\lambda }^{\dagger } .$};
\draw (411,50) node [anchor=north west][inner sep=0.75pt]   [align=left] {\small Difference equations obeyed \\ \small by vortex partition functions\\ \small with $\displaystyle \Omega _{\epsilon }$-deformation.};
\draw (241,200) node [anchor=north west][inner sep=0.75pt]   [align=left] { \small $0$-difference modules \& \\ \small Cherkis--Kapustin\\ \small spectral variety $\displaystyle \mathcal{L}$.};
\draw (134,162) node [anchor=north west][inner sep=0.75pt]    {\small $\lambda \rightarrow 0$};
\draw (261,89) node [anchor=north west][inner sep=0.75pt]   [align=left] 
{\small Conformal limit:\\
\small $\,\,\,\lambda \rightarrow 0,\, L \rightarrow 0$,\\
\small $\lambda /L = \epsilon \text{ constant}$};
\draw (461,162) node [anchor=north west][inner sep=0.75pt]    {\small$\epsilon \rightarrow 0$};
\end{tikzpicture}

\caption{The types of spectral data considered in Section~\ref{sec:spectral_data_1} and their relation via limits.}\label{fig:triangle_limits}
\end{figure}

\paragraph{Hemisphere partition functions \& difference equations.}

Although having novel difference equations satisfied by physical D-brane amplitudes is already satisfying, it is of course difficult to compute these observables and in particular to check our result beyond the free chiral case, as the brane amplitudes are not supersymmetric. Thus, in Section~\ref{sec:conf_vortex}, we move on to inspect a particular limiting case, investigated previously in \cite{Hori:2000ck, Cecotti:2013mba}. We take the \textit{conformal limit}: $\lambda \rightarrow 0$ and $L \rightarrow 0$, where $L$ is the length of the $S^1$ on which the theory is quantised, but take $\lambda/L \coloneqq \ep$ constant. In this limit, the amplitudes are expected to reduce to hemisphere, or vortex partition functions, which can be exactly computed via localisation \cite{Hori:2013ika, Honda:2013uca,Sugishita:2013jca, Fujimori:2015zaa}. The parameter $\ep$ is identified with the $\Omega$-deformation parameter in the vortex partition function, or the inverse radius for the hemisphere partition function.

Our arguments then imply that these partition functions must satisfy novel difference equations that themselves reduce in the $\ep\rightarrow 0$ limit to the Cherkis--Kapustin spectral curve. The relations are schematically illustrated in Figure~\ref{fig:triangle_limits}. In the conformal limit we are actually able to show a bit more than \eqref{eq:intro_difference_equation} and \eqref{eq:intro_deformation}, namely:
\begin{equation}\label{eq:intro_matrix_difference_vortex}
    \hat{p}\, \mathcal{Z}_D[\cO_a, m ]=  \widetilde{G}_{ab}(m,\ep) \mathcal{Z}_D[\cO_b, m ], \qquad
    \lim_{\ep \rightarrow 0 }  \widetilde{G} = F,
\end{equation}
where $\hat{p}$ is a difference operator which shifts the complex mass appearing in the hemisphere partition functions by $m \mapsto m+\ep$. Thus, the difference equations provide a quantisation of the action of the operators $p$, $w$ on $QH^\bullet_T(X)$. In particular, we obtain difference equations that quantise the spectral curve and are solved by hemisphere partition functions. Since we are in the calculable realm of supersymmetric partition functions, these assertions are eminently verifiable.

Pleasingly, as hemisphere and vortex partition functions admit direct calculation via localisation, this yields a recipe to construct solutions to difference equations which arise as quantisations of the spectral varieties. These objects naturally involve hypergeometric functions, and coincide with equivariant Gromov--Witten invariants of the Higgs branches \cite{Bonelli:2013mma}. We stress again that all of these constructions generalise to $n>1$. In particular, we present the examples of the free chiral ($X=\bC$), SQED$[2]$ ($X=\mathbb{CP}^1$) and in appendix \ref{appendix:sqed3}, SQED$[3]$ ($X=\mathbb{CP}^2$). The latter is of course an example for $n=2$.

\paragraph{Holomorphic filtrations and equivariant K--theory.}

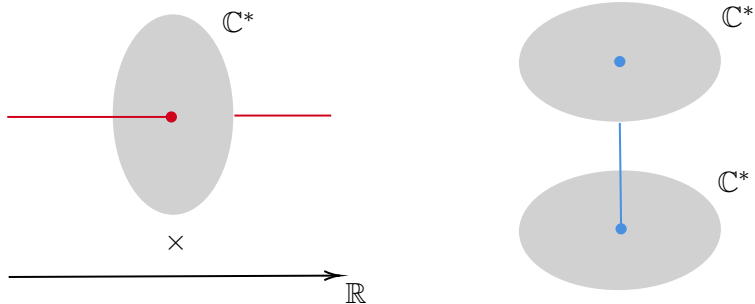
\begin{figure}
    \centering

\tikzset{every picture/.style={line width=0.75pt}} 

\begin{tikzpicture}[x=0.75pt,y=0.75pt,yscale=-0.7,xscale=0.7]

\draw  [draw opacity=0][fill={rgb, 255:red, 155; green, 155; blue, 155 }  ,fill opacity=0.46 ] (167.9,15.3) .. controls (191.64,15.1) and (211.04,47.18) .. (211.22,86.95) .. controls (211.4,126.72) and (192.29,159.12) .. (168.55,159.32) .. controls (144.81,159.52) and (125.41,127.44) .. (125.23,87.67) .. controls (125.05,47.9) and (144.15,15.49) .. (167.9,15.3) -- cycle ;
\draw    (51,204) -- (285,203.01) ;
\draw [shift={(287,203)}, rotate = 179.76] [color={rgb, 255:red, 0; green, 0; blue, 0 }  ][line width=0.75]    (10.93,-3.29) .. controls (6.95,-1.4) and (3.31,-0.3) .. (0,0) .. controls (3.31,0.3) and (6.95,1.4) .. (10.93,3.29)   ;
\draw [color={rgb, 255:red, 208; green, 2; blue, 27 }  ,draw opacity=1 ]   (50,89) -- (167,89) ;
\draw [color={rgb, 255:red, 208; green, 2; blue, 27 }  ,draw opacity=1 ]   (212,88) -- (281,88) ;
\draw  [draw opacity=0][fill={rgb, 255:red, 208; green, 2; blue, 27 }  ,fill opacity=1 ] (163,89) .. controls (163,86.79) and (164.79,85) .. (167,85) .. controls (169.21,85) and (171,86.79) .. (171,89) .. controls (171,91.21) and (169.21,93) .. (167,93) .. controls (164.79,93) and (163,91.21) .. (163,89) -- cycle ;
\draw  [draw opacity=0][fill={rgb, 255:red, 155; green, 155; blue, 155 }  ,fill opacity=0.46 ] (558.89,48.06) .. controls (559.43,71.8) and (527.63,91.65) .. (487.87,92.41) .. controls (448.1,93.17) and (415.43,74.54) .. (414.89,50.8) .. controls (414.34,27.06) and (446.14,7.21) .. (485.9,6.45) .. controls (525.67,5.69) and (558.34,24.32) .. (558.89,48.06) -- cycle ;
\draw  [draw opacity=0][fill={rgb, 255:red, 74; green, 144; blue, 226 }  ,fill opacity=1 ] (486.75,45.22) .. controls (488.96,45.19) and (490.78,46.95) .. (490.81,49.16) .. controls (490.84,51.37) and (489.08,53.19) .. (486.87,53.22) .. controls (484.66,53.25) and (482.85,51.48) .. (482.81,49.28) .. controls (482.78,47.07) and (484.55,45.25) .. (486.75,45.22) -- cycle ;
\draw  [draw opacity=0][fill={rgb, 255:red, 155; green, 155; blue, 155 }  ,fill opacity=0.46 ] (558.89,169.06) .. controls (559.43,192.8) and (527.63,212.65) .. (487.87,213.41) .. controls (448.1,214.17) and (415.43,195.54) .. (414.89,171.8) .. controls (414.34,148.06) and (446.14,128.21) .. (485.9,127.45) .. controls (525.67,126.69) and (558.34,145.32) .. (558.89,169.06) -- cycle ;
\draw [color={rgb, 255:red, 74; green, 144; blue, 226 }  ,draw opacity=1 ]   (486.81,93.22) -- (487.74,169.4) ;
\draw  [draw opacity=0][fill={rgb, 255:red, 74; green, 144; blue, 226 }  ,fill opacity=1 ] (487.68,165.4) .. controls (489.89,165.37) and (491.7,167.14) .. (491.74,169.35) .. controls (491.77,171.56) and (490,173.37) .. (487.79,173.4) .. controls (485.59,173.44) and (483.77,171.67) .. (483.74,169.46) .. controls (483.71,167.25) and (485.47,165.44) .. (487.68,165.4) -- cycle ;

\draw (160,170.4) node [anchor=north west][inner sep=0.75pt]    {$\times $};
\draw (201,11.4) node [anchor=north west][inner sep=0.75pt]    {$\mathbb{C}^{*}$};
\draw (289,206.4) node [anchor=north west][inner sep=0.75pt]    {$\mathbb{R}$};
\draw (557.15,7.33) node [anchor=north west][inner sep=0.75pt]  [rotate=-358.91]  {$\mathbb{C}^{*}$};
\draw (554.15,126.36) node [anchor=north west][inner sep=0.75pt]  [rotate=-358.91]  {$\mathbb{C}^{*}$};

\end{tikzpicture}

    \caption{The spectral data considered in Section~\ref{sec:spectral_data_2} for $\lambda=1$ (left) and its relation to equivariant K--theory (right). On the left, the red line supports sections of the vector bundles $\cE^{t_0}\rightarrow \mathbb{C}^*$ that decay in both $t_0\rightarrow \pm \infty$ directions (bound states). This is a twistor--like spectral data. The support of the line on $\mathbb{C}^*$ corresponds to the point where the sheets of the equivariant K--theory variety are glued together, as shown on the right.}
    \label{fig:intro-K-theory}
\end{figure}

Finally, in Section~\ref{sec:spectral_data_2} we proceed to describe the second algebraic description encoding aspects of the monopole solution that emerges when $|\lambda | = 1$, as well as its consequences. For simplicity, we fix $\lambda=1$, but the same reasoning goes through for $|\lambda|=1$. As mini--complex manifolds, we now have
\begin{equation}
 M^{\lambda=1} \cong \mathbb{R} \times \mathbb{C}^*
\end{equation}
where the real direction can be parameterised by an alternative local mini--complex coordinate $t_0=\Im(\bar{w})$, accompanied by a complex one $\beta_0$. (These can be thought of as alternative coordinates to $(t_1,\beta_1)$ on the same mini--complex manifold). We can then consider scattering along the real, non--compact line defined by $t_0$, as in Figure~\ref{fig:intro-K-theory}.

Sections of the holomorphic bundle $\mathcal{E}^{t_0}\rightarrow \mathbb{C}^*$ that decay at $t_0 \rightarrow \pm \infty$ can be filtered by their fall--off rate, which is determined by the twisted superpotential. Bound states (that is, sections that demonstrate exponential decay in both directions) can only be supported on a codimension--one locus $\Delta \in \mathbb{C}^*$. This locus is the equivalent of the Hitchin or twistor spectral curve for monopoles in $\mathbb{R}^3$ (see \cite{Cherkis:2007xs}). The last contribution of this paper is to show that in the case of an NLSM with target $X$, this locus is precisely the one where $\mathbb{C}^*$ sheets glue together to form the \emph{equivariant K--theory variety} of $X$, which in this simple, rank $1$ example (with $N=2$ vacua) takes the form:
\begin{equation}
    \mathrm{Spec} (K_T(X)) = \left( \mathbb{C}^* \sqcup \mathbb{C}^* \right) / \Delta 
\end{equation}
This constitutes a direct dimensional reduction of the physical construction of elliptic cohomology provided by one of the authors \cite{Bullimore:2021rnr}, as well as Dedushenko--Nekrasov \cite{Dedushenko:2021mds}.

We conclude this overview by remarking that the fact that \emph{two} algebraic descriptions of Berry connections related to \emph{two} different spectral curves is a remarkable physical incarnation of the announced Riemann--Hilbert correspondence of Kontsevich and Soibelman \cite{Kontsevich_Soibelman_1, Kontsevich_Soibelman_2}. To the best of our knowledge, the fact that in theories with a target space interpretation this results in a relation between two generalised cohomology theories has not previously appeared in the literature.

\subsection{Further Research}

It is convenient to already offer here some directions for future research. 

\paragraph{Two dimensions.}

Firstly, it is not clear to the authors whether \textit{all} generalised periodic monopoles of GCK--type can be engineered as the supersymmetric Berry connection of a 2d GLSM, and therefore can be interpreted in light of the results established in this work. It would be interesting to explore this further. 

It would also  be interesting to explore the action of T--duality, and the gauging of global symmetries (corresponding to the Nahm transform on the Berry connection), on the structures we have uncovered. The former would necessitate a deeper study of Landau--Ginzburg models in this framework.  On the side of the difference modules investigated in Section~\ref{sec:difference_modules}, it should prove enlightening to investigate D-brane amplitudes and vortex partition functions with insertions of twisted chiral ring elements in a different basis. Mathematically, there is a well--known and natural basis in the hierarchy of (quantum) equivariant cohomology, K--theory and elliptic cohomology known as the stable envelopes \cite{maulik2012quantum, Aganagic:2016jmx}. They have been investigated physically from a variety of perspectives in \cite{Bullimore:2021rnr, Dedushenko:2021mds, Ishtiaque:2023acr, Tamagni:2023wan, Haouzi:2023doo}. In our context, we expect that working in these bases will be convenient for the matrix difference equations we uncover. On the side of spectral data provided by holomorphic filtrations of vector bundles investigated in Section~\ref{sec:filtrations}, it is natural to ask how the constructions extend to non--GKM varieties. 

\paragraph{Three dimensions.} Finally, all of the structures should lift to interesting counterparts in 3d $\cN=2$ theories, from which our results should be obtainable via dimensional reduction. We aim to develop this further in upcoming work. Berry connections for 3d theories have been studied in \cite{Cecotti:2013mba, Bullimore:2021rnr, Dedushenko:2021mds, Dedushenko:2022pem}, and localisation techniques for the computation of partition functions in \cite{Yoshida:2014ssa, Dimofte:2017tpi, Bullimore:2018jlp}, which also have interesting enumerative interpretations.  Of course, in three dimensions there is also the added element of 3d mirror symmetry, manifesting mathematically as symplectic duality, to explore.

In particular, the so--called $A$-polynomial for 3d $\cN=2$ theories \cite{Dimofte:2011jd, Dimofte:2011py, Dimofte:2017tpi}, which is a Lagrangian submanifold of $(\bC^*)^{2 N}$, where $N$ is the number of flavour symmetries, corresponds to the analogue of the Cherkis--Kapustin spectral variety considered in this work. This is clear from their determination by the effective twisted superpotential. This gives an interpretation of this previously studied object in terms of the supersymmetric ground states of the 3d $\cN=2$ theory on $T^2$. The 3d lift of the main results of this paper would imply  a linear, \textit{matrix} $q$-difference equation for 3d brane amplitudes, which in the conformal limit imply a linear matrix $q$-difference equation for holomorphic blocks \cite{Beem:2012mb} or alternatively 3d vortex partition functions in an Omega background. By taking the determinant of this linear matrix difference equation, one would obtain a set of difference equations obeyed by each holomorphic block, resulting in the polynomial difference identities studied in \cite{Beem:2012mb}, and yielding a quantisation of the A--polynomial/Cherkis--Kapustin spectral variety. 

The polynomial difference equations studied in \textit{e.g.} \cite{Beem:2012mb} also have an interpretation as line operator identities obeyed by holomorphic blocks. Along these lines, in 2d, it would be interesting to derive our results from the setup considered in \cite{Bullimore:2016hdc}, \textit{i.e.} as Ward identities of monopole operators in 3d $\cN=4$ theories sandwiched between $(2,2)$ boundary conditions. 

Geometrically, the 3d hemisphere partition functions correspond \cite{Crew:2023tky, Dedushenko:2023qjq} to the vertex functions in enumerative geometry of \cite{Okounkov:2015spn}. The aforementioned $q$-difference equations are the qKZ equations for those theories with an integrable system interpretation, in the sense of Nekrasov--Shatashvili \cite{Nekrasov:2009uh, Nekrasov:2009ui}. Our difference equations in the 2d setting can interpreted as a dimensional reduction thereof. In \cite{Aganagic:2017smx}, a physical origin of these difference equations is provided via various compactifications of little string theory. Our results give another physical perspective, which is a purely two (or three) dimensional construction via the Berry connection, $tt^*$ geometry and D-branes of these theories. They therefore also apply to theories which are not obtainable via such compactifications. Further, it would be interesting to explore connections to geometric representation--theoretic considerations of the above references from our constructions. 

Finally, we also expect that the lift of the interpretation of the Riemann--Hilbert correspondence between constructions related to $QH^\bullet_T(X)$ and $K_T(X)$ will lift to one between the equivariant quantum K--theory $QK_T(X)$ and the equivariant elliptic cohomology $\mathrm{Ell}_T(X)$.


\section{Preliminaries}\label{sec:setup}

In this work, we will be interested in 2d $(2,2)$ gauged linear sigma models quantised on the Euclidean cylinder $\mathbb{R}\times S^1$. To set up our notation, we first consider such theories on the Minkowski $x^0-x^1$ plane. 
The supersymmetry algebra is then given by:
\begin{equation}\label{eq:init_supercharge_algebra}
\begin{split}
&\quad q_{+}^2 = q_{-}^2 = \bar{q}_{+}^2 = \bar{q}_{-}^2=0,\\
&\qquad\{q_{\pm}, \bar{q}_{\pm}\} = P_0 \pm P_1,\\
&\{\bar{q}_{+}, \bar{q}_{-}\} = Z,\quad \{q_{+}, q_{-}\}  = Z^*,\\
&\{q_{-}, \bar{q}_{+}\} = \tilde{Z},\quad \{q_{+}, \bar{q}_{-}\}  = \tilde{Z}^*
\end{split}
\end{equation}
where we are denoting complex conjugation by $*$. We will also generically be interested in the A--twist of such theories, and thus set chiral parameters to zero, so that $Z=0$. This allows us to preserve the vector R--symmetry $R_V$, which is necessary for the A--twist. The twisted central charge is given by
\begin{equation}
\tilde{Z}= -i(\sigma_1 + i \sigma_2) \cdot J_G
\end{equation}
where $(\sigma_1$, $\sigma_2)$ are real scalars, and $J_G$ is the generator of the gauge symmetry $G$. In the combination above they form the complex scalar in a $\mathcal{N}=(2,2)$ vector multiplet.

We will henceforth Wick--rotate to Euclidean space, setting $x^0 = -ix^2$, and compactify this direction to an $S^1$ of length $L$. Thus, we obtain coordinates $(x^1,x^2)$ on $\mathbb{R}\times S^1$, and
\begin{equation}\label{eq:ang_mom}
P_0 = D_{0} = \partial_{0}-i A_0 = i (\partial_{2}-i A_2 )= iP_2.
\end{equation}
We will use the complex coordinate $z= x^1+ix^2$, and define conjugation with respect to time evolution in the $x^1$ direction, so that the Hamiltonian is $H = P_1$. This implies~\cite{Dedushenko:2021mds}
\begin{equation}\label{conjugation}
q_{\pm}^{\dagger} = \pm \bar{q}_{\pm}\,.
\end{equation}
Thus, when we Wick--rotate we must replace $\tilde{Z}^* \mapsto -\tilde{Z}^*$ in \eqref{eq:init_supercharge_algebra}. From the point of view of the quantum mechanics, $P_2$ is a central charge. 

We will consider theories with an abelian flavour symmetry $T \cong U(1)^n$, for which turning on a generic (complex) twisted mass deformation $w_1+iw_2 \in \mathfrak{t}_{\mathbb{C}}$ leaves only isolated, topologically trivial vacua. The complex mass arises as a non--zero vacuum expectation value for a complex scalar in a background vector multiplet for $T$, and turning it on shifts $\tilde{Z}$ by an extra contribution $-i(w_1+i w_2) \cdot J_T$, where $J_T$ is the generator for $T$. Doing so breaks the axial $R_A$ symmetry. We may also turn on a flat connection, with holonomy $t$, for $T$ around the $S^1$, which simply shifts $A_2$ in \eqref{eq:ang_mom}. Thus, as a real manifold, the space of deformation parameters $T \times \mathfrak{t}_\mathbb{C}$ will be diffeomorphic to
\begin{equation}\label{eq:space-def}
    M := (S^1 \times \mathbb{R}^2 )^n,
\end{equation}
with the $S^1$ parametrising holonomies and $\mathbb{R}^2$ complex masses.

\subsection{A family of \texorpdfstring{$\cN=(2,2)$}{} quantum mechanics}

One of the main themes of this work is that by giving different representations of the ground states of a 2d $(2,2)$ model in terms of the ground states of a one--parameter family of 1d $(2,2)$ SQMs (or the cohomology of one of its supercharges), one obtains interesting relations between algebraic structures characterising Berry connections for these spaces of ground states. In the case of GLSMs which flow to NLSMs, the cohomologies can be related to the geometry of the target.

The one--parameter family that we are interested in will contain the following supercharges:
\begin{equation}
 Q_A := \bar{q}_+ + q_- ,\quad Q:=   \frac{1}{2}(\bar{q}_{+}+q_{-} + q_{+}+\bar{q}_{-} ).
\end{equation}
Here, $Q_A$ is the A--type supercharge, whose cohomology corresponds to the twisted chiral ring. Therefore, in a GLSM it encodes the quantum cohomology of the target. The supercharge $Q$ is the one obtained via dimensional reduction from the supercharge used in 3d in \cite{Bullimore:2021rnr, Dedushenko:2021mds}, whose cohomology can be described in terms of the equivariant elliptic cohomology $E_T(X)$ of the Higgs branch $X$ of the corresponding GLSM. As we shall see in Section~\ref{sec:ktheory}, in this 2d context we will recover from its cohomology the equivariant K--theory of the target. 

We now identify an appropriate one--parameter family of 1d $(2,2)$ SQMs. To do so, let us first consider a basis of supercharges of an SQM that contains the supercharge $Q_A$. This is
\begin{equation}\label{eq:supercharges_A_basis}
\begin{split}
Q_{A} &= \bar{q}_{+}+q_{-}, \quad \bar{Q}_A = q_{+}+\bar{q}_{-}, \\
Q_{A}^{\dagger} &= q_{+} - \bar{q}_{-},\quad \bar{Q}_A^{\dagger} = \bar{q}_{+}-q_{-},
\end{split}
\end{equation}
satisfying:
\begin{equation}
\begin{split}
&\{Q_A,Q_A^{\dagger}\}=\{\bar{Q}_A,\bar{Q}_A^{\dagger}\}= 2H,\\ 
&\qquad\quad \{Q_A,\bar{Q}_A\} = \tilde{Z}_t ,\\
&\quad Q_A^2 = \tilde{Z}_w, \quad \bar{Q}_A^2 = -\tilde{Z}_w^{*}.
\end{split}
\end{equation}
with all other anti--commutators vanishing. Here we are assuming we are acting on gauge--invariant states and operators (and thus have dropped $\sigma, A_2$ from this expression), and have introduced
\begin{equation}
\tilde{Z}_w = -iw \cdot J_T
\end{equation}
where $w=w_1 + iw_2$ is the complex mass. Moreover,
\begin{equation}
\tilde{Z}_t =   2 i \partial_2 + 2t \cdot J_T,
\end{equation}
where $t$ is the holonomy for $T$ around the circle.

With this basis as a starting point, we can introduce the following twistor family of supercharges~(\emph{cf.}~\cite{Gaiotto:2016hvd}, appendix B)
\begin{equation}\label{eq:lambda_supercharges}
\begin{split}
Q_{\lambda} = \frac{1}{\sqrt{1+|\lambda|^2}}(Q_A +\lambda \bar{Q}_A), \qquad \bar{Q}_{\lambda} = \frac{1}{\sqrt{1+|\lambda|^2}}(\bar{Q}_A - \bar{\lambda } Q_A), \\
Q_{\lambda}^{\dagger} = \frac{1}{\sqrt{1+|\lambda|^2}}(Q_A^{\dagger} +\bar{\lambda} \bar{Q}_A^{\dagger}), \qquad \bar{Q}_{\lambda}^{\dagger} = \frac{1}{\sqrt{1+|\lambda|^2}}(\bar{Q}_A^{\dagger} - \lambda  Q_A^{\dagger})
\end{split}
\end{equation}
depending on a parameter $\lambda$. This family satisfies
\begin{equation}\label{eq:modified_algebra}
\begin{split}
&\{Q_{\lambda}, Q_{\lambda}^{\dagger}\} = \{\bar{Q}_\lambda , \bar{Q}_\lambda^\dagger \} = 2H  \\
&\qquad\quad \{Q_{\lambda},\bar{Q}_{\lambda}\} =  \tilde{Z}_{t_0}\\
&\quad Q_{\lambda}^2 = \tilde{Z}_{\beta_0}, \quad \bar{Q}_{\lambda}^2 =  -\tilde{Z}^*_{\beta_0}.
\end{split}
\end{equation}
Here we have introduced
\begin{equation}
\begin{split}
    &\tilde{Z}_{t_0} = \frac{1-|\lambda|^2}{1+|\lambda|^2} (2i \partial_2) + 2 t_0\cdot J_T \\
    &\tilde{Z}_{\beta_0} =  \frac{2i \lambda \partial_2}{1+|\lambda|^2}  -i\beta_0 \cdot J_T
\end{split}
\end{equation}
with
\begin{equation}\label{eq:par-dep}
(t_0, \beta_0) = \frac{1}{1+|\lambda|^2} \left((1-|\lambda|^2)t+ 2 \text{Im}(\lambda \bar{w}),  w+\lambda^2 \bar{w}+2 i \lambda t ) \right).
\end{equation}
Notice that at any given $\lambda$, this corresponds to the algebra of a 1d $(2,2)$ SQM. We will be interested in the space of supersymmetric ground states, defined as follows. First, we require the vanishing of the central charges on the space of ground stats
\begin{equation}
\tilde{Z}_{t_0} = \tilde{Z}_{\beta_0} = 0.
\end{equation}
This will generically imply that ground states are uncharged under $J_T$, and have no KK modes along the circle. Then, on the states satisfying this condition, we can further impose
\begin{equation}
    \{Q_\lambda , Q_\lambda^\dagger \}= 2 H = 0.
\end{equation}
Consider fixed values of deformation parameters $(t,w)$ and denote the vector space of supersymmetric ground states by $E$ as a subspace of the states of the theory $S$. Whenever the system is gapped we have a cohomological description
\begin{equation}
    E \cong H^0( S|_{\tilde{Z}_{t_0} = \tilde{Z}_{\beta_0}=0}, Q_\lambda).
\end{equation}
As desired, $Q_\lambda$ interpolates between the supercharges $Q_A$ and $Q$
\begin{equation}
    Q_{\lambda=0} = Q_A,\quad Q_{\lambda=1} = Q .
\end{equation}

\subsection{Mini--complex structures}
\label{subsec:def-par}

In the $n=1$ case, the parameters $(t_0,\beta_0)$ precisely correspond to the first one--parameter family of coordinates on $S^1\times \mathbb{R}^2$ introduced by Mochizuki~\cite{mochizuki2017periodic} in his study of monopoles. As we now explain, they determine a so--called mini--complex structure on the parameter space of deformations $M=S^1 \times \mathbb{R}^2$ introduced in~\eqref{eq:space-def}. For $n>1$ each copy of $(S^1\times \mathbb{R}^2)$ will simply be endowed with the same mini--complex structure.

What is meant by a mini--complex structure on a three--manifold (\emph{c.f.}~\cite{mochizuki2017periodic}) is a maximal collection of local charts $U_\lambda : M \rightarrow \mathbb{R}\times \mathbb{C}$ for $M$ such that
\begin{itemize}
\item $M$ is covered by these charts.
\item transition functions $(F_t(t,w),F_w(t,w))$ are so that $\partial_t F_t>0$ (preserve the orientation of the real coordinate) and $F_w$ is constant in $t$ and holomorphic in $w$.
\end{itemize}
This ensures that one can define mini--holomorphic functions, which are smooth functions locally constant along $\mathbb{R}$ and holomorphic along $\mathbb{C}$.

The simplest case to understand is $\lambda=0$, so that $(t_0 , \beta_0) = (t,w)$. It will be useful to think of $t$ as an $\mathbb{R}$-valued variable first (by taking a lift and abusing notation slightly) and then take a quotient to enforce its periodicity. If $t$ is valued in $\mathbb{R}$, then clearly the pair $(t,w)$ are coordinates on $\mathbb{R}\times \mathbb{C}$, which is the prototypical mini--complex manifold. Therefore, so is
\begin{equation}\label{eq:mini-complex}
    M^{\lambda = 0} := (\mathbb{R} \times \mathbb{C}) / \Gamma
\end{equation}
where $\Gamma$ is the $\mathbb{Z}$-action 
\begin{equation}
    \Gamma_{\lambda = 0}: (t,w) \mapsto (t+L,w)
\end{equation}
because this action does not change the vector fields $\partial_{t}$ and $\partial_{\bar{w}}$. In~\eqref{eq:mini-complex} as well as below the superscript in $M^\lambda$ simply denotes the chosen mini--complex structure on $M$.

The same reasoning applies at different values of $\lambda$, but with an important qualitative difference. We again start from the obvious mini--complex manifold $\mathbb{R}\times \mathbb{C}$ now parameterised by $(t_0,\beta_0)$, but the quotient by the same $\mathbb{Z}$-action shifts $\beta_0$ as well
\begin{equation}
    \Gamma_\lambda :
	(t_0, \beta_0 ) \mapsto (t_0, \beta_0 ) + \frac{L}{1+|\lambda|^2} \left(1-|\lambda|^2 , 2i \lambda  \right).
\end{equation}
As a result, as a mini--complex manifold we in general no longer have $M^{\lambda} \cong S^1\times \mathbb{C}$. The most dramatic change happens at $|\lambda| = 1$, where the action of $\Gamma$ degenerates in the $t_0$ direction and we get
\begin{equation}
    M^{\lambda} \cong \mathbb{R}\times \mathbb{C}^*, \quad |\lambda|=1.
\end{equation}
For instance, at $\lambda =1 $ we have the coordinates 
\begin{equation}
    \frac{1}{2}(-\Im (\bar w ), 2\Re (w) +2it).
\end{equation}

Thus, we will qualitatively distinguish between two cases:
\begin{itemize}
    \item The first case, also known as the product case, is characterised by $\lambda = 0$ so that $\Gamma \subset \mathbb{R} \times \{0\} \subset (\mathbb{R}_{t_0} , \mathbb{C}_{\beta_0} ) $. There is an isometry $M^{\lambda = 0 } \cong S^1_L \times \mathbb{C}$.
    \item The second case, also known as the non--product case, is characterised by $\lambda \neq 0 $ so that the $\Gamma$ action shifts the $\beta_0$ coordinate by $2i\lambda L/(1+|\lambda|^2)$.
\end{itemize}
Within the non--product case, the locus $|\lambda|=1$ is precisely the locus at which we will discuss the emergence of two kinds of algebraic for monopoles on $M^\lambda$, which are expected to be related to a Riemann--Hilbert correspondence.

Finally, we remark that although the explicit expressions for $(t_0, \beta_0)$ might appear a little contrived at first sight, these can be illuminated by considering $S^1 \times \mathbb{R} $ as a quotient of $\mathbb{R}^4$ by the group $\mathbb{R} \times \mathbb{Z} $ acting on the first and second copy of $\mathbb{R}$ by translations and shifts by $L$, respectively~\cite{mochizuki2017periodic}. The manifold $\mathbb{R}^4\cong \mathbb{C}^2\times \mathbb{C}^2$ is hyperk\"ahler and therefore enjoys a $\mathbb{P}^1$-family of complex structures with holomorphic coordinates. What we have here is a dimensional reduction of this more familiar setup.\footnote{In greater detail, on $\bR^4$ we can take complex coordinates
\begin{equation}
	(\xi , \eta ) =  (z + \lambda \bar{w}, w-\lambda \bar{z}),
\end{equation}
where $(z,w)$ are the standard complex parameters on $\mathbb{C}^2$. The parameters $(t_0, \beta_0)$ can be obtained from a suitable combination of $(\xi , \eta)$ that behaves conveniently with the quotient. Set
\begin{equation}
	(\alpha_0 , \beta_0 ) = \frac{1}{1+|\lambda|^2} (\xi - \bar{\lambda} \eta , \eta + \lambda \xi),
\end{equation}
so that by construction the transitive $\mathbb{R}$-action on the first factor of $\mathbb{R}^1\times \mathbb{R}^3$ only shifts $\alpha_0$. Then we have a family of mini--complex coordinate systems on $\mathbb{R}^3 \cong (\mathbb{R}_t \times \mathbb{C}_w)$, denoted $(\mathbb{R}_t \times \mathbb{C}_w)^\lambda$ that agree with the ones we obtained above: $(t_0, \beta_0) 	= (\Im (\alpha_0), \beta_0 )$.
}

Mochizuki also introduces a second set of mini--complex coordinates that are closely related to $(t_0,\beta_0)$. For later convenience, we report them here
\begin{equation}\label{eq:mochizukisecondcoords}
(t_1,\beta_1) = (t_0 + \text{Im}(\bar{\lambda}\beta_0),  (1+|\lambda|^2)\beta_0) = (t+ \text{Im}(\lambda\bar{w}), w + 2 i \lambda t + \lambda^2 \bar{w}).
\end{equation}
One has:
\begin{equation}
\begin{aligned}
\partial_{t_1} &= \partial_{t_0} \\
\partial_{\bar{\beta_1}} &= \frac{\lambda}{1+|\lambda|^2} \frac{1}{2i} \partial_{t_0} + \frac{1}{1+|\lambda|^2} \partial_{\bar{\beta}_0}.
\end{aligned}
\end{equation}
It is easy to see that the map $(t_1(t_0,\beta_0),t_1(t_0,\beta_0))$ is a mini--holomorphic transition function as defined above. In particular, functions are mini--holomorphic in these mini--complex coordinates if and only if they are in the other. This is therefore simply an additional coordinate system on the same mini--complex manifold $M^{\lambda}$. However, these mini--complex coordinates have the additional feature that the induced action of $\Gamma$ on the respective $\mathbb{R}\times \mathbb{C}$ does not degenerate at $|\lambda|=1$. We will rely upon this fact later on.


\section{Berry Connections \& Asymptotics}

In this section, we introduce the vector bundle of supersymmetric ground states
\begin{equation}
    E \rightarrow M
\end{equation}
over the space of deformation parameters $M$. We first review the features of $tt^*$ geometry for a twisted mass deformation that we utilise throughout the remainder of this work. In particular, we review the fact that Berry connections and the adjoint Higgs field satisfy the Bogomolny equations. We also identify the asymptotic behaviour of the Berry connection and the adjoint Higgs field with the values of the twisted central charge in each of the massive vacua of the theory, which is computed for the GLSMs of our interest in terms of the effective twisted superpotential.

For simplicity, we will henceforth take the circle on which the theory is quantised to have length $L=1$, re--introducing it where necessary. We will work with coordinates $x=(t,w)$ and define $x_i =(t_i, w_i)$, $i=1,\ldots,n$.

\subsection{\texorpdfstring{$tt^*$ }{}geometry}

For theories with $N$ vacua, there is a $U(N)$ Berry connection on the rank-$N$ vector bundle of supersymmetric ground states over the parameter space $M$. This is the Berry connection for a twisted mass deformation and accompanying holonomy, as studied in \cite{Cecotti:2013mba}. The connection itself is defined in the usual way, where if $\ket{\alpha(x)}$ denotes an orthonormal basis of ground states at parameter value $x$
\begin{equation}
    \frac{\partial}{\partial x_{i}} \ket{\alpha(x)} = (A_{i})_{\alpha}{}^{\beta} \ket{\beta(x)}.
\end{equation}
The $tt^*$ equations obeyed by the Berry connection are then~\cite{Cecotti:1991me}
\begin{gather}
\left[\bar{D}_i,C_j\right] = 0 = \left[D_i,\bar{C}_j\right]\label{eq:tt*equation_1} \\
\,\left[D_i,D_j\right] = 0 = \left[\bar{D}_i, \bar{D}_j\right] \\
    \left[D_i,C_j\right] = \left[D_j,  C_i\right], \quad \left[\bar{D}_i, \bar{C}_j\right] = \left[\bar{D}_j, \bar{C}_i\right] \label{eq:tt*equation_3}  \\
\left[D_i,\bar{D}_j\right] = -\left[C_i,\bar{C}_j\right] \label{eq:tt*equation_4} 
\end{gather}
where
\begin{equation}
    D_i = \partial_{w_i}-A_{w_i},\quad \bar{D}_i = \bar{\partial}_{\bar{w}_i}-\bar{A}_{\bar{w}_i},
\end{equation}
but now the structure constants of the (twisted chiral) ring are replaced by~\cite{Cecotti:2013mba}
\begin{equation}
C_i = D_{t_i} - i\phi_i,\qquad \bar{C}_i= -D_{t_i} - i\phi_i,
\end{equation}
where $\phi_i$ is an anti--Hermitian adjoint Higgs field.

For the case of a rank one flavour symmetry, the $tt^*$ equations may be equivalently written as:
\begin{equation}
\begin{aligned}
&[\bar{D}_{\bar{w}},D_t - i\phi ]  = 0 = [D_{w},D_t + i\phi ] \\
&\qquad\quad 2 [D_w, \bar{D}_{\bar{w}}] = i [D_t,\phi]\,,\label{bogomolny}
\end{aligned}
\end{equation}
which are simply the Bogomolny equations on $\mathbb{R}^2\times S^1$
\begin{equation}\label{eq:bogomolny}
	F(D) = \star D \phi
\end{equation}
where $F(D)$ is the curvature of the connection. That the Berry connection satisfies Bogomolny equations was also derived as a consequence of a theory reducing to a 1d $\cN=(2,2)$ quantum mechanics in \cite{Pedder:2007ff,Sonner:2008be,Sonner:2008fi,Wong:2015cnt}. The higher--rank equations can be viewed as generalised Bogomolny equations on $(\mathbb{R}^2\times S^1)^n$. 

In summary, the structure of the supersymmetric ground states over a twisted mass deformation and accompanying holonomy can be packaged into a tuple $(E,h,D ,\phi)$ consisting of a vector bundle $E$ of ground states, with Hermitian metric $h$ determined by the inner product, a connection $D$ that is unitary with respect to $h$ and an anti--Hermitian endomorphism $\phi$ of $(E,h)$. The tuple satisfies the Bogomolny equation \eqref{eq:bogomolny}. For $n=1$ the solutions may literally be regarded as periodic monopoles over $\mathbb{R}^{3}$.

\subsection{Asymptotics} 

In this work, we will be concerned only with periodic monopoles of  generalised Cherkis--Kapustin (from now on, GCK) type \cite{Cherkis:2000cj, Cherkis:2000ft}, as coined in \cite{mochizuki2017periodic}. It is shown therein that they are in one--to--one correspondence with certain difference modules, which will play a central role in our work. 

A monopole is of GCK--type if it has Dirac--type singularities at a discrete finite subset $Z\subset M$ and satisfies the following conditions
\begin{equation}\label{eq:GCK_condition}
| \phi_x| = O(\log (d(x,x_0))),~|F(D)_x|\rightarrow 0
\end{equation}
for some reference point $x_0$ as $x$ goes to infinite distance. 

Such conditions are satisfied for the basic Dirac monopole \cite{Cherkis:2000cj} of charge $k$, for which the Higgs field $\phi$ satisfies the Laplace equation, and
\begin{gather}\label{eq:dirac_monopole_asymptotics}
     \phi =   -i c_1 -  \frac{ i \gamma k }{2} - \frac{i k }{2} \sum_{n \in \bZ}'\left[ \frac{1}{\sqrt{|w|^2+(t-n)^2}} - \frac{1}{|n|}\right] \rightarrow -i c_1 - i  k\log \left|\frac{iw}{2}\right|+o(1),\\
     A_t \rightarrow  i c_2 + i k \arg \left(\frac{iw}{2}\right) + o(1),  \quad A_w \rightarrow \frac{b}{w} + o(1/w)\nonumber
\end{gather}
as $|w| \rightarrow \infty$. Here $\gamma$ is the Euler constant, and the prime on the sum means that for $n=0$ the second term in the summand is omitted. Further, $b,c_1$ and $c_2$ are real constants. For $b = -\frac{1}{4}$ and $ c_1 = c_2 = 0$ this is the Berry connection for the free $\mathcal{N}=(2,2)$ chiral \cite{Cecotti:2013mba}, over the parameter space for its $U(1)$ flavour symmetry.

The GCK conditions \eqref{eq:GCK_condition} are further satisfied for the Berry connections of the theories we consider in this work. This is because we have made the assumption that as $w \rightarrow \infty$ in a generic direction in $\mathfrak{t}_{\bC} \cong \bC^n$, the theory is fully gapped with massive topologically trivial vacua. Further, the theory will fail to be gapped only at a discrete finite subset of points in the parameter space, which corresponds to $Z$ above. Thus, asymptotically in $w$, the $U(N)$ vector bundle $E$ splits (as a real vector bundle) into a direct sum of $U(1)$ bundles $\bigoplus_{\alpha} E_{\alpha}$, each corresponding to a decoupled sector for the effective theory of massive chiral multiplets parametrising perturbations around a massive vacuum. It follows that the solution is asymptotically gauge-equivalent to an abelian solution $(A^{\alpha}, \phi^{\alpha})$ of Dirac monopole solutions with particular moduli determined by the theory. 

The asymptotic value of $A_{t_i}^{\alpha} + i \phi_i^{\alpha}$ is determined by the dependence of the twisted central charge $\tilde{\mathcal{Z}}_{\alpha}$ in the vacuum $\alpha$ on $w_i$. For a GLSM, this can be evaluated as the effective (twisted) superpotential $W_{\text{eff}}$, which is the twisted superpotential appearing in the low--energy theory in the Coulomb branch~\cite{Witten:1993yc}, in the vacuum $\alpha$~
\begin{equation}\label{eq:asymptotics}
    A_{t_i}^{\alpha} + i \phi_i^{\alpha}\sim -2 i \partial_{w_i} W_{\text{eff}}^{(\alpha)} \quad \text{as} \quad { |w| \rightarrow \infty.}
\end{equation}
This was demonstrated for LG theories in \cite{Cecotti:2013mba}, where $W_{\text{eff}}^{(\alpha)}$ can be evaluated as just the usual superpotential at the critical point $\alpha$. The statement for GLSMs is simply the mirror dual of this. This was also demonstrated in for $\mathcal{N}=(2,2)$ supersymmetric quantum mechanics where the twisted central charge is given by VEVs of the moment map for the corresponding flavour symmetry~\cite{Sonner:2008fi}. In 2d, these receive quantum corrections, and hence one must consider the effective (twisted) superpotential. 

\subsubsection{Example: free chiral} For the free chiral, one has an effective twisted superpotential
\begin{equation}\label{eq:superpotential_chiral}
    W_{\text{eff}} = m \left(\log \frac{m}{\mu} -1\right)
\end{equation}
in the unique vacuum, where $ m  = iw/2$, and $\mu$ is the RG scale. Thus
\begin{equation}
    -2i \partial_w W_{\text{eff}} = \partial_m W_{\text{eff}} =  \log( m/\mu),
\end{equation}
matching the asymptotics of the charge--one Dirac monopole solution \eqref{eq:dirac_monopole_asymptotics}.

\subsubsection{Example: supersymmetric QED \& $\mathbb{CP}^1$ $\sigma$-model}\label{subsubsec:sqed2_asymptotics}

As a running example throughout this work, we will take supersymmetric QED with two chiral multiplets, which engineers the $\mathbb{CP}^1$ $\sigma$-model in the IR. This is a $U(1)$ GLSM with two chiral multiplets $\Phi_1$, $\Phi_2$ of charges $(+1,+1)$ and $(+1,-1)$ under $G\times T$, where $G$ is the gauge group and $T$ the flavour symmetry. We will turn on a mass $m = i w /2$ for $T$, and study the Berry connection over $m$ and the associated holonomy $t$. It is a smooth $SU(2)$ monopole solution \cite{Cecotti:2013mba}.

The effective twisted superpotential of the theory is given by
\begin{equation}\label{eq:w_eff_sqed2}
W_{\text{eff}} = -2 \pi i \tau(\mu)  \sigma + (\sigma + m)\left(  \log  \left(\frac{\sigma + m}{\mu}\right)  -1 \right) +(\sigma - m)\left(  \log  \left(\frac{\sigma - m}{\mu}\right)  -1 \right)
\end{equation}
Here $\tau(\mu)$ is the renormalised complex FI parameter
\begin{equation}
    \tau(\mu)= \tau_0 + \frac{2}{2\pi i }\log(\Lambda_0/\mu),
\end{equation}
where $\Lambda_0$ is some fixed UV energy scale, $\mu$ is the RG scale and $\tau_0$ is the bare complexified FI parameter
\begin{equation}
    \tau_0 = \frac{\theta}{2\pi} + i r_0.
\end{equation}
where $\theta$ is the instanton angle and $r$ is the bare real FI parameter. For convenience, we will define the renormalisation invariant quantity
\begin{equation}
    q=   \Lambda_0^{\,2} \,  e^{2 \pi i\tau_0}.
\end{equation}

The vacuum (Bethe) equations are 
\begin{equation}
1 = e^{\frac{\partial W}{\partial \sigma}} = q^{-1}(\sigma+m)(\sigma-m),
\end{equation}
which yield solutions $\sigma = \pm\sqrt{m^2 + q}$ corresponding to the two vacua. These can be substituted into $2 i \partial_w W_{\text{eff}} = \partial_m W_{\text{eff}}$ to obtain:
\begin{equation}\label{eq:w_eff_sqed2_derivatives}
\begin{aligned}
\frac{\partial W_{\text{eff}}}{\partial m} &= \log\left(\frac{\sigma+m}{\sigma-m}\right)  = \log\left( \frac{\pm m +\sqrt{m^2 + q} }{\mp m + \sqrt{m^2 + q} }\right).
\end{aligned}
\end{equation}
Labelling the vacua as above requires a choice of branch cut for the square root, so we will find it instructive to instead label the vacua by $\alpha = 1,2$ where
\begin{equation}
\text{As } q \rightarrow 0: \qquad
\begin{aligned}
    (1): \quad \sigma &=+ m + \frac{q}{2m}+ O\left(\frac{1}{m^2}\right),\\
    (2): \quad \sigma &= -m - \frac{q}{2m}+ O\left(\frac{1}{m^2}\right).
\end{aligned}
\end{equation}
These correspond to the two fixed points of the $\mathbb{CP}^1$ sigma model, which to leading order in $q$ are the classical values $\sigma$ must take in order for $\Phi_1$ or $\Phi_2$ to acquire a VEV.

We therefore have:
\begin{equation}
\text{As } |m| \rightarrow \infty: \qquad
\begin{aligned}
    \partial_m W_{\text{eff}}^{(1)} &\rightarrow +2\log(2m /\Lambda_0 )-2\pi i \tau_0+O(m^{-2}),\\
    \partial_m W_{\text{eff}}^{(2)}&\rightarrow -2\log(2m /\Lambda_0)+2\pi i \tau_0+O(m^{-2}).\\
\end{aligned}
\end{equation}
The asymptotics are those of two Dirac monopoles with charges $k= \pm2$ and $c_1+ic_2 = \mp 2\pi i (\tau_0 + \frac{2}{2\pi i } \log \Lambda_0)$, where the argument of the log is now twice $m$. This is no surprise because, for large mass, the vacua of the theory correspond to the north $(1)$ and south $(2)$ pole of the $\mathbb{CP}^1$. The effective theory in the neighbourhood of either vacuum is that of a chiral of effective mass $\pm 2m$ and charge $\pm 2$ parametrising the tangent spaces $T_{(1)}\mathbb{CP}^1$ and $T_{(2)}\mathbb{CP}^1$.


\section{Spectral Data I -- Difference Modules}\label{sec:spectral_data_1}

In this section we explain how one can extract certain difference modules from the bundles of supersymmetric ground states in the cohomology of $Q_\lambda$. In the case of a $U(1)$ flavour symmetry ($n=1$), these modules correspond those that Mochizuki~\cite{mochizuki2017periodic} used to construct a Hitchin--Kobayashi correspondence between GCK monopole solutions and so--called polystable, parabolic and filtered difference modules. In this work, we mainly ignore the additional structures (represented by the additional adjectives) that make the correspondence work. Instead, we focus on difference modules that can be obtained from a solution of the (generalised) Bogomolny equations for all $n \geq 1$, relate them to the physical data of the theories, and draw interesting consequences.

As we reviewed in Section~\ref{subsec:def-par}, the parameter space of deformations $M^\lambda$ is endowed with a certain mini--complex structure parameterised by $\lambda$. In Section~\ref{sec:mini-hol} we show that the vector bundle of supersymmetric ground states $E$ 
\begin{equation}
    E \rightarrow M^\lambda,
\end{equation}
viewed as states in the cohomology of $Q_\lambda$, acquires what is called a mini--holomorphic structure as well as a collection of complexified flat connections. This will be the main starting point to define the difference modules, which we do in subsequent sections. We treat the cases $\lambda = 0$ and $\lambda \neq 0$ in Sections~\ref{sec:product_case}~and~\ref{sec:difference_modules} respectively, where we also relate the modules to the Cherkis--Kapustin spectral varieties and quantisations thereof. The quantisations are represented by difference equations satisfied by brane amplitudes. Since brane amplitudes are not easy to explicitly write down, in Section~\ref{sec:conf_vortex} we inspect the so--called conformal limit and derive difference equations for exactly calculable hemisphere partition functions, which arise from the brane amplitudes in this limit. Throughout the section we will comment on the meaning of these structures in the case of a GLSM that flows to a K\"ahler target $X$: difference modules will be related to the quantum equivariant cohomology of $X$ ($\lambda=0$) and a quantisation thereof ($\lambda \neq 0$).

\subsection{Mini--holomorphic vector bundles and complexified flat connections}\label{sec:mini-hol}

One fundamental question for the bundle of supersymmetric ground states is how the supercharges behave with respect to changes in deformation parameters. In the language of~\cite{Gaiotto:2016hvd}, $Q_{\lambda}$ is a B--type supercharge with respect to $\beta_0$ and A--type with respect to $t_0$. This means that the supercharges $Q_{\lambda}$ have the following explicit dependencies
\begin{equation}\label{eq:supercharge_dependencies}
\partial_{\bar{\beta}_{0},i}Q_{\lambda} = 0,\quad 
\partial_{t_{0},i} Q_{\lambda} - i[\phi_i,Q_{\lambda}] = 0,
\end{equation}
where $\phi_i$ is the Higgs field. The above follow from the dependencies of $Q_A, \bar{Q}_A$ on $(w,t)$, which are simply the above equations evaluated at $\lambda=1$ and $\lambda = \infty$. This is equivalent to saying that for the A--supercharge basis \eqref{eq:supercharges_A_basis}, the twisted mass deformation is of BAA--type.\footnote{See appendix B.2 of \cite{Gaiotto:2016hvd} for further details. Concretely, $Q_1,Q_2$ there are mapped to $Q_A$ and $\bar{Q}_A$ here.}

The first equation in \eqref{eq:supercharge_dependencies} implies that the anti--holomorphic derivatives commute with the supercharge and thus descend to a holomorphic Berry connection  $\partial_{E,\bar{\beta_0}}$ on supersymmetric ground states, with components
\begin{equation}
    \partial^{(i)}_{E,\bar{\beta}_{0}}  := \frac{1}{1+|\lambda|^2} \left( \lambda i D_{t_i} +\lambda^2 D_{w_i} +  \bar{D}_{\bar{w}_i}\right).   
\end{equation}
At a fixed value $t^*_0$ of the real parameter, this operator endows the bundle $E|_{t_0 = t_0^* }$ with the structure of a holomorphic vector bundle. 

The second equation implies the existence of complexified flat connections
\begin{equation}
    \partial^{(i)}_{E, t_{0}} := D_{t_{0,i}}  - i\phi_i
\end{equation}
for supersymmetric ground states, where
\begin{equation}
    D_{t_{0,i}} = \frac{1}{1+|\lambda|^2} \left( (1-|\lambda|^2)D_{t_i} -2 i \lambda D_{w_i} + 2 i \bar{\lambda} \bar{D}_{\bar{w}_i}\right),
\end{equation}
which also commutes with $Q_{\lambda}$. The $tt^*$ equations~\eqref{eq:tt*equation_1}--\eqref{eq:tt*equation_4} imply that these operators commute
\begin{equation}\label{eq:commutator}
\begin{aligned}
\left[\partial^{(i)}_{E, t_{0}} , \partial^{(j)}_{E, t_{0}}  \right] &= \left[ \partial^{(i)}_{E,\bar{\beta}_{0}} , \partial^{(j)}_{E,\bar{\beta}_{0}} \right] = 0,\\
\qquad\quad\left[\partial^{(i)}_{E, t_{0}} , \partial^{(j)}_{E,\bar{\beta}_{0}} \right] &= 0.
\end{aligned}
\end{equation}
These equations imply that the bundle has a mini--holomorphic structure in the sense of~\cite{mochizuki2017periodic}.

Identical statements hold if we work with the mini--complex coordinates $(t_1,\beta_1)$ for $M^{\lambda}$ instead. More concretely, $Q_\lambda$ is still B--type with respect to $\beta_1$ and A--type with respect to $t_1$, as can be easily checked. Moreover, the operators
\begin{equation}\label{eq:mochizukioperators}
\begin{aligned}
&\partial^{(i)}_{E,t_{1}} = \partial^{(i)}_{E,t_{0}} = D_{t_{0,i}} - i \phi,\\
&\partial^{(i)}_{E,\bar{\beta}_{1}} = \frac{\lambda}{1+|\lambda|^2} \frac{1}{2i} \partial^{(i)}_{E, t_{0}} + \frac{1}{1+|\lambda|^2} \partial^{(i)}_{E,\bar{\beta}_{0}}.
\end{aligned}
\end{equation}
commute due to (\ref{eq:commutator})
\begin{equation}\label{eq:commutator2}
\begin{aligned}
\left[\partial^{(i)}_{E, t_{1}} , \partial^{(j)}_{E, t_{1}}  \right] &= \left[ \partial^{(i)}_{E,\bar{\beta}_{1}} , \partial^{(j)}_{E,\bar{\beta}_{1}} \right] = 0,\\
\left[\partial^{(i)}_{E, t_{1}} , \partial^{(j)}_{E,\bar{\beta}_{1}} \right] &= 0.
\end{aligned}
\end{equation}
Thus, ground states are annihilated by the operators $\partial^{(i)}_{E,t_0}$ and $\partial^{(i)}_{E,\beta_0}$ if and only if they are annihilated by $\partial^{(i)}_{E,t_1}$ and $\partial^{(i)}_{E,\beta_1}$. Further, all of these operators are well--defined on $Q_{\lambda}$-cohomology.

To make contact with the difference module constructions, we will work with the local coordinates $(t_1,\beta_1)$.

\subsection{\texorpdfstring{$\lambda=0$}{}: product case \& Cherkis--Kapustin spectral variety}\label{sec:product_case}

We now consider the product ($\lambda=0$) case and explain how we can obtain certain $0$-difference modules from the space of supersymmetric ground states in $Q_A$-cohomology. Given that $\lambda=0$, we will be working with the respective mini--complex coordinates $(t,w)$. For expository purposes, we first set $n=1$, and return to $n>1$ in due course.

More precisely, what we shall obtain in the $n=1$ case is a $0$-difference $\mathbb{C}(w)$-module. This is a finite--dimensional $\mathbb{C}(w)$-module $V$ together a $\mathbb{C}(w)$-linear automorphism
\begin{equation}
  F : V\rightarrow V.
\end{equation}
Here, $\mathbb{C}(w)$ denotes the field of rational functions of $w$. The construction of this module forms the initial part of the remarkable Hitchin--Kobayashi correspondence established by Mochizuki~\cite{mochizuki2017periodic}, which we can approximately state as follows (see Corollary 9.1.4 in \emph{loc. cit.}). There is a bijective correspondence between isomorphism classes of: 
\begin{itemize}
    \item Periodic monopoles of GCK--type on $S_L^1\times \mathbb{C}$.
    \item Polystable, parabolic, filtered $0$-difference modules.
\end{itemize}
In particular, the modules arising from monopole solutions of GCK--type have additional properties --polystability, a parabolic structure, and filtrations-- that turn out to be sufficient to establish a bijective correspondence between isomorphism classes. The roles of these properties are:
\begin{itemize}
    \item Polystability: this is an instance of replacements of real equations with stability conditions of algebraic objects as is usual in Hitchin--Kobayashi correspondences.
    \item Parabolic structure: this encodes the information concerning the allowed Dirac singularities at finite distance in $\mathbb{C}_w$.
    \item Filtrations: these represent the allowed behaviour at infinity in $\mathbb{C}_w$.
\end{itemize}
In the following, we will touch upon these properties only superficially, as they are not crucial to our discussion. It is however relevant to keep in mind the remarkable fact that \emph{monopole solutions, and therefore Berry connections, can in principle be uniquely parameterised by such modules.}

For simplicity, we start by discussing the case of smooth monopoles. Consider the differential operators \eqref{eq:mochizukioperators} at $\lambda=0$:
\bea
	\partial_{E,t} &:= D_t - i\phi, \\
	\partial_{E,\bar{w}} &:= \bar{D}_{\bar{w}}.
\eea
As we mentioned in the previous section, for each  $0 \leq t^* \leq L$ we can define a holomorphic vector bundle $\mathcal{E}^{t^*}$ on $\{t^*\}\times \mathbb{C}_w$ determined by the operator $\partial_{E,\bar{w}}$, which we will denote as follows
\begin{equation}
    \cE^{t^*} := (E|_{\{t^*\} \times\bC_{w}}, \partial_{E,\bar{w}} ).
\end{equation}
By definition, the holomorphic sections $\ket{\alpha}$ of this bundle satisfy
\begin{equation}
 \partial_{E,\bar{w}}\ket{\alpha (t=t^*,w)} = 0.
\end{equation}
Our first approximation for the module will be the space of holomorphic sections at $t=0$, that is
\begin{equation}\label{eq:naive-mod}
    V := H^0 (\mathbb{C}_w,\mathcal{E}^0) \otimes_{\mathbb{C}[w]} \mathbb{C}(w).
\end{equation}
Notice that above we emphasise the parametrisation by $w$ of $\mathbb{C}$, by writing $\mathbb{C}_w$.

Our next goal is to define an automorphism of this sheaf of holomorphic sections that determines a $0$-difference $\mathbb{C}(w)$-module structure. As usual in the context of a Hitchin--Kobayashi correspondence, the strategy to assign algebraic data to solutions of the $tt^*$ equations~\eqref{eq:tt*equation_1}-\eqref{eq:tt*equation_4} is to first consider equations~\eqref{eq:tt*equation_1}-\eqref{eq:tt*equation_3} (thus neglecting the real, `D--term' equation~\eqref{eq:tt*equation_4}) for the complexified gauge group $GL(N,\mathbb{C})$. The remaining equations imply the key, complex equation
\begin{equation}\label{eq:comm-bog}
	[\partial_{E,t} , \partial_{E,\bar{w}}] = 0.
\end{equation}
This equation is the only one we consider. As rigorously demonstrated (for $n=1$) by Mochizuki~\cite{mochizuki2017periodic}, the real equation~\eqref{eq:tt*equation_4} can be traded by polystability conditions imposed on solutions to~\eqref{eq:comm-bog}.

The complex equation~\eqref{eq:comm-bog} implies the parallel transport of a state $\ket{\alpha}$ along the $S^1_L$ direction can be performed whilst preserving holomorphicity of the section. In particular, by parallel transporting around the full circle (from $0$ to $L$) we can define the desired $\mathbb{C}(w)$-automorphism of $V$
\begin{equation}
    F: V \rightarrow V,
\end{equation}
In other words, the automorphism comes from considering the holonomy of the connection $\partial_{E,t}$ around $S^1_L$. In local coordinates, we may write
\begin{equation}\label{eq:physical-F-def}
   F(w) = \exp{\oint_{S^1_L} dt \left(A_{t}+i\phi\right)}.
\end{equation}
Note that $F(w)$ is holomorphic in $w$, as follows from equation \eqref{eq:comm-bog}, and so it defines (in particular) a $\mathbb{C}(w)$-module structure.

The pair $( V,F)$ is the first candidate for our $0$-difference module. It however suffers from the deficiency that it is not an \emph{acceptable}, well--defined algebraic object. In particular, we have not yet prescribed the behaviour at large $w$; this makes these objects transcendental rather than algebraic. As demonstrated in~\cite{mochizuki2017periodic}, this deficiency can be amended by considering compactifications of $\mathbb{C}_w$ and filtering the space of section by their growth at infinity.\footnote{
To prescribe the behaviour at infinity compatible with the GCK condition, one can consider a compactification $\mathbb{R}_t \times \mathbb{P}^1_w$ of $\mathbb{R}_t \times \mathbb{C}_w$. The $\Gamma$-action extends to this compactification, and we obtain a manifold $S_t^1 \times \mathbb{P}_w^1$ that inherits the mini--complex structure from the quotient. Notice that the compactification adds the set $S^1_L \times \{\infty \}$. We can then filter holomorphic sections based on their behaviour around  $\{\infty \}\in \mathbb{P}^1$. That is, for an open neighbourhood $U\subset \mathbb{P}^1$ such that $\{ w \} \in U$, set 
\begin{equation}
    \mathcal{P}_a \mathcal{E}^t(U) :=\{s \in \mathcal{E}^t(U\setminus \{\infty\}) ~|~|s|_h = O(|w|^{a+\epsilon}) ~\forall \epsilon > 0 \},
\end{equation}
(here we are abusing notation and denoting by $\cE^t$ the sheaf of sections of $\cE^t$, and we are allowing poles at $U\{\infty\}$). This is an increasing sequence of $\mathcal{O}_{\mathbb{P}^1}$-modules, and one can consider
\begin{equation}
    \mathcal{P}_*\mathcal{E}^t =  \bigcup_{a\in \mathbb{R}} \mathcal{P}_a \mathcal{E}^t.
\end{equation}
It is a non--trivial result that this is also a sheaf of $\mathcal{O}_{\mathbb{P}^1}$-modules. The automorphism $F$ induces an automorphism of $\mathcal{P}_*\mathcal{E}^0$.}

Let us now briefly describe how the presence of singularities affects the representations of the solutions. Since we dealing with GCK monopoles, these are all of Dirac type and therefore supported at a discrete set of points $Z \subset M \cong S^1_L \times \mathbb{C}_w$. Let $D$ be the projection of $Z$ on $\mathbb{C}_w$. Given $Q\in D$, we can assign a set of `times' $t_{Q}^a$, $Q\in D$ at which the singularities are located (see~Fig.~\ref{fig:Dirac-sing}).
\begin{figure}
\centering

\tikzset{every picture/.style={line width=0.75pt}} 

\begin{tikzpicture}[x=0.6pt,y=0.6pt,yscale=-1,xscale=1]

\draw  [dash pattern={on 4.5pt off 4.5pt}]  (299.17,151.5) -- (299.17,193.5) ;
\draw  [dash pattern={on 4.5pt off 4.5pt}]  (298.17,60) -- (298.17,132.5) ;
\draw  [dash pattern={on 4.5pt off 4.5pt}]  (452.25,62.25) -- (452.25,134.75) ;
\draw   (268,66) .. controls (268,54.95) and (316.17,46) .. (375.58,46) .. controls (435,46) and (483.17,54.95) .. (483.17,66) .. controls (483.17,77.05) and (435,86) .. (375.58,86) .. controls (316.17,86) and (268,77.05) .. (268,66) -- cycle ;
\draw   (268,200) .. controls (268,188.95) and (316.17,180) .. (375.58,180) .. controls (435,180) and (483.17,188.95) .. (483.17,200) .. controls (483.17,211.05) and (435,220) .. (375.58,220) .. controls (316.17,220) and (268,211.05) .. (268,200) -- cycle ;
\draw  [draw opacity=0][fill={rgb, 255:red, 155; green, 155; blue, 155 }  ,fill opacity=0.26 ][dash pattern={on 4.5pt off 4.5pt}] (268,135) .. controls (268,123.95) and (316.17,115) .. (375.58,115) .. controls (435,115) and (483.17,123.95) .. (483.17,135) .. controls (483.17,146.05) and (435,155) .. (375.58,155) .. controls (316.17,155) and (268,146.05) .. (268,135) -- cycle ;
\draw    (219,221) -- (219.16,41.5) ;
\draw [shift={(219.17,39.5)}, rotate = 90.05] [color={rgb, 255:red, 0; green, 0; blue, 0 }  ][line width=0.75]    (10.93,-3.29) .. controls (6.95,-1.4) and (3.31,-0.3) .. (0,0) .. controls (3.31,0.3) and (6.95,1.4) .. (10.93,3.29)   ;
\draw    (212,91) -- (226.17,91) ;
\draw    (213,135) -- (227.17,135) ;
\draw    (212,178) -- (226.17,178) ;
\draw  [draw opacity=0][fill={rgb, 255:red, 208; green, 2; blue, 27 }  ,fill opacity=1 ] (293,90.75) .. controls (293,87.85) and (295.35,85.5) .. (298.25,85.5) .. controls (301.15,85.5) and (303.5,87.85) .. (303.5,90.75) .. controls (303.5,93.65) and (301.15,96) .. (298.25,96) .. controls (295.35,96) and (293,93.65) .. (293,90.75) -- cycle ;
\draw  [draw opacity=0][fill={rgb, 255:red, 208; green, 2; blue, 27 }  ,fill opacity=1 ] (293,177.75) .. controls (293,174.85) and (295.35,172.5) .. (298.25,172.5) .. controls (301.15,172.5) and (303.5,174.85) .. (303.5,177.75) .. controls (303.5,180.65) and (301.15,183) .. (298.25,183) .. controls (295.35,183) and (293,180.65) .. (293,177.75) -- cycle ;
\draw  [draw opacity=0][fill={rgb, 255:red, 208; green, 2; blue, 27 }  ,fill opacity=1 ] (447,103.75) .. controls (447,100.85) and (449.35,98.5) .. (452.25,98.5) .. controls (455.15,98.5) and (457.5,100.85) .. (457.5,103.75) .. controls (457.5,106.65) and (455.15,109) .. (452.25,109) .. controls (449.35,109) and (447,106.65) .. (447,103.75) -- cycle ;
\draw  [dash pattern={on 4.5pt off 4.5pt}]  (369.17,154.5) -- (369.17,212.5) ;
\draw  [dash pattern={on 4.5pt off 4.5pt}]  (452.17,150.5) -- (451.17,200) ;
\draw  [dash pattern={on 4.5pt off 4.5pt}]  (369.17,74.5) -- (369.17,147) ;
\draw  [draw opacity=0][fill={rgb, 255:red, 208; green, 2; blue, 27 }  ,fill opacity=1 ] (364,201.75) .. controls (364,198.85) and (366.35,196.5) .. (369.25,196.5) .. controls (372.15,196.5) and (374.5,198.85) .. (374.5,201.75) .. controls (374.5,204.65) and (372.15,207) .. (369.25,207) .. controls (366.35,207) and (364,204.65) .. (364,201.75) -- cycle ;

\draw (200,126.4) node [anchor=north west][inner sep=0.75pt]    {$0$};
\draw (179,80.4) node [anchor=north west][inner sep=0.75pt]    {$t_{Q,1}$};
\draw (177,167.4) node [anchor=north west][inner sep=0.75pt]    {$t_{Q,2}$};
\draw (300,191.4) node [anchor=north west][inner sep=0.75pt]    {$Q$};

\end{tikzpicture}

\caption{Dirac singularities of the Berry connection in an open subset $U\times I \subset \mathbb{C}_w \times S^1_L$, where $ U $ is a small open subset $U \subset \mathbb{C}_w$, and $I \subset S^1_L$ a small interval. The module $V$ is constructed by considering sections of the bundle of supersymmetric ground states restricted to $\mathbb{C}_w \times \{0\}$ (which intersects $U\times I$ in the shaded area) that are meromorphic, with poles located along the projections (in the $t$ direction, dashed lines) of the Dirac singularities to $\mathbb{C}_w \times \{0\} $. In particular, two of the Dirac singularities (located at $(Q , t_{Q,1})$, $(Q , t_{Q,2})$) may result in poles at ${Q} \in \mathbb{C}_w $ in the elements of $V$. The locations $t_{Q,1}$, $t_{Q,2}$ enter the parabolic data of the module.}
\label{fig:Dirac-sing}
\end{figure} 
The modules are then modified as follows.
\begin{itemize}
    \item We consider sections \emph{meromorphic} along $D$. To emphasise this, we sometimes denoted $\mathcal{E} (\star D)$ the sheaf of sections.
    \item The modules are endowed with lattices $\mathcal{L}_{a,Q}$ of $\mathcal{E}_Q$ encoding the modifications of the spaces of meromorphic sections due to the Dirac singularities (Hecke modifications).
\end{itemize}
The pair $(V,F)$ where $V$ is endowed with a polystability condition, filtration at infinity and lattices is the $0$-difference module that~\cite{mochizuki2017periodic} proves to be in bijective correspondence with smooth monopoles. As we mentioned above, we will not describe these structures, which are not crucial for our purposes, in any further detail. 

If $n>1$ the fundamental ideas that lead to a difference module can be easily generalised. Recall that the deformation space corresponds to $n$ copies of $S^1 \times \mathbb{R}^2$, each endowed with a mini--complex structure that identifies it with $S^1\times \mathbb{C}^2$ as a mini--complex manifold. The vector bundle of supersymmetric ground states restricted to fixed values of the real coordinates still has the structure of a holomorphic bundle on $\mathbb{C}_w^{2n}$, with the holomorphic structure induced by the Dolbeault operator $\partial_{E,\bar{w}}$ with components $\bar{D}_{\bar{w}_i}$. In principle, we can similarly consider the space of meromorphic sections with prescribed behaviour at the singularities as well as at infinity. Although formulating these conditions precisely is beyond the scope of this paper, let us denote by $V$ the space of such sections. There are then $n$ complexified flat connections $\partial^{(i)}_{E,t}$ defined in \eqref{eq:mochizukioperators} that we can consider the parallel transport by. By the generalised Bogomolny (or $tt^*$) equations~\eqref{eq:tt*equation_1}-\eqref{eq:tt*equation_4}, the parallel transport operations preserve meromorphicity and moreover commute with each other. We therefore have $n$ commuting automorphisms
\begin{equation}\label{eq:automorphisms}
    F^{(i)} (w) : V \rightarrow V,
\end{equation}
which turn $V$ into a $\mathbb{C}(w_i)$ modules for all $i$. 

\subsubsection{Cherkis--Kapustin spectral variety}

For the difference module $(V, \{F^{(i)}\})$, the associated \textit{spectral variety} (see \emph{e.g.}~\cite{Cecotti:2013mba}) is the Lagrangian submanifold of $(\bC^*)^n \times \bC^n$ defined by the simultaneous eigenvalues of the automorphisms $F^{(i)}(w)$ given in~\eqref{eq:automorphisms}
\begin{equation}\label{eq:spectral_curve}
    \mathcal{L} = \left\{ (p_1,\ldots,p_n, w_1,\ldots w_n) \in  (\bC^*)^n \times \bC^n \,|\, \exists\, v \in \bC^N \backslash \{0\} \text{ s.t. } F^{(i)}(w) v = p v \right\}.
\end{equation}
It is Lagrangian with respect to the holomorphic symplectic form $\sum_i \frac{dp_i}{p_i} \wedge dw_i$. For $n=1$ this reduces to
\begin{equation}\label{eq:rank_1_spectral_curve}
    \mathcal{L} = \left\{ (p,w) \,|\, \text{det}( p\mathbf{1} - F(w)) = 0 \right\},
\end{equation}
which is the spectral \emph{curve} first defined by Cherkis and Kapustin~\cite{Cherkis:2000cj, Cherkis:2000ft}. We therefore refer to the variety~\eqref{eq:spectral_curve} for $n\geq 1$ as the Cherkis--Kapustin spectral \emph{variety}.  

The spectral variety $\mathcal{L}$ is an $N$-sheeted cover of $\mathfrak{t}_{\bC} \cong \bC^n$ (which is spanned by $w_i$). It is equipped further with a coherent sheaf $\mathcal{M}$, whose stalks are the eigenspaces spanned by $v$. The pushforward of $\mathcal{M}$ under the projection $\pi: \mathcal{L} \rightarrow \mathfrak{t}_{\bC}$ is a rank $N$ holomorphic vector bundle, and coincides with the space of supersymmetric ground states restricted at $t=0$ $(\mathcal{E}^0,\partial_{E,\bar{w}})$. The corresponding values of $p_i$ on $\cL$ encode the parallel transport with respect to the complexified connection $D_{t_i} - i \phi_i$.

\subsubsection{Physical construction of difference modules and spectral varieties}\label{sec:physical_examples}

We now describe specifically how one may recover the above structures (the $0$-difference module and associated spectral variety) physically, for a GCK monopole arising as the supersymmetric Berry connection for a GLSM.

Let us consider the states $\ket{a}$ obtained on the boundary $S^1$ of an A--twisted cigar, by inserting an operator $\cO_a$ in $Q_A$-cohomology (an element of the twisted chiral ring). These states are in $Q_A$-cohomology, and can be projected onto ground states via stretching the topological path integral, implementing a Euclidean time evolution $e^{-\beta H}$ with $\beta \rightarrow \infty$. One can generate a basis for the space of ground states via a basis of the twisted chiral ring in this way, and working with respect to such a basis is often called working in \textit{topological gauge} \cite{Cecotti:2013mba}. In particular, it is a standard result \cite{Cecotti:1991me} of $tt^*$ that in this basis $(\bar{A}_{\bar{w}_i})_a{}^{b}=0$ and thus
\begin{equation}
    \partial_{E,\bar{w}}^{(i)} \ket{a(w)} = 0.
\end{equation}
Thus, such states can be identified as generating a basis of the module $V$.

The automorphisms $F^{(i)}$ also admit a clean interpretation. Recall the origin of $A_{t_i} + i \phi_i$ in the $tt^*$ equations as the chiral ring matrix, describing the action of the $tt^*$-dual operator to $w_i$ (the operator to which $w_i$ couples in the action) on the ground states. As $w_i$ is the complex scalar component of a background vector multiplet for $U(1)_i \leq T$, this is the defect operator inserting a unit of flux for the $U(1)_i$ gauge field, or alternatively winding the corresponding holonomy. The action of $F^{(i)}$ on $V$ corresponds precisely to the action of such defects, which due to topological invariance can be localised to a local operator.

There is another way of seeing this, which further allows an explicit computation of $F^{(i)}(w)$. Consider an effective description of the theory as an abelian theory in the IR after integrating out all the chiral multiplets \cite{Witten:1993yc}. This theory is determined by $W_{\text{eff}}(\sigma, w)$, the effective twisted superpotential, with $\sigma_a$ parametrising the Cartan of the complex scalar in the vector multiplet of the GLSM. In this description, the twisted chiral ring is represented by gauge--invariant polynomials in $\sigma_a$, subject to the ring relations $\exp{\partial_{\sigma_a}W_{\text{eff}}}=1$. From this perspective, the dual operator to $w_i$ is simply $-2 i \partial_{w_i} W_{\text{eff}}$, and from the form of the effective action, see \textit{e.g.} section 7.1.2 of \cite{Closset_2019}, the operator
\begin{equation}\label{eq:flux_operator}
    p_i = e^{-2 i \frac{ \partial W_{\text{eff}}(\sigma, w)}{ \partial  w_i }}
\end{equation}
corresponds precisely to the insertion of a unit of flux for $U(1)_i \leq T$ in the path integral. 

In more details, the field strength $f_{z \bar{z}}^{(i)}$ in a background vector multiplet with scalar component $w_i$ appears in the effective action in a term:
\begin{equation}
    - 2 i  \int  d^2x \sqrt{g} \, f_{z \bar{z}}^{(i)} \, \frac{\partial W_{\text{eff}}}{\partial w_i}.
\end{equation}
Here, the integral is over the cigar, and the background field strength obeys GNO quantisation:
\begin{equation}
    \int  d^2x \sqrt{g} \, f_{z \bar{z}}^{(i)} \in \bZ
\end{equation}
It is thus clear that an insertion of \eqref{eq:flux_operator} corresponds to an insertion of unit flux.

To compute $F^{(i)}$ using this description, write $\ket{a} = \cO_a \ket{1}$, where $\ket{1}$ is the state generated by the $A$--twisted cigar path integral with no insertions, and $\cO_a$ is a polynomial in $\sigma$. This notation makes sense because in the twist $\cO_a$ may be brought to act on the boundary. We suppress the $w$-dependence for clarity. The action of $F^{(i)}$ may now be derived by multiplying $\cO_a$ by $p_i$ in \eqref{eq:flux_operator}. This naively yields an operator rational in $\sigma$, but by consistency must be able to be brought back into the $ \{ \cO_a \}$ basis by identifications using the vacuum equations $\exp{\partial_{\sigma_a}W_{\text{eff}}}=1$. Performing these, we have
\begin{equation}\label{eq:0_difference}
    p_i \, \cO_a \ket{1} = (F^{(i)})_a{}^b \, \cO_b \ket{1},
\end{equation}
yielding the automorphism $F(w)$ in the basis $\ket{a}$ generated by twisted chiral ring elements $\cO_a$. We see an example of this below. Note that the above arguments can also be made also via Coulomb branch localisation, where the chiral ring insertions concretely take the form of polynomial insertions in a contour integral over $\sigma$.

Let us now show how to derive the spectral variety in terms of the physical data, which does not require performing the above substitutions. We in particular describe the eigenvalues of the automorphisms $F^{(i)}$. We first note that \eqref{eq:comm-bog} is independent of the radius $L$ on which the theory is quantised, which simply rescales $D_{t_j} - i\phi_j $. Thus, the eigenvalues of $F^{(i)}(w)$ can be computed in the $L \rightarrow \infty $, \textit{i.e.} flat space limit. There, outside of codimension-1 loci in $w$ space, the ground states are simply the massive vacua of the theory. 
In this basis, $A_{t_i}+i\phi_i$ is given by the VEVs of the aforementioned defect operator for the flavour symmetry $U(1)_i \leq T$ in the massive vacua $\{\alpha\}$, which may in turn be expressed via the low energy effective twisted superpotential:
\begin{equation}\label{eq:0_diff_module}
    A_{t_i}+i\phi_i = \text{diag}_{\{\alpha\}}\left( -2 i L \partial_{w_i} W_{\text{eff}}^{(\alpha)} \right) = \text{diag}_{\{\alpha\}}\left( L^{-1}\partial_{m_i} W_{\text{eff}}^{(\alpha)} \right).
\end{equation}
Here we have introduced the redefined coordinates $m_i = i w_i/2L^2$.\footnote{Note that $m$ has length dimension $-1$ as usual for a mass in 2d, because $w$ has dimension $+1$, as can be seen from consistency for the formulae for $\beta_0$ or $\beta_1$ in section \ref{subsec:def-par}.} Notice that this was also demonstrated for LG models in \cite{Cecotti:2013mba}. Thus in the $L\rightarrow \infty$ limit
\begin{equation}
    F^{(i)}(w) = \text{diag}_{\{\alpha\}}\left( e^{\partial_{m_i} W_{\text{eff}}^{(\alpha)}} \right).
\end{equation}
Therefore, in the case of GLSMs, the spectral variety equations can be written as:
\begin{equation}\label{eq:glsm_spectral_curve}
\begin{aligned}
    e^{\frac{\partial W_{\text{eff}}(\sigma, m)}{\partial_{m_i} }} &= p_i, \quad i=1\ldots n,\\
    e^{\frac{\partial W_{\text{eff}}(\sigma, m)}{\partial_{\sigma_a} }} &= 1, \quad\, a=1\ldots r.
\end{aligned}
\end{equation}
For non-abelian gauge groups, one most quotient the set of solutions by the Weyl group. Eliminating $\sigma$ from the combined system \eqref{eq:glsm_spectral_curve} yields the variety $\cL$. The spectral curve has been studied for LG models \cite{Cecotti:2013mba} and in a different context for 3d theories \cite{Dimofte:2011ju,Dimofte:2011jd, Bullimore:2014awa}.

\subsubsection{Relation to equivariant quantum cohomology}\label{subsec:reln_quantum_cohom}

The chiral ring, which appears in~\eqref{eq:0_difference}, is known to reproduce the \textit{quantum equivariant cohomology} ring $QH^\bullet_T(X)$ of the Higgs branch $X$ of the GLSM~\cite{Vafa:1991uz}. This is a quantisation of the normal cohomology ring via the contribution of higher degree pseudo--holomorphic curves to correlation functions.

In the description of the theory as an IR effective abelian theory, the twisted chiral ring operators are gauge--invariant combinations of complex scalars $\sigma$ in the Cartan of the gauge group. They are subject to the relations in the first line of \eqref{eq:glsm_spectral_curve} involving the effective twisted superpotential, which are the vacuum equations. Since the twisted chiral ring is protected under renormalisation, these relations coincide with the relations in $QH^\bullet_T(X)$ \cite{Nekrasov:2009uh, Nekrasov:2009ui}.

Our analysis above therefore shows that the difference module $(V,\{F^{(i)}\})$ endows $QH^\bullet_T(X)$, via \eqref{eq:0_difference}, with the structure of a module for the action of the algebra of functions $\bC[p_i^{\pm1}, w_i]$ on $T_{\bC} \times \mathfrak{t}_{\bC} \cong (\bC^*_p \times \bC_w)^n$. Geometrically, the module defines a sheaf over $(\bC^*_p \times \bC_w)^n$ with holomorphic Lagrangian support $\cL$ precisely given by~\eqref{eq:glsm_spectral_curve}, the Cherkis--Kapustin spectral variety. We expect that this result is related to work by Teleman \cite{Teleman:2018wac}, realising 3d Coulomb branch algebras via their actions on quantum cohomologies of K\"ahler targets, which we will return to in upcoming work. Thus, we can summarise our main results of these sections as follows.

\begin{mdframed}[roundcorner=10pt, linewidth=0.3mm, linecolor=gray]
  Suppose that a monopole arises as the Berry connection of a GLSM with Higgs branch $X$, or an NLSM with target $X$ (and in particular it is of GCK--type). Then the $0$-difference module $(V,F)$ can be interpreted as describing the action of defect operators for the flavour symmetry on the space of states generated by inserting operators in the twisted chiral ring on an A--twisted cigar. Identifying the twisted chiral ring with $QH_T^\bullet(X)$, this endows $QH^\bullet_T(X)$ with the structure of a module over  the algebra of functions on $T_{\bC} \times \mathfrak{t}_{\bC}$. The Cherkis--Kapustin spectral variety of the monopole corresponds to the Lagrangian support of a sheaf over $T_{\bC} \times \mathfrak{t}_{\bC}$ determined by this action.
\end{mdframed}

We will see momentarily that the $2i\lambda$-difference modules provide a quantisation of the above action, and that the Cherkis--Kapustin spectral variety is quantised by difference equations solved by D--brane amplitudes and hemisphere or vortex partition functions. In the latter case, the quantisation coincides with the insertion of an Omega background. Before we move on, we spell out the above remarks for our two main examples.

\subsubsection{Examples}

We now compute the spectral curve for our two examples, taking $L=1$ for simplicity.

\paragraph{Free chiral.}

For the free chiral one has $W_{\text{eff}} = m (\log \frac{m}{\mu} -1)$, so that the spectral curve is simply
\begin{equation}\label{eq:CK_free}
    p- \frac{m}{\mu} = 0.
\end{equation}
This is $QH^\bullet_T(\bC)$, the quantum equivariant cohomology of $\bC$. This clearly reproduces the spectral curve for a Dirac monopole of charge 1 centred at the origin \cite{Cherkis:2000cj,Cherkis:2000ft}.

\paragraph{$\mathbb{CP}^1$ $\sigma$-model.}

For supersymmetric QED with $2$ chirals, from \eqref{eq:w_eff_sqed2}, the vacuum equations are:
\begin{equation}\label{eq:bethe}
1 = e^{\frac{\partial W_{\text{eff}}}{\partial \sigma}} = q^{-1} (\sigma+m)(\sigma-m),
\end{equation}
where $m \equiv iw/2$. This describes the quantum equivariant cohomology  $QH^\bullet_T(\mathbb{CP}^1)$. 

To obtain the automorphism $F(w)$ on a basis of $V$, $\{\ket{\mathbf{1}}, \sigma \ket{\mathbf{1}}\}$ generated by the twisted chiral ring basis $\{\mathbf{1},\sigma\}$, note
\begin{equation}\label{eq:0_diff_cp1}
    p =  e^{\frac{\partial W_{\text{eff}}}{\partial m}}  = \frac{\sigma+m}{\sigma-m} \quad \Rightarrow \quad p 
    \begin{pmatrix} 
    \mathbf{1} \\ 
    \sigma 
    \end{pmatrix} 
    = 
    F(m) 
    \begin{pmatrix} 
    \mathbf{1} \\ 
    \sigma 
    \end{pmatrix} 
\end{equation}
where
\begin{equation}\label{eq:sqed2_matrix_0_diff}
    F(m) = 
    \begin{pmatrix}
    1+2m^2 q^{-1} & 2m q^{-1} \\
    2m (1+ m^2q^{-1}) & 1+ 2m^2 q^{-1}
    \end{pmatrix}.
\end{equation}
The equality in the second equation in \eqref{eq:0_diff_cp1} should be considered up to the ring relation \eqref{eq:bethe}.

To derive the spectral curve, one can simply take the characteristic polynomial of the above, or alternatively solving for $\sigma$ in $p =  e^{\frac{\partial W_{\text{eff}}}{\partial m}} $ gives 
\begin{equation}
\sigma = \frac{m(p+1)}{p-1},
\end{equation}
and substituting into (\ref{eq:bethe}) we obtain
\begin{equation}\label{eq:spectralcurve}
\cL(m,p) \coloneqq p^2 - 2(1+2 m^2q^{-1} )p+ 1 = 0.
\end{equation}
It is easy to check the action of $p$ on $V$ (\textit{i.e.} $QH^\bullet_T(\mathbb{CP}^1)$) defined by \eqref{eq:0_diff_cp1} obeys \eqref{eq:spectralcurve}.

\subsection{\texorpdfstring{$\lambda \neq 0$}{}: branes, difference equations \& variety quantisation}\label{sec:difference_modules}

We now move on to consider the $\lambda \neq 0 $ case. In physical terms, this corresponds to viewing the space of supersymmetric ground states as classes in $Q_\lambda$ cohomology, where we recall from~\eqref{eq:lambda_supercharges} that
\begin{equation}
Q_{\lambda} = \frac{1}{\sqrt{1+|\lambda|^2}}(Q_A +\lambda \bar{Q}_A).
\end{equation}
We first review how, in the work of Mochizuki \cite{mochizuki2017periodic}, the $0$-difference modules we discussed in the previous section are replaced at generic $\lambda$ by genuine, $2i \lambda $-difference modules. In the context of our paper, this corresponds to a single rank $U(1)$ flavour symmetry. We return to $n \geq 1$ momentarily. We also continue to set $L=1$ in this subsection.

Recall that $\lambda$ parametrises mini--complex structures on $S^1\times \bR$, and that these can be constructed by means of some $\lambda$-dependent mini--complex structures on a lift $\mathbb{R}\times \mathbb{C}$. It follows from the above remarks, and in particular equation~\eqref{eq:supercharge_dependencies}, that the operators $\partial_{\bar{\beta}_1}$, $\partial_{t_1}$ descend to Berry connections $\partial_{E, \bar{\beta_1}}$, $\partial_{E, t_1}$ on the space of supersymmetric ground states, as given in equation \eqref{eq:mochizukioperators}. By restricting to a constant value $t_1 = t^*_1$, we can then define the holomorphic vector bundle on $\bC_{\beta_1}$
\begin{equation}\label{eq:diff_module_bundle}
    \cE^{t^*_1} := (E|_{\{t^*_1\} \times\bC_{\beta_1}}, \partial_{E,\bar{\beta}_1} ).
\end{equation}
We can then consider the complex Bogomolny equation in these variables (\emph{cf.}~\eqref{eq:comm-bog} for the analogous key equation in the $\lambda=0$ case)
\begin{equation}
    [\partial_{E,t_1}, \partial_{E,\bar{\beta}_1}] = 0.
\end{equation}
The difference operator of Mochizuki is now defined as:
\begin{equation}
    \Phi_V^* = \Phi_1^* \circ F
\end{equation}
where
\begin{equation}
    F: \cE^0 \rightarrow \cE^1
\end{equation}
is, as before, the endomorphism given by parallel transport with respect to $\partial_{E,t_1}$, and 
\begin{equation}
\Phi_1^*(\mathcal{E}^1) \cong \mathcal{E}^0
\end{equation}
is the pullback induced by the automorphism $\Phi_1 : \bC_{\beta_1} \rightarrow \bC_{\beta_1}$ given by $\Phi_1(\beta)= \beta_1+2i\lambda $. This pullback is necessary due to the coordinate identifications $(t_1,\beta_1) = (t_1+1, \beta_1 +2 i \lambda )$. 

Let $V$ be the $\bC(\beta_1)$-module of holomorphic sections of $\cE^0$, which in the smooth case we can schematically as in~\eqref{eq:naive-mod}:
\begin{equation}
    V:= H^0 (\mathbb{C}_{\beta_1},\mathcal{E}^0)\otimes_{\mathbb{C}[\beta_1]}\mathbb{C}(\beta_1).
\end{equation}
As before, this can be turned into a rigorous algebraic object provided one considers the behaviour of the sections at infinity. It is a highly non--trivial result of \cite{mochizuki2017periodic} that if one does so, the space of sections becomes an acceptable algebraic object.
In the presence of singularities, we must in general allow for meromorphic sections with poles along $D$ and prescribe a corresponding parabolic structure. We will not review this here, but instead emphasise that the existence of a finite--dimensional, polystable parabolic $\mathbb{C}(\beta_1)$-module $V$ has been proven by Mochizuki in \cite{mochizuki2017periodic}.

The pair $(V, \Phi_1^*)$ of a $\mathbb{C}(\beta_1)$-module $V$ together with the automorphism $\Phi_1^*$ constitutes a $2i\lambda$-difference module. This means that if we have $f \in  \mathbb{C}(\beta_1)$ and $s \in V$, we must have:
\begin{equation}\label{eq:differencemodule}
\Phi_V^*(fs) = \Phi_1^*(f) \Phi_V^*(s).
\end{equation}
This follows from the commutativity of the operators $\partial_{E,t_1}$ and $\partial_{E,\bar{\beta}_1}$ due to the Bogomolny equations, see~\eqref{eq:commutator2}.

In the rest of the paper we will consider generalisations of the above modules to $n > 1$, which can be obtained precisely as outlined at the end of Section~\ref{sec:product_case}. In particular, we may na\"ively define $\cE^0$ as in~\eqref{eq:diff_module_bundle}, where the Dolbeault operator $\partial_{E,\bar{\beta}_1}$ inducing the holomorphic structure has components $\partial_{E,\bar{\beta}_1}^{(i)}$ given in \eqref{eq:mochizukioperators}, but now for a generalised periodic monopoles on a compactification of $\mathbb{C}^{2n}$. We now have $n$ parallel transport operators $F^{(i)}$ from $t_{1,i}~\mapsto~t_{1,i}+1$. We may similarly define the pullback $\Phi_{1,i}^*$, which is is simply pre--composition with $\beta_{1,i} \rightarrow \beta_{1,i} + 2 i \lambda $, and therefore also the automorphisms $\Phi_{V,i}^* := \Phi_{1,i}^* \circ F^{(i)} $ for $i=1,\ldots n$.
Taking $V$ together with $\Phi_V^*$ (we have suppressed indices here), yields a natural generalisation of the equation \eqref{eq:differencemodule} and thus the difference modules of Mochizuki. Such generalisations are under active study in the mathematical community, see \textit{e.g.} \cite{Kontsevich_Soibelman_1, Kontsevich_Soibelman_2}.

\subsubsection{Branes \& states}

We now turn to the relation between ground states of the SQM along $\mathbb{R}$ of a cigar geometry and the $2i\lambda$-difference modules of Mochizuki. In the next few sections we will derive from these states and the module action, novel difference equations for D--brane amplitudes, and by taking a suitable limit, hemisphere and vortex partition functions. These difference equations can be viewed as providing a quantisation of the spectral variety and thus, by the discussion in section \ref{subsec:reln_quantum_cohom}, an action of an algebra of functions on $T_{\bC} \times \mathfrak{t}_C$ on $QH^\bullet_T(X)$ where $X$ is the target of the GLSM.

Consider a cigar configuration that is A--twisted in the bulk. We want to consider states generated by D--branes. For our purposes, a D--brane is a half--BPS boundary condition for this configuration preserving $R_V$, and two supercharges:
    \begin{equation}\label{eq:branesupercharges}
        \bar{q}_+ + \lambda \bar{q}_-, \qquad q_{+}+\bar{\lambda} q_-.
    \end{equation}
In the above, $\lambda$ lies initially on the unit circle. We note that by taking linear combinations, such branes also preserve the supercharges $Q_{\lambda}$ and $\bar{Q}^{\dagger}_{\lambda}$, and thus in particular generate a harmonic, albeit not necessarily normalisable representative of a state in $Q_{\lambda}$ cohomology. Note that when $\lambda =1$, the first supercharge in \eqref{eq:branesupercharges} is precisely the usual B--type supercharge, and the corresponding D--branes are usually referred to as B--branes.

\begin{figure}
    \centering
    \tikzset{every picture/.style={line width=0.75pt}} 
\begin{tikzpicture}[x=0.75pt,y=0.75pt,yscale=-.8,xscale=.8]

\draw  [draw opacity=0][dash pattern={on 0.84pt off 2.51pt}] (450,160) .. controls (438.95,160) and (430,142.09) .. (430,120) .. controls (430,97.91) and (438.95,80) .. (450,80) -- (450,120) -- cycle ; \draw  [color={rgb, 255:red, 208; green, 2; blue, 27 }  ,draw opacity=1 ][dash pattern={on 0.84pt off 2.51pt}] (450,160) .. controls (438.95,160) and (430,142.09) .. (430,120) .. controls (430,97.91) and (438.95,80) .. (450,80) ;  
\draw  [draw opacity=0] (450,80) .. controls (450,80) and (450,80) .. (450,80) .. controls (450,80) and (450,80) .. (450,80) .. controls (461.05,80) and (470,97.91) .. (470,120) .. controls (470,142.09) and (461.05,160) .. (450,160) -- (450,120) -- cycle ; \draw  [color={rgb, 255:red, 208; green, 2; blue, 27 }  ,draw opacity=1 ] (450,80) .. controls (450,80) and (450,80) .. (450,80) .. controls (450,80) and (450,80) .. (450,80) .. controls (461.05,80) and (470,97.91) .. (470,120) .. controls (470,142.09) and (461.05,160) .. (450,160) ;  
\draw  [draw opacity=0] (236,160) .. controls (236,160) and (236,160) .. (236,160) .. controls (169.73,160) and (116,142.09) .. (116,120) .. controls (116,97.91) and (169.73,80) .. (236,80) -- (236,120) -- cycle ; \draw   (236,160) .. controls (236,160) and (236,160) .. (236,160) .. controls (169.73,160) and (116,142.09) .. (116,120) .. controls (116,97.91) and (169.73,80) .. (236,80) ;  
\draw    (370,80) -- (450,80) ;
\draw    (370,160) -- (450,160) ;
\draw  [line width=3]  (116,120.5) .. controls (116,120.22) and (116.22,120) .. (116.5,120) .. controls (116.78,120) and (117,120.22) .. (117,120.5) .. controls (117,120.78) and (116.78,121) .. (116.5,121) .. controls (116.22,121) and (116,120.78) .. (116,120.5) -- cycle ;
\draw  [dash pattern={on 4.5pt off 4.5pt}]  (236,80) -- (370,80) ;
\draw  [dash pattern={on 4.5pt off 4.5pt}]  (240,160) -- (370,160) ;

\draw (484,112.4) node [anchor=north west][inner sep=0.75pt]  [color={rgb, 255:red, 208; green, 2; blue, 27 }  ,opacity=1 ]  {$D$};
\draw (88,112.4) node [anchor=north west][inner sep=0.75pt]    {$\cO_a$};
\draw (271,172.4) node [anchor=north west][inner sep=0.75pt]    {$\braket{a|\textcolor[rgb]{0.82,0.01,0.11}{D}}$};
\end{tikzpicture}
    \caption{The D--brane amplitude given by the overlap between the state $\ket{D}$ generated by the brane, and the state $\bra{a}$ generated by the path integral with an insertion of a (twisted) chiral ring operator at the tip.}
    \label{fig:brane_amplitude}
\end{figure}
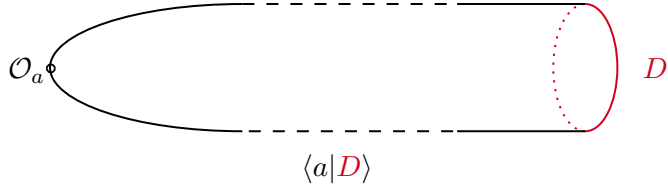

We denote by $\Pi[D]$ the projection of $\ket{D}$ onto the space of supersymmetric ground states. This can be done by taking inner products (via computing the path integral on the infinite cigar), yielding brane amplitudes. For example, we can consider the overlap
\begin{equation}\label{eq:overlap}
    \Pi[D,a] = \braket{a|D}
\end{equation}
where $\ket{a}$ is the ground state generated by the path integral for the topologically twisted theory on an indefinitely long cigar, with a (twisted) chiral ring operator $\cO_a$ labelled by $a$ inserted at the tip. This is shown in Figure~\ref{fig:brane_amplitude}. The path integral implements a Euclidean time evolution $e^{-\beta H}$ with $\beta \rightarrow \infty$ and it generates a ground state at the boundary; the states $\ket{a}$ generated in this way will by construction be holomorphic with respect to $\partial_{E,\bar{w}}$~\cite{Cecotti:2013mba}.

As functions of $\lambda$, it is known that the brane amplitudes can then be analytically continued to the whole of $\bC\backslash\{0,\infty\}$ \cite{Cecotti:1992rm,Gaiotto:2011tf}. Further, it is a classic result in the context of $tt^*$ equations that they are flat sections of the Lax connection, which in the twisted chiral ring basis means
\begin{equation}\label{eq:lax_equations}
\begin{aligned}
   \big( D_{w_i} + \frac{i}{2\lambda}(D_{t_i}-i\phi_i)\big)_a^{\,\,b} \,\Pi[D,b] = 0,\qquad
   \big( \bar{D}_{\bar{w}_i} + \frac{i\lambda}{2}(D_{t_i}+i\phi_i) \big)_a^{\,\,b}\, \Pi[D,b] = 0.
\end{aligned}
\end{equation}
We could however expand this in any basis we like.

The Lax equations can straightforwardly be related to the defining equations of the difference modules for $n=1$, and to a generalisation thereof to $n>1$. To show this, we will define $L_{w,t}^{(i)} := D_{w_i} + \frac{i}{2\lambda}(D_{t_i}-i\phi_i)$ and $L_{\bar{w},t}^{(i)}:=  \bar{D}_{\bar{w}_i} + \frac{i\lambda}{2}(D_{t_i}+i\phi_i)$. The $tt^*$ equations are equivalent to:
\begin{equation}
    [L_{w,t}^{(i)}, L_{w,t}^{(j)}] = [L_{w,t}^{(i)}, L_{\bar{w},t}^{(j)}] = [L_{\bar{w},t}^{(i)}, L_{\bar{w},t}^{(j)}] = 0.
\end{equation}
The key insight is then the following: one may rewrite the generalisation to $ n\geq 1$ of the holomorphic covariant derivative, and parallel transport operator associated with the difference modules as
\begin{equation}
\begin{aligned}
    \partial_{E,t_1}^{(i)} &= \frac{2i}{1+|\lambda|^2} \left(\bar{\lambda} \bar{L}_{\bar{w},t}^{(i)} - \lambda L_{w,t}^{(i)}\right), \\
    \partial_{E,\bar{\beta}_1}^{(i)} &= \frac{1}{1+|\lambda|^2} \bar{L}_{\bar{w},t}^{(i)}. \\
\end{aligned}
\end{equation}
Thus, holomorphic sections that are parallel transported correspond to flat sections of the Lax connection.

Although the expansion in the basis $\ket{a}$ is quite natural from the point of view of the $tt^*$ geometry (see \emph{e.g.}~\cite{Ooguri:1996ck}), in the following it will be useful to introduce a basis $\ket{a^\lambda}$ such that
\begin{itemize}
    \item $\ket{a^\lambda (t_1=0,\beta_1)}$ is a holomorphic section of $\cE^0$
    \item{$\lim_{\lambda \rightarrow 0 }\ket{a^\lambda} = \ket{a}$.}
\end{itemize}
Locally, we can always find a basis holomorphic sections, and so a basis $\ket{a^\lambda}$ satisfying the first bullet point. We also know that as $\lambda \rightarrow 0$, the chiral ring basis is holomorphic. Therefore, without loss of generality, we can assume that the second bullet point holds.

We can then define states
\begin{equation}
\Pi[D] := \Pi[D,b] \eta^{ab} \ket{a^\lambda}.
\end{equation}
where $\eta^{ab}$ is the inverse matrix of $\eta_{ab} = \braket{a^\lambda|b^\lambda}$ and the sum over $\ket{a^\lambda}$, $\ket{b^\lambda}$ is understood. $\Pi[D]$ can understood as the projection of the state $\ket{D}$ generated by the D--brane onto a holomorphic basis for the ground states subspace. 

Then, whenever they are well--defined, the flatness of the D--brane amplitudes under the Lax connection \eqref{eq:lax_equations} imply that restricting to $t_1= 0$, $\Pi[D]|_{t_1=0}$ is holomorphic in $\beta_1$, and thus $\Pi[D]|_{t_1=0}$ can in principle be identified with elements of the difference modules $V$.\footnote{A subtle point is admittedly the behaviour at $\beta_1 \rightarrow \infty$: in order to be a genuine element of the module as defined by Mochizuki, the growth must be at most polynomial, and we are assuming this for the (twisted) chiral ring basis.} Moreover, recall that we also know that
\begin{equation}
\partial_{E,t_1}^{(i)}  \Pi[D] = 0.
\end{equation}
Thus, at least formally, we see that these brane states are solutions to the parallel transport equations. This means that one can implement the automorphism $F^{(i)}$ simply by evaluating $\Pi[D]$ at $t_{1,i}=1$. Therefore, the automorphism $\Phi_{V,i}^* := \Phi_{1,i}^* \circ F^{(i)} $ is implemented on the brane amplitudes by a shift
\begin{equation}
\Phi_{V,i}^* : \Pi[D](t_1, \beta_1, \bar{\beta}_1) \rightarrow \Pi[D](t_1+ e_i, \beta_1+2i\lambda  e_i, \bar{\beta}_1-2i\bar{\lambda}  e_i). 
\end{equation}
where we have suppressed indices and $e_i$ is the $i^{\text{th}}$ unit vector in $\bR^n$. 

Suppose now that brane amplitudes are globally defined functions of $t$ (that is, they are periodic). This seems to be a reasonable assumption. It was shown to hold for the free chiral, as well as general LG theories in  \cite{Cecotti:2013mba}. More generally, any non--trivial behaviour under shifting $t$ by an element of $\bZ^n$ will arise due to 't Hooft anomalies involving the corresponding symmetry $T$. For the GLSMs we consider, the only such anomalies are mixed $T-R_A$ anomalies where $R_A$ is the axial R--symmetry. Since there is no non--trivial background for $R_A$, the shift in $t$ cannot produce any non--trivial phase in the partition function/brane amplitude. Then, in terms of the original coordinates $(t,w,\bar{w})$, we can see from \eqref{eq:mochizukisecondcoords} that this becomes simply
\begin{equation}\label{eq:brane_amplitude_transport}
\Phi_{V,i}^* : \Pi[D](t, w, \bar{w}) \rightarrow \Pi[D](t+ e_i, w, \bar{w}) = \Pi[D](t, w, \bar{w}).
\end{equation}
This means that the D--brane states are invariants of the module action. Therefore, under these assumptions (that the D--brane states are genuine elements of the difference modules, and that they are global functions of $t$) computing a basis of brane amplitudes would be equivalent to determining the module associated to the monopole representing the Berry connection. In fact, if we were able to find a basis of brane amplitudes for $V$, a general section $s$ of $\mathcal{E}^0$ could be expanded in terms of a $\mathbb{C}(\beta_1)$ linear combination of the brane amplitudes, then the action of the automorphism(s) $\Phi_{V,i}^*$ on would be trivial to compute.

The problem is of course that it is in general very difficult to compute the brane amplitudes (and therefore brane states) explicitly and to evaluate their properties. This is because they are non--supersymmetric: they are A--twisted in the bulk yet preserve $Q_{\lambda},\bar{Q}_{\lambda}^{\dagger}$ at the boundary. An exception is represented by Calabi--Yau GLSMs and NLSMs, where the brane amplitudes coincide with the hemisphere partition functions (an argument is presented in \textit{e.g.} \cite{hori2019notes}), which can be computed exactly using localisation techniques. For more general K\"ahler GLSMs, such as the ones of interest in this paper vortex, or hemisphere, partition functions are expected to be recovered in the so--called conformal limit.

There \emph{is}, however, something that we know about them, namely their asymptotic behaviour at $\lambda\rightarrow 0$. In this limit we would expect to recover the $0$-difference modules, and therefore the Cherkis--Kapustin spectral variety. Our strategy to inspect the expected behaviour for D--brane states is then the following: in the next section, we demonstrate that~\eqref{eq:brane_amplitude_transport} implies certain novel difference equations for D--brane amplitudes, and explain how these recover the spectral variety equations in the $\lambda\rightarrow 0$ limit. In Section~\ref{sec:conf_vortex} we further corroborate this claim by taking the conformal limit and deriving difference equations that are explicitly solved by hemisphere partition functions.

\subsubsection{Difference equations for brane amplitudes \& variety quantisation}

In this section we derive, from our previous considerations, difference equations for brane amplitudes. We further demonstrate that in the $\lambda \rightarrow 0$ limit we can recover from these equations the Cherkis--Kapustin spectral variety discussed in Section~\ref{sec:product_case}.

Note that the automorphism $\Phi_{V}^*$ sends an element of $V$, \textit{i.e.} a holomorphic section of $\cE^0$, to another element of $V$. Therefore we can expand the action of  $\Phi_V^*$  on the basis $\{\ket{b^\lambda}\}$, the twisted chiral ring, by:
\begin{equation}
    \Phi_{V,i}^*\ket{a^{\lambda}(0,\beta_1)} = {M^{(i),b}_{a}}(\beta_1) \ket{b^\lambda(0,\beta_1)}
\end{equation}
where ${M^{(i)}_{ab}}$ must be holomorphic in $\beta_1$. There is an implicit sum over $b$. Using equation \eqref{eq:brane_amplitude_transport}, we have:
\begin{equation}
\begin{aligned}
    \braket{b^{\lambda}|D}\eta^{ab}\ket{a^\lambda} &= \Phi_{V,i}^*\left[ \braket{b^\lambda|D}\eta^{ab}\ket{a^\lambda}\right] \\
    &= (\Phi_{1,i}^*\braket{b^\lambda|D} \eta^{ab})\Phi_V^* \ket{a^\lambda} \\
    &= \Phi_{1,i}^*(\braket{b^\lambda|D}\eta^{ab}){M^{(i),c}_{a}}\ket{c^\lambda} 
\end{aligned}
\end{equation}
where in the above, all objects are evaluated at $t_1=0$ and arbitrary $(\beta_1,\bar{\beta}_1)$. We conclude that:
\begin{equation}\label{eq:explicit_matrix_difference}
    (\Phi_{1,i}^{*})^{-1}\braket{a^\lambda|D} =  G^{(i),b}_{a} \braket{b^\lambda|D}
\end{equation}
where for convenience we have defined
\begin{equation}
    G_{a}^{(i),b} :=  (\Phi_{1,i}^*)^{-1}\left( \eta_{ad} M^{(i),d}_{c}\right)\eta^{cb} .
\end{equation}
Note that this holds for any D--brane, not just the thimble branes we have used here. Also in the above and in the following, the brane amplitudes are computed at $t_1=0$. The above equation is a matrix difference equation: if we regard $\braket{\cdot | D}$ as an $N$-vector with components $\braket{ a^\lambda | D}$ we can write this as
\begin{equation}\label{eq:matrix_difference_equation}
    (\Phi_{1,i}^{*})^{-1} \braket{\cdot | D} =  G^{(i)} \, \braket{\cdot | D}.
\end{equation}
To the best of our knowledge, this difference equation is novel.\footnote{To date, it seems as though only \textit{differential} equations, often known as Pichard--Fuchs equations,  arising from the $tt^*$ geometry associated with the K\"ahler (Fayet--Iliopoulos) parameter have been studied, see \textit{e.g.} \cite{Blok:1991bi, Lerche:1991wm, cadavid1991picard, Morrison:1991cd}} 

We will now show that the difference equations provide a quantisation of the Cherkis--Kapustin spectral variety (we will be able to derive stronger results more directly for hemisphere partition functions in the next subsection). First of all, notice that the operators $(\Phi_{1,i}^*)^{-1}$ and the operators acting as multiplication by $\beta_{1,i}$ satisfy
\begin{equation}
    [(\Phi_{1,j}^*)^{-1},\beta_{1,i}] = - 2i\lambda  \delta_{ij} (\Phi_{1,j}^*),
\end{equation}
and that $\lim_{\lambda\rightarrow 0} \beta_1 = w$. Thus, these operators can genuinly be viewed as quantisations of the operators $p$ and $w$, respectively.

To investigate the limit as $\lambda \rightarrow 0$ of $G^{(i)} $, we make use of a particularly nice set of brane amplitudes, namely \textit{thimble branes} $D_\alpha$, whose boundary amplitudes give a fundamental basis of flat sections for the $tt^*$ Lax connection \cite{Cecotti:1992rm,Gaiotto:2011tf}. For LG models, they are Lagrangian submanifolds projecting to straight lines in the $W$-plane beginning at critical points $\alpha$ of $W$. For GLSMs which flow in the IR to NLSMs, they are the holomorphic Lagrangian submanifolds of $X$ corresponding to attracting submanifolds of fixed points (\textit{i.e.} vacua $\{\alpha\}$)  for the Morse flow generated by $w_2$. Such boundary conditions were analysed explicitly for massive $(2,2)$ theories in \cite{Hori:2000ck, Gaiotto:2015aoa} and for 3d $\mathcal{N}=4$ theories in \cite{Bullimore:2016nji, Bullimore:2020jdq, Bullimore:2021rnr, Crew:2023tky}. 

Note that the difference equation \eqref{eq:explicit_matrix_difference} holding for any B--brane $D$ is equivalent to it holding for each of the thimble branes. This is because any brane amplitude can be written as a $\bZ$-linear combination of the $\{D_a\}$ amplitudes 
\begin{equation}\label{eq:thimbles_basis}
    \Pi[D] = \sum_{\alpha} n_{\alpha} \Pi[D_{\alpha}]
\end{equation}
where $n_{\alpha}$ are the framed BPS degeneracies \cite{Gaiotto:2011tf}.

A key fact we will make extensive use of is that the asymptotic behaviour in $\lambda$ of the thimble brane amplitudes is known:
\begin{equation}\label{eq:thimble_brane_asymptotic}
    \braket{b^\lambda|D_\alpha}\sim\braket{b|D_\alpha} \sim e^{\frac{W_{\text{eff}}^{(\alpha)}}{\lambda}} \cO_b|_{\alpha},\qquad\text{as } \lambda \rightarrow 0.
\end{equation}
In the above, the effective twisted superpotential is computed at an RG scale $\mu = \lambda$. Here, the effective twisted superpotential $W_{\text{eff}}^{(\alpha)}$ is evaluated at the vacuum solution for $\sigma$ specified by requiring that in the limit where the (exponential of) the FI parameters goes to zero it reduces to its value in the classical vacuum configuration $\alpha$. This is discussed for supersymmetric QED in Section~\ref{subsubsec:sqed2_asymptotics}.

Let us now see in what sense this provides a quantisation of the $\lambda=0$ spectral variety. If we denote $\braket{\mathbf{1}|D_{\alpha}}$ the thimble brane amplitude with the trivial (no) operator insertion, then from \eqref{eq:thimble_brane_asymptotic} we note that:
\begin{equation}
    \lim_{\lambda \rightarrow 0 }\frac{ (\Phi_{1,i}^{*})^{-1}\braket{a^{\lambda}|D_{\alpha}}}{\braket{\mathbf{1}|D_{\alpha}}} = e^{-2 i \frac{\partial W_{\text{eff}}^{(\alpha)}}{\partial w_i} } \cO_a|_{\alpha}.
\end{equation}
In the above, we have traded $(\Phi_{1,i}^{*})^{-1}$ for a shift $w_i \rightarrow w_i-2i\lambda  $, which is valid in the $\lambda \rightarrow 0$ limit due to the coordinate definitions \eqref{eq:mochizukisecondcoords}. Here $\cO_a|_{\alpha}$ denotes the evaluation of the operator $\cO_a$ in the vacuum $\alpha$. Note that $\exp{-2 i \partial_{w_i} W_{\text{eff}}^{(\alpha)}} $ for $\alpha =1,\ldots,N$ are precisely the solutions for $p_i$ in the spectral variety \eqref{eq:spectral_curve}.

In an abuse of notation let $\mathcal{L}_i$ be the $n$ holomorphic functions on $(\bC \times {\bC^*})^n$ such that
\begin{equation}
    \cL_i( w,p)=0
\end{equation}
cuts out the spectral variety~\eqref{eq:spectral_curve}, found by eliminating $\sigma$ from the combined system \eqref{eq:glsm_spectral_curve}. We therefore know that
\begin{equation}
    \lim_{\lambda \rightarrow 0 } \frac{\cL_i(w, (\Phi_{1,j}^{*})^{-1}) \braket{\cdot |D_{\alpha}}}{\braket{\mathbf{1}|D_{\alpha}}} = 0.
\end{equation}
Using \eqref{eq:matrix_difference_equation}, this allows to conclude that
\begin{equation}
    \lim_{\lambda \rightarrow 0 } \frac{\cL_i(w, \{G^{(j)}\}) \braket{\cdot |D_{\alpha}}}{\braket{\mathbf{1}|D_{\alpha}}} = 0
\end{equation}
Since this holds for the basis of thimble amplitudes $\braket{\cdot|D_{\alpha}}$, we conclude that:
\begin{equation}\label{eq:brane-amp-quant}
   \lim_{\lambda \rightarrow 0}  \cL_i(w, G) =0.
\end{equation}
Thus, by Cayley--Hamilton, as $\lambda \rightarrow 0$ the eigenvalues of $G^{(i)}$ tend to $\exp{-2 i \partial_{w_i} W_{\text{eff}}^{(\alpha)}} $ for $\alpha =1,\ldots,N$.

\begin{mdframed}[roundcorner=10pt, linewidth=0.3mm, linecolor=gray]

In conclusion, we can view the operators $\Phi_{V,i}^* , (\Phi_{V,i}^*)^{-1}$ and multiplication by $\beta_{1,i}$ as quantisations of the generators $p_i^{\pm},w_i$. The module $(V,F)$ defines a module for the algebra of these quantised operators. Moreover, we have derived novel difference equations for brane amplitudes, which provide a quantisation (parameterised by $\lambda$) of the spectral variety of the monopole associated to $0$-difference modules.  
\end{mdframed}

\subsection{Difference equations for hemisphere partition functions}\label{sec:conf_vortex}

We have remarked above that D--brane amplitudes are difficult to compute in general. This is because they are \textit{not} BPS objects, as the supercharges $Q_\lambda$, $\bar{Q}^\dagger_{\lambda}$ preserved at the boundary of the cigar and the $Q_A$ supercharge preserved in the bulk due to the topological twist  differ. However, we have also remarked that in the so--called conformal limit these are expected to degenerate into hemisphere partition functions \cite{Cecotti:2013mba}. The conformal limit corresponds to taking 
\begin{equation}\label{eq:conf-lim}
   \mathrm{lim}_c :\quad \lambda \rightarrow 0,\quad L \rightarrow 0,\quad \frac{\lambda}{L} = \ep.
\end{equation}
Here $\ep$ is an arbitrary constant, and $L$ is the radius of the circle of the cylinder on which our system is quantised. Thus, we will explicitly re--introduce this length scale $L$ in this section.

With $L$ made explicit, we define the complex mass $m$ and normalised holonomy $x$ (with period $1$) such that $w = -2iL^2 m$ and $t = L x$. This ensures the dimensions of the summands in
\begin{gather}
    \beta_1 = -2 i L^2 m + 2 i \lambda L x + 2 i \lambda^2 L^2 \bar{m} \\
    t_1 = Lx + \text{Im}(  2i \lambda L^2 \bar{m})
\end{gather}
are consistent. Thus, the conformal limit of the brane amplitudes is:
\begin{equation}
    \mathrm{lim}_c\, \Pi[D,a](t_1=0,\beta_1,\bar{\beta}_1) = \mathcal{Z}_D[\cO_a, m-\epsilon x].
\end{equation}
Here $\mathcal{Z}_D[\cO_a]$ denotes the hemisphere partition function with boundary condition $D$ on $S^1 = \partial HS^2$, and a twisted chiral ring operator $\cO_a$ inserted at the tip of the hemisphere. The radius of the hemisphere is given by $\epsilon^{-1}$. For a certain choice of boundary condition (all chiral multiplets equipped with Neumann boundary conditions on $\partial HS^2$), they are also equivalent to the vortex partition functions on $\bR^2_{\ep}$ computed in an Omega background \cite{Fujimori:2015zaa}, with the Omega deformation parameter $\ep$.

In our conventions, $m-\epsilon x$ appears in place of the usual complex mass deformation $m$ in the hemisphere partition function of \cite{Honda:2013uca}, as we shall see in our examples. We will replace this combination simply by $m$ in the following. The hemisphere partition functions \textit{are} BPS objects that can in fact be computed explicitly using localisation techniques~\cite{Hori:2013ika, Honda:2013uca,Sugishita:2013jca}, and are holomorphic in the parameter $m$. 

Let us briefly recap why the D--brane amplitudes are expected to degenerate into the hemisphere partition functions in this limit. In the conformal limit, the Lax operators become
\begin{equation}
     L_{w,t} \rightarrow D_w + \frac{i}{2 \epsilon}(D_t-i\phi),\qquad \bar{L}_{\bar{w},t} \rightarrow \partial_{\bar{w}}.
\end{equation}
In the second limit we have used the fact we are working in topological gauge $(A_{\bar{w}})_a{}^{b}=0$.\footnote{This follows via the usual argument, as $(A_{\bar{w}})_a{}^{b} = \bra{a} \partial_{\bar{w}} \ket{b}$, where $\ket{a}$ are the states generated on a circle represented by a path--integral on a right cigar, with the insertion of $\cO_a$. In the mirror (T--dual) picture, acting with  $\partial_{\bar{w}}$ brings down insertions of $d^2 \bar{\theta}$ integrals of periodic twisted chiral fields $Y_i$ (integrals of twisted chiral ring operators over half the odd part of superspace). Thus in the GLSM frame, it brings down integrals of the supersymmetric completion of twist operators for the winding of the phase of chiral multiplets. These insertions are $Q_A$-exact, and the $Q_A$ can then be brought to act on $\bra{a}$, which is generated by a left cigar path--integral. $\bra{a}$ is a ground state and thus annihilated by $Q_A$, so we conclude that $(A_{\bar{w}})_a{}^{b}=0$.} This is consistent with the holomorphy of hemisphere partition functions in the complex mass. For LG models, the solutions to such equations are given by period integrals \cite{Hori:2000ck}, which have been shown to equal the results of localisation for their mirror dual GLSMs \cite{Fujimori:2012ab}. Later in this section we also compute the hemisphere (vortex) partition functions for some examples, and verify they satisfy the conformal limit of the difference equations noted above for D--brane amplitudes. This gives further support for the claim that D--brane amplitudes reduce to the hemisphere partition functions in the conformal limit. We refer the reader to \cite{Cecotti:2013mba} for further evidence, in particular the explicit example of the free chiral. 

\subsubsection{Finite--difference equations for hemisphere partition functions}\label{sec:finite_difference_HPS}

We now write out explicitly difference equations that are obeyed by hemisphere or vortex partition functions, which emerge by taking the conformal limit of~\eqref{eq:explicit_matrix_difference}. Analogously to the brane amplitudes case, to the best of our knowledge these difference equations are also novel. Let us substitute $m-\ep x \rightarrow m$, which is the usual complex mass appearing in the Lagrangian, superpotential \textit{etc}. Denoting
\begin{equation}\label{eq:diff_operators}
    \hat{p}_i= e^{\epsilon \partial_{m_i}},\qquad\hat{m}_i = \times m_i,
\end{equation}
one has
\begin{equation}
[\hat{p}_i,\hat{m}_j ]= \epsilon \delta_{ij} \hat{p}_j \qquad\Rightarrow\qquad \hat{p}_i: m_i \rightarrow m_i+\epsilon.
\end{equation}
That is, $\hat{p}_i$ is a difference operator for $m_i$. We see that $(\hat{p}, \hat{m})$ provide a quantisation of the algebra of functions on $T_{\bC}\times \mathfrak{t}_{\bC}$.

Note that $\hat{p}_i$ coincides with the operator $(\Phi_{1,i}^{*})^{-1}:\beta_1 \rightarrow \beta_1-2i\lambda L$ in the conformal limit. Therefore
\begin{equation}\label{eq:matrix_difference_vortex}
    \hat{p}_i\, \mathcal{Z}_D[\cO_a, m ] = \mathcal{Z}_D[\cO_a, m +\ep e_i] =  \widetilde{G}^{(i)}_{ab}(m,\ep) \mathcal{Z}_D[\cO_b, m ].
\end{equation}
Here, $\widetilde{G}^{(i)}= \mathrm{lim}_c\,G^{(i)}$, where $G^{(i)}$ is the matrix appearing in the difference equations for brane amplitudes \eqref{eq:matrix_difference_equation}. 

It is straightforward to show that an equation similar to~\eqref{eq:brane-amp-quant} (but with $\lambda$ replaced by $\epsilon$) must hold for hemisphere partition functions, and so that also in the conformal limit we obtain equations quantising the Cherkis-Kapustin spectral variety. Due to \eqref{eq:thimbles_basis}, the difference equations~\eqref{eq:matrix_difference_vortex} hold for any B--type boundary condition or brane $D$ if and only if it holds for the hemisphere partition functions equipped with each of the thimble branes. The $\epsilon \rightarrow 0$ behaviour of the hemisphere partition functions equipped with the thimble brane boundary conditions $\{D_{\alpha}\}$ can be derived from the asymptotic behaviour of the respective thimble brane amplitudes \eqref{eq:thimble_brane_asymptotic}:
\begin{equation}
    \cZ_{D_{\alpha}}[\cO_b,m] \sim e^{\frac{W_{\text{eff}}^{(\al)}}{\ep}} \cO_b, \quad \text{as } \ep \rightarrow 0.
\end{equation}
This is consistent with the limit for thimble brane amplitudes as upon reintroducing the circle length $L$ the superpotential is rescaled $W \rightarrow L W$. Thus, the same arguments as before lead to:
\begin{equation}
    \lim_{\ep \rightarrow 0 } \frac{\cL_i(\hat{m}, \hat{p}) \, \cZ_{D_{\alpha}}[\,\cdot\,,m]}{\cZ_{D_{\alpha}}[\mathbf{1},m]} = 0,
\end{equation}
In the above we have substituted $w=-2i L^2 m$ and replaced $m$ and $p$ with the difference operators \eqref{eq:diff_operators}, and we recall that the equations $\cL_i(w,p)=0$ for $i=1,\ldots n$ cut out the Cherkis-Kapustin spectral variety. This implies, similarly as for the brane amplitudes, that:
\begin{equation}\label{eq:vortex_spectral_curve}
    \lim_{\ep \rightarrow 0} \cL_i(\{m_j\}, \{\widetilde{G}^{(j)}\}) = 0,
\end{equation}
and by Cayley--Hamilton that $\widetilde{G}^{(i)}$ has eigenvalues $\exp \partial_{m_i} W_{\text{eff}}^{(\alpha)}$.

In the present case of hemisphere partition functions, we can however derive more stringent, basis--dependent results. This is because we can import localisation formulae \cite{Honda:2013uca, Hori:2013ika, Sugishita:2013jca} that express $\mathcal{Z}_D[\cO_a, m ]$ as a contour integral over the Coulomb branch scalars $\sigma$, and $\cO_a$ is represented by a polynomial in $\sigma$. The integrand scales as $\ep \rightarrow 0$ as $e^{W_{\text{eff}}[\sigma,m]/\ep}$, and so in the integral and in the limit,  $\hat{p}_i$ acts precisely as multiplication by $e^{\partial_{m_i} W_{\text{eff}}[\sigma,m]}$, recovering its action in the $0$-difference module case, as described in section \ref{sec:physical_examples}. In fact, we can verify that:
\begin{equation}\label{eq:limit_matrix}
    \lim_{\ep \rightarrow 0 }\widetilde{G}^{(i)}(m,\ep) = F^{(i)}(m)
\end{equation}
where $F$ is the automorphism appearing in the $0$-difference module \eqref{eq:0_difference}. Thus~\eqref{eq:matrix_difference_vortex} exhibits $QH^\bullet_T(X)$ as a module for the quanitsed algebra of functions $\bC[\hat{p}^{\pm1}, \hat{w}]$. In summary:

\begin{mdframed}[roundcorner=10pt, linewidth=0.3mm, linecolor=gray]
   The hemisphere (vortex) partition functions obey difference equations, which demonstrate that the insertions of twisted chiral ring elements \textit{i.e.} elements of $QH^\bullet_T(X)$, in these partition functions, furnish a module for the quantised algebra of functions on $T_{\bC} \times \mathfrak{t}_{\bC}$. The difference equations provide a quantisation of the spectral variety.
   
\end{mdframed}

Beautifully, due to the calculable nature of hemisphere and vortex partition functions, this gives a recipe, arising from 2d GLSMs, to construct solutions (often involving hypergeometric functions) to difference equations arising as deformed spectral varieties (which in turn also correspond to quantum equivariant cohomologies of K\"ahler varieties). Note that hemisphere partition functions can be interpreted as equivariant Gromov--Witten invariants of the Higgs branches \cite{Bonelli:2013mma}.  

We now exhibit examples of the constructions above. In the main body, in the interest of brevity, we include the examples of the free hyper and SQED$[2]$ for which the Higgs branches are $\mathbb{C}$ and $\mathbb{CP}^1$ respectively. In appendix \ref{appendix:sqed3}, we present the example of SQED$[3]$, for which $X= \mathbb{CP}^2$. This is notably a rank-$2$ example: the Berry connection is an connection for an $SU(3)$ bundle solving the generalised Bogomolny equations over $(\mathbb{C}\times S^1)^2$, and there are two difference equations the hemisphere partition functions satisfy.

\subsubsection{Example: free chiral}

We start with the simplest example of the free chiral multiplet, \textit{i.e.} an NLSM with target $\bC$. As usual we introduce a mass $m$ for the single $U(1)$ flavour symmetry rotating the chiral. There is a single vacuum $\mathbf{0}$ at the origin of $\bC$, with effective twisted superpotential given by \eqref{eq:superpotential_chiral}. We set the energy scale $\mu = \ep$.

For the vacuum $\mathbf{0}$, there are two possible thimble branes corresponding to the attracting Lagrangians for a choice of chamber of $\text{Re}(m) = -w_2/2$:
\begin{equation}
    \mathfrak{C}_{+} = \{m > 0 \}, \qquad \mathfrak{C}_{-} = \{m < 0 \}.
\end{equation}
These are either the whole of $\bC$, or just the origin. As boundary conditions these correspond to either Neumann $(N)$ or Dirichlet $(D)$ boundary conditions for the chiral. The only twisted chiral ring operator is simply the identity. 

The hemisphere partition functions for these boundary conditions can be found in \cite{Honda:2013uca}:
\begin{equation}
    \mathcal{Z}_{N} = \Gamma\left[\frac{m}{\ep}\right],\qquad
    \mathcal{Z}_{D} = \frac{-2\pi i e^{\frac{i \pi m}{\ep}}}{\Gamma\left[1-\frac{m}{\ep}\right]},
\end{equation}
where $\Gamma$ is Gamma function. Using the identity $\Gamma(x+1)=x\Gamma(x)$, it is easy to see that:
\begin{equation}
   \left( \hat{p}-\frac{m}{\ep}\right) \mathcal{Z}_{N}[m] = 0, \qquad \left( \hat{p}-\frac{m}{\ep}\right) \mathcal{Z}_{D}[m] = 0.
\end{equation}
Thus the hemisphere partition function (in either chamber) provides a quantisation of the Cherkis--Kapustin spectral variety \eqref{eq:CK_free}.

\subsubsection{Example: $\mathbb{C}\mathbb{P}^1$}

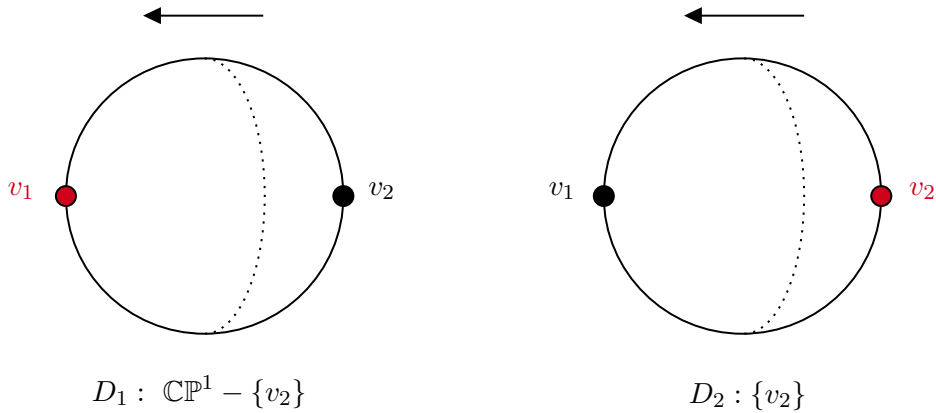
\begin{figure}   \centering

  
\tikzset {_4em7x9s1j/.code = {\pgfsetadditionalshadetransform{ \pgftransformshift{\pgfpoint{89.1 bp } { -128.7 bp }  }  \pgftransformscale{1.32 }  }}}
\pgfdeclareradialshading{_284suey81}{\pgfpoint{-72bp}{104bp}}{rgb(0bp)=(1,1,1);
rgb(0bp)=(1,1,1);
rgb(25bp)=(0.82,0.01,0.11);
rgb(400bp)=(0.82,0.01,0.11)}
\tikzset{_ar3t0txhx/.code = {\pgfsetadditionalshadetransform{\pgftransformshift{\pgfpoint{89.1 bp } { -128.7 bp }  }  \pgftransformscale{1.32 } }}}
\pgfdeclareradialshading{_jhb5n2thn} { \pgfpoint{-72bp} {104bp}} {color(0bp)=(transparent!0);
color(0bp)=(transparent!0);
color(25bp)=(transparent!40);
color(400bp)=(transparent!40)} 
\pgfdeclarefading{_evzegvolg}{\tikz \fill[shading=_jhb5n2thn,_ar3t0txhx] (0,0) rectangle (50bp,50bp); } 

  
\tikzset {_vgnoi74zt/.code = {\pgfsetadditionalshadetransform{ \pgftransformshift{\pgfpoint{89.1 bp } { -128.7 bp }  }  \pgftransformscale{1.32 }  }}}
\pgfdeclareradialshading{_ukzk2nnf3}{\pgfpoint{-72bp}{104bp}}{rgb(0bp)=(1,1,1);
rgb(1.607142857142857bp)=(1,1,1);
rgb(25bp)=(0.61,0.61,0.61);
rgb(400bp)=(0.61,0.61,0.61)}
\tikzset{_61sun0l58/.code = {\pgfsetadditionalshadetransform{\pgftransformshift{\pgfpoint{89.1 bp } { -128.7 bp }  }  \pgftransformscale{1.32 } }}}
\pgfdeclareradialshading{_53cfrpsln} { \pgfpoint{-72bp} {104bp}} {color(0bp)=(transparent!0);
color(1.607142857142857bp)=(transparent!0);
color(25bp)=(transparent!44.99999999999999);
color(400bp)=(transparent!44.99999999999999)} 
\pgfdeclarefading{_gnx4dp8jq}{\tikz \fill[shading=_53cfrpsln,_61sun0l58] (0,0) rectangle (50bp,50bp); } 
\tikzset{every picture/.style={line width=0.75pt}} 

\begin{tikzpicture}[x=0.75pt,y=0.75pt,yscale=-1,xscale=1]

\path  [shading=_284suey81,_4em7x9s1j,path fading= _evzegvolg ,fading transform={xshift=2}] (101.5,130.75) .. controls (101.5,92.5) and (132.5,61.5) .. (170.75,61.5) .. controls (209,61.5) and (240,92.5) .. (240,130.75) .. controls (240,169) and (209,200) .. (170.75,200) .. controls (132.5,200) and (101.5,169) .. (101.5,130.75) -- cycle ; 
 \draw  [color={rgb, 255:red, 0; green, 0; blue, 0 }  ,draw opacity=1 ] (101.5,130.75) .. controls (101.5,92.5) and (132.5,61.5) .. (170.75,61.5) .. controls (209,61.5) and (240,92.5) .. (240,130.75) .. controls (240,169) and (209,200) .. (170.75,200) .. controls (132.5,200) and (101.5,169) .. (101.5,130.75) -- cycle ; 

\path  [shading=_ukzk2nnf3,_vgnoi74zt,path fading= _gnx4dp8jq ,fading transform={xshift=2}] (370,130.75) .. controls (370,92.5) and (401,61.5) .. (439.25,61.5) .. controls (477.5,61.5) and (508.5,92.5) .. (508.5,130.75) .. controls (508.5,169) and (477.5,200) .. (439.25,200) .. controls (401,200) and (370,169) .. (370,130.75) -- cycle ; 
 \draw   (370,130.75) .. controls (370,92.5) and (401,61.5) .. (439.25,61.5) .. controls (477.5,61.5) and (508.5,92.5) .. (508.5,130.75) .. controls (508.5,169) and (477.5,200) .. (439.25,200) .. controls (401,200) and (370,169) .. (370,130.75) -- cycle ; 

\draw  [color={rgb, 255:red, 0; green, 0; blue, 0 }  ,draw opacity=1 ][fill={rgb, 255:red, 208; green, 2; blue, 27 }  ,fill opacity=1 ] (96.5,130.75) .. controls (96.5,127.99) and (98.74,125.75) .. (101.5,125.75) .. controls (104.26,125.75) and (106.5,127.99) .. (106.5,130.75) .. controls (106.5,133.51) and (104.26,135.75) .. (101.5,135.75) .. controls (98.74,135.75) and (96.5,133.51) .. (96.5,130.75) -- cycle ;
\draw  [fill={rgb, 255:red, 0; green, 0; blue, 0 }  ,fill opacity=1 ] (235,130.75) .. controls (235,127.99) and (237.24,125.75) .. (240,125.75) .. controls (242.76,125.75) and (245,127.99) .. (245,130.75) .. controls (245,133.51) and (242.76,135.75) .. (240,135.75) .. controls (237.24,135.75) and (235,133.51) .. (235,130.75) -- cycle ;
\draw  [fill={rgb, 255:red, 0; green, 0; blue, 0 }  ,fill opacity=1 ] (365,130.75) .. controls (365,127.99) and (367.24,125.75) .. (370,125.75) .. controls (372.76,125.75) and (375,127.99) .. (375,130.75) .. controls (375,133.51) and (372.76,135.75) .. (370,135.75) .. controls (367.24,135.75) and (365,133.51) .. (365,130.75) -- cycle ;
\draw  [fill={rgb, 255:red, 208; green, 2; blue, 27 }  ,fill opacity=1 ] (503.5,130.75) .. controls (503.5,127.99) and (505.74,125.75) .. (508.5,125.75) .. controls (511.26,125.75) and (513.5,127.99) .. (513.5,130.75) .. controls (513.5,133.51) and (511.26,135.75) .. (508.5,135.75) .. controls (505.74,135.75) and (503.5,133.51) .. (503.5,130.75) -- cycle ;
\draw  [draw opacity=0][dash pattern={on 0.84pt off 2.51pt}] (170.75,61.5) .. controls (187.32,61.5) and (200.75,92.5) .. (200.75,130.75) .. controls (200.75,169) and (187.32,200) .. (170.75,200) -- (170.75,130.75) -- cycle ; \draw  [dash pattern={on 0.84pt off 2.51pt}] (170.75,61.5) .. controls (187.32,61.5) and (200.75,92.5) .. (200.75,130.75) .. controls (200.75,169) and (187.32,200) .. (170.75,200) ;  
\draw    (200,40) -- (143,40) ;
\draw [shift={(140,40)}, rotate = 360] [fill={rgb, 255:red, 0; green, 0; blue, 0 }  ][line width=0.08]  [draw opacity=0] (8.93,-4.29) -- (0,0) -- (8.93,4.29) -- cycle    ;
\draw    (470,40) -- (413,40) ;
\draw [shift={(410,40)}, rotate = 360] [fill={rgb, 255:red, 0; green, 0; blue, 0 }  ][line width=0.08]  [draw opacity=0] (8.93,-4.29) -- (0,0) -- (8.93,4.29) -- cycle    ;
\draw  [draw opacity=0][dash pattern={on 0.84pt off 2.51pt}] (440,61.5) .. controls (440,61.5) and (440,61.5) .. (440,61.5) .. controls (456.57,61.5) and (470,92.5) .. (470,130.75) .. controls (470,169) and (456.57,200) .. (440,200) -- (440,130.75) -- cycle ; \draw  [dash pattern={on 0.84pt off 2.51pt}] (440,61.5) .. controls (440,61.5) and (440,61.5) .. (440,61.5) .. controls (456.57,61.5) and (470,92.5) .. (470,130.75) .. controls (470,169) and (456.57,200) .. (440,200) ;  

\draw (71,122.4) node [anchor=north west][inner sep=0.75pt]  [color={rgb, 255:red, 208; green, 2; blue, 27 }  ,opacity=1 ]  {$v_{1}$};
\draw (251,122.4) node [anchor=north west][inner sep=0.75pt]    {$v_{2}$};
\draw (341,122.4) node [anchor=north west][inner sep=0.75pt]    {$v_{1}$};
\draw (521,122.4) node [anchor=north west][inner sep=0.75pt]  [color={rgb, 255:red, 208; green, 2; blue, 27 }  ,opacity=1 ]  {$v_{2}$};
\draw (113,219.4) node [anchor=north west][inner sep=0.75pt]    {$D_{1} :\ \mathbb{CP}^{1} -\{v_{2}\}$};
\draw (411,222.4) node [anchor=north west][inner sep=0.75pt]    {$D_{2} :\{v_{2}\}$};

\end{tikzpicture}
    \caption{The support of the thimble boundary conditions for vacua $v_1$ and $v_2$ for supersymmetric QED with two chirals, \textit{i.e.} the $\mathbb{CP}^1$ sigma model. The arrow indicates the direction of Morse flow.}
    \label{fig:sqed2_thimbles}
\end{figure}

We now turn to supersymmetric QED with two chirals, which flows to a non--linear sigma model to $\mathbb{CP}^1$ in the IR. For the sake of brevity, we work in a fixed chamber for the real part of the mass parameter: $\text{Re}(m) > 0$. We will denote the vacua $v_1$ and $v_2$, for which the thimble branes should be supported (in the NLSM) on:
\begin{equation}
    D_1:\quad \mathbb{CP}^1 - \{v_2\}, \qquad D_2:\quad \{v_2\}.
\end{equation}
These are illustrated in figure \ref{fig:sqed2_thimbles}.

If we denote the two chirals $\Phi_1$ and $\Phi_2$, which have charges $(+1,+1)$ and $(+1,-1)$ under $G\times G_F$ respectively, only $\Phi_1$ obtains a VEV in vacuum $v_1$, and $\Phi_2$ in $v_2$. In the UV GLSM, the thimble branes are engineered by assigning the following boundary conditions to the chirals:
\begin{equation}
    D_1\,:\quad \Phi_1,\Phi_2 \text{ Neumann}, \qquad 
    D_2\,:\quad \Phi_1 \text{ Dirichlet}, \Phi_2 \text{ Neumann}.
\end{equation}
In the opposite chamber, boundary conditions are the same but with $1\leftrightarrow 2$ in the above. The twisted chiral ring of supersymmetric QED is generated by $\{\mathbf{1},\sigma\}$ and is subject to the relation \eqref{eq:bethe}.

We now proceed to compute the hemisphere partition functions equipped with these boundary conditions, and demonstrate the quantisation \eqref{eq:matrix_difference_vortex}~--~\eqref{eq:vortex_spectral_curve} of the spectral curve \eqref{eq:spectralcurve}. Hemisphere partition functions were computed via localisation in \cite{Honda:2013uca,Hori:2013ika,Sugishita:2013jca}. We primarily use \cite{Honda:2013uca}, as it includes the case with Dirichlet boundary conditions for chiral multiplets.

\paragraph{Vacuum 1.}

We start with the vacuum $v_1$, for which:
\begin{equation}\label{eq:sqed2_hps}
\begin{aligned}
    \mathcal{Z}_{D_1}[\mathbf{1}] &= 
    \oint_{\mathcal{C}_1} \frac{d\sigma}{2\pi i \ep}
    e^{-\frac{2\pi i \sigma \tau}{\ep}} \,\,\Gamma\left[\frac{\sigma+m}{\ep}\right]
    \Gamma\left[\frac{\sigma-m}{\ep}\right],\\
    \mathcal{Z}_{D_1}[\sigma] &= 
    \oint_{\mathcal{C}_1} \frac{d\sigma}{2\pi i \ep}
    e^{-\frac{2\pi i \sigma \tau}{\ep}} \,\,\Gamma\left[\frac{\sigma+m}{\ep}\right]
    \Gamma\left[\frac{\sigma-m}{\ep}\right]\sigma.
\end{aligned}
\end{equation}
The contour $\mathcal{C}_1$ encloses the poles of $\Gamma[\frac{\sigma + m}{\ep}]$ at $\sigma = -\ep k - m$, where $k \in \mathbb{N}_0$. Here, $\tau = \tau(\ep)= \tau_0 + \frac{2}{2\pi i }\log(\Lambda_0/\ep)$ is the renormalised FI parameter at energy $\mu=\ep$. Note that the first line in \eqref{eq:sqed2_hps} coincides with the vortex partition function computed on the Omega background on $\bR_{\ep}^2$ \cite{Fujimori:2015zaa}.

To compute these integrals we change variables using $x= \frac{\sigma+m}{\ep}$, and use the fact that
\begin{equation}
    \oint \frac{dx}{2\pi i} \Gamma[x]f(x) = \sum_{k=0}^{\infty} \frac{(-1)^k}{k!} f(-k),
\end{equation}
where the contour encloses all of the poles of $\Gamma[x]$, at $x=-k$ where $k\in \mathbb{N}_0$. We further use the definition of the generalised hypergeometric function
\begin{equation}
    {}_0F_1[a;z]= \sum_{k=0}^{\infty} \frac{z^k}{(a)_k k!},
\end{equation}
where $(a)_k := (a)(a+1)\ldots (a+k-1)$. Then it is easy to see that
\begin{equation}
\begin{gathered}
    \mathcal{Z}_{D_1}[\mathbf{1}] = e^{\frac{2\pi i m \tau}{\epsilon}}\Gamma\left[-\frac{2m}{\epsilon}\right] {}_{0}F_1\left[1+\frac{2m}{\epsilon} ; e^{2\pi i \tau}\right], \\ 
    \mathcal{Z}_{D_1}[\sigma] = -m \mathcal{Z}_{D_1}[\mathbf{1}]  + \ep \, e^{2\pi i \tau} e^{\frac{2\pi i m \tau}{\ep}} \Gamma\left[-1-\frac{2m}{\ep}\right]{}_0F_1\left[2+\frac{2m}{\ep}; e^{2\pi i \tau}\right].\\ 
\end{gathered}
\end{equation}
Using the standard identity
\begin{equation}\label{eq:hypergeom_identity}
    {}_{0}F_1(b,z) = {}_{0}F_1(b+1,z) + \frac{z}{b(b+1)} {}_{0}F_1(b+2,z),
\end{equation}
it is not hard to check that, after some algebra
\begin{equation}\label{eq:matrix_diff_sqed2}
    \hat{p} 
    \begin{pmatrix} 
    \mathcal{Z}_{D_1}[\mathbf{1}] \\ 
    \mathcal{Z}_{D_1}[\sigma]
    \end{pmatrix} 
    = 
    \widetilde{G}(m,\ep) 
    \begin{pmatrix} 
    \mathcal{Z}_{D_1}[\mathbf{1}] \\ 
    \mathcal{Z}_{D_1}[\sigma]
    \end{pmatrix}
\end{equation}
where
\begin{equation}\label{eq:sqed2_matrix}
    \widetilde{G}(m,\ep) = 
    \begin{pmatrix}
    1+m(2m+\ep)q^{-1} & (2m+\ep)q^{-1} \\
    (2m+\ep)(1+ m(m+\ep)q^{-1}) & 1+ (m+\ep)(2m+\ep)q^{-1}
    \end{pmatrix}.
\end{equation}
We have defined as before $q= \ep^2 e^{2 \pi i \tau}$.

\paragraph{Vacuum 2.}

For the second vacuum, recall that $\Phi_1$ is now assigned a Dirichlet boundary condition. The hemisphere partition functions are given by:
\begin{equation}
\begin{aligned}
    \mathcal{Z}_{D_2}[\mathbf{1}] &= 
    \oint_{\mathcal{C}_2} 
    \frac{d\sigma}{2\pi i \ep}
    e^{-\frac{2\pi i \sigma \tau}{\ep}} \,\,
    \frac{(-2\pi i)e^{ \frac{\pi i (\sigma+m)}{\ep}}}{\Gamma\left[1-\frac{\sigma+m}{\ep}\right]}
    \Gamma\left[\frac{\sigma-m}{\ep}\right],\\
    \mathcal{Z}_{D_2}[\sigma] &= 
    \oint_{\mathcal{C}_2} 
    \frac{d\sigma}{2\pi i \ep}
    e^{-\frac{2\pi i \sigma \tau}{\ep}} \,\,
    \frac{(-2\pi i)e^{ \frac{\pi i (\sigma+m)}{\ep}}}{\Gamma\left[1-\frac{\sigma+m}{\ep}\right]}
    \Gamma\left[\frac{\sigma-m}{\ep}\right] \sigma,
\end{aligned}
\end{equation}
where the contour $\mathcal{C}_2$ encloses the poles at $\sigma = -\ep k + m$. Note that $\Gamma[z]^{-1}$ is an entire function, and that as long as $\frac{\sigma+m}{\ep} \notin \bZ$, which is true on $\mathcal{C}_2$ for generic $m$, one can use the Euler reflection formula to write
\begin{equation}\label{eq:gamma_euler_reflection}
    \frac{(-2\pi i)e^{ \frac{\pi i (\sigma+m)}{\ep}}}{\Gamma\left[1-\frac{\sigma+m}{\ep}\right]}
    =
    \left( 1-e^{\frac{2\pi i (\sigma +m)}{\ep}}\right) \Gamma\left[\frac{\sigma +m}{\ep}\right].
\end{equation}
The factor $1-e^{\frac{2\pi i (\sigma +m)}{\ep}}$ evaluates to $1-e^{\frac{4\pi i m}{\ep}}$ on all of the poles, which is itself invariant under $\ep$-shifts of $m$. The remaining integrand is then identical to the one for $D_1$ in \eqref{eq:sqed2_hps} but with $m \rightarrow -m$ in the choice of contour. Thus:
\begin{equation}
\begin{aligned}
    \mathcal{Z}_{D_2}[\mathbf{1}] 
    &= \left(1-e^{\frac{4\pi i m}{\ep}}\right) \mathcal{Z}_{D_1}[\mathbf{1}]|_{m\rightarrow -m} \\
    &= \left(1-e^{\frac{4\pi i m}{\ep}}\right) e^{-\frac{2\pi i m \tau}{\epsilon}}\Gamma\left(\frac{2m}{\epsilon}\right) {}_{0}F_1\left[1-\frac{2m}{\epsilon}, e^{2\pi i \tau}\right].
\end{aligned}
\end{equation}
The same holds for $\cZ_{D_2}[\sigma]$, which after an application of  \eqref{eq:hypergeom_identity} can be expressed as
\begin{equation}
    \mathcal{Z}_{D_2}[\sigma] = m \mathcal{Z}_{D_2}[\mathbf{1}] + \left(1-e^{\frac{4\pi i m}{\ep}}\right) \ep e^{2\pi i \tau} e^{-\frac{2\pi i m \tau}{\ep}}    \Gamma\left[\frac{2m}{\ep}-1\right] {}_{0}F_{1}\left[2-\frac{2m}{\ep};e^{2\pi i \tau}\right].
\end{equation}
Acting with $\hat{p}$ and further extensive use of \eqref{eq:hypergeom_identity}, one finds that
\begin{equation}\label{eq:matrix_diff_sqed2_2}
    \hat{p} 
    \begin{pmatrix} 
    \mathcal{Z}_{D_2}[\mathbf{1}] \\ 
    \mathcal{Z}_{D_2}[\sigma]
    \end{pmatrix} 
    = 
    \widetilde{G}(m,\ep) 
    \begin{pmatrix} 
    \mathcal{Z}_{D_2}[\mathbf{1}] \\ 
    \mathcal{Z}_{D_2}[\sigma]
    \end{pmatrix}
\end{equation}
where $\widetilde{G}(m,\ep) $ is the \textit{same} matrix as \eqref{eq:sqed2_matrix}, as predicted in Section~\ref{sec:finite_difference_HPS}.

\paragraph{Spectral curve quantisation.}

Notice that $\widetilde{G}(m,0) = F(m)$ where $F(m)$ is the corresponding automorphism in the $0$-difference case \eqref{eq:sqed2_matrix_0_diff}. Thus \eqref{eq:matrix_diff_sqed2} and \eqref{eq:matrix_diff_sqed2_2} provide a quantisation of the corresponding action on  $QH_T(\mathbb{CP}^1)$. Notice also that
\begin{equation}
   \lim_{\ep \rightarrow 0} \cL( m, \widetilde{G}(m,\ep)) 
   = 0
\end{equation}
where $\cL$ defines the monopole spectral curve \eqref{eq:spectralcurve}.

Additionally one can check by acting twice with $\hat{p}$, and eliminating $\cZ_{D_1}[\sigma]$ for $\hat{p} \cZ_{D_1}[\mathbf{1}]$ and $\cZ_{D_1}[\mathbf{1}]$, that \textit{e.g.}:
\begin{equation}
    \left[\hat{p}^2 + 2\left(\frac{1+\frac{\ep}{2m}}{1+\frac{\ep}{2m}}+ (2m+3\ep)(m+\ep) q^{-1}\right)\hat{p}+ \frac{1+\frac{3\ep}{2m}}{1+\frac{\ep}{2m}}\right]\mathcal{Z}_{D_1}[\mathbf{1}] = 0,
\end{equation}
which is an explicit quantisation of the spectral curve. This can be written more symmetrically via acting on the left with $\hat{p}^{-1}$
\begin{equation}
    \cL_{\ep}(\hat{m},\hat{p}) \mathcal{Z}_{D_1}[\mathbf{1}] = 0
\end{equation}
where
\begin{equation}
    \cL_{\ep}(\hat{m},\hat{p}) = \left(1-\frac{\epsilon}{2m}\right)\hat{p}- 2\left(1+2m^2 e^{-2\pi i \tau_0}\left(1-\frac{\epsilon^2}{4m^2}\right)\right)+ \left(1+\frac{\epsilon}{2m}\right)\hat{p}^{-1}.
\end{equation}


\section{Spectral Data II -- Filtrations}\label{sec:spectral_data_2}

Consider a monopole solution occurring as the Berry connection of some 2d $(2,2)$ GLSM with Higgs branch $X$, or NLSM with target $X$. In the previous section we discussed how the solution can be encoded into difference modules, and how the modules can be constructed physically in terms of supersymmetric ground states in $Q_\lambda$ cohomology. We have further described how for $\lambda = 0$ these modules can be identified with the quantum equivariant cohomology $QH^\bullet_T(X)$ of $X$ as a module for a certain algebra of functions. We have described how the Cherkis-Kapustin spectral variety coincides with the support of this module. Moreover, we have discussed how this structure is deformed at $\lambda \neq 0$. The aim of this section is to describe yet \emph{another} algebraic object encoding aspects of the monopole solution that emerges whenever $|\lambda|=1$, and to relate it to another generalised cohomology theory of $X$: the equivariant K--theory $K_T(X)$.

For example, suppose that $\lambda=1$, $n=1$ and $N=2$. Then we have $M^{\lambda=1}\cong \mathbb{R}\times \mathbb{C}^*$ as a mini--complex manifold. One considers the space of supersymmetric ground states at each fixed real value as a vector bundle on $\mathbb{C}^*$. The fall off rate of the sections of these bundles as we scatter towards at $\pm \infty$ determine two filtrations. The support of sections falling with exponential rate in both directions (that is, matching these filtrations) carves out a set of points $\Delta \subset \mathbb{C}^*$ which are the analogue of the twistor spectral curve of Hitchin for monopoles in $\mathbb{R}^3$ (but this time considering only lines parallel to the real direction). In this section, we show how these filtrations are determined by physical data, and show that in the case of a GLSM with target space $X$, the locus corresponds to the gluing points of the equivariant K--theory variety of $X$, which in this looks like
\begin{equation}
    K_T(X) \cong \left(\mathbb{C}^* \sqcup \mathbb{C}^* \right) / \Delta.
\end{equation}
In principle, we expect to be able to reconstruct the monopole from a refined version of this algebraic data.\footnote{In particular, one ought to compactify $\mathbb{C}^*\subset M^{\lambda=1}$ to $\mathbb{P}^1$ and pay attention to the behaviour of the sections at $\{0,\infty\}\in \mathbb{P}^1$, which in similarity to the spectral data at $\lambda=0$ discussed in Section~\ref{sec:product_case} will also lead to pairs of filtrations, We will ignore these subtleties in this work.}

\subsection{Note on Riemann--Hilbert correspondences} 

What we have claimed above is that whenever $|\lambda|=1$ a periodic monopole solution possesses two distinct algebraic descriptions, which must therefore encode the same data. There is a natural mathematical context for the phenomenon we are observing, namely a Riemann--Hilbert correspondence that holds in the more general case $\lambda \neq 0$. Although we have not yet found a detailed published account of this correspondence for periodic monopoles (see~\cite{mochizuki2017periodic} for some preliminary remarks, and~\cite{Kontsevich_Soibelman_1} for an unpublished account), in the case of doubly periodic monopoles such correspondences have been established in a series of works, whose results are worth briefly recalling.

In~\cite{mochizuki2019doubly}, Mochizuki constructs from a doubly periodic monopole on $\mathbb{R}\times \Sigma$ where $\Sigma$ is an elliptic curve, a $q^{\lambda'}$-difference module with parabolic structure. Here $\lambda'$ is again a twistor parameter. When $|q^{\lambda'}|\neq 1$, by work of van der Put \& Reversat \cite{van2005galois} and Ramis, Sauloy \& Zhang \cite{ramis2009local},  and in the global case by Kontsevich \& Soibelman \cite{Kontsevich_Soibelman_1, Kontsevich_Soibelman_2}, there exists a Riemann--Hilbert correspondence that relates these difference modules (the so--called ``de Rahm" objects) to a locally free sheaf with some filtrations on the elliptic curve obtained from $\mathbb{C}^*$ by quotienting by the action of $(q^{\lambda})^\mathbb{Z}$ (the so--called ``Betti" object). The lift of our setup to 3d appears to be an interesting physical arena where to study this Riemann--Hilbert correspondence arising from doubly periodic monopoles, and we leave the study of these fascinating aspects to future work. Similarly, we hope that the results of this section, combined with our results presented in Section~\ref{sec:spectral_data_1}, present a natural and fruitful physical setup where the Riemann--Hilbert correspondence for periodic monopoles can be embedded. For example, to the best of our knowledge, the fact that in an NLSM with target $X$ this correspondence leads to a relation between a quantisation of an action on the equivariant quantum cohomology of $X$ and the equivariant K--theory of $X$ has not appeared in the literature. We expect this fact to lift in three dimensions to relations between elliptic cohomology and quantum K--theory.

\subsection{Filtrations}\label{sec:filtrations}

We now demonstrate how to recover some of the algebraic structures mentioned above via physical constructions in 2d $(2,2)$ gauged linear sigma models. We first describe how the vacuum structure of the theory assigns holomorphic filtrations to vector bundles on $(\bC^*)^n$, recovering aspects of the Betti--type spectral data, which is a higher rank generalisation of the spectral data described by Mochizuki \cite{mochizuki2019doubly} in the doubly periodic monopole case.\footnote{Such monopoles are in fact supersymmetric Berry connections for 3d $\cN=2$ theories, which we will explore in future work.} In the next subsection, we then demonstrate how to build the equivariant K--theory variety of the Higgs branch of the GLSM from an incidence variety for this set of filtrations for the Berry connection.\footnote{This construction is quite similar to that of the elliptic cohomology variety of the Higgs branches of 3d theories as in \cite{Bullimore:2021rnr}.}

Let us work in a theory with a rank $n$ abelian flavour symmetry $T \cong U(1)^n$, indexed by $i=1,\ldots,n$. Recall that for $|\lambda|=1$, the parameter space splits as:
\begin{equation}\label{eq:phase_parameter_splitting}
    (t_0,\beta_0) =\left(\frac{1}{2} (\sin\theta\, w_1 - \cos\theta\, w_2), e^{i\theta} ( \cos\theta \,w_1 + \sin\theta \,w_2 + i t)\right),
\end{equation}
where $\lambda \coloneqq e^{i\theta}$. The parameter space can thus be regarded as $(\bR \times \bC^*)^n$. Concretely, each $\bC$ in the original product space $\bC^n$ spanned by the $w$ coordinates is separated into two orthogonal coordinates, $(\sin\theta\, w_1 - \cos\theta\, w_2)_i$ and $( \cos\theta \,w_1 + \sin\theta \,w_2)_i$. The latter is paired with the coordinate $t$ on $(S^1)^n$ to form (after exponentiation) a $(\bC^*)^n$ parameter.

\paragraph{Notation.} In the interest of clarity, we will set $\lambda = 1$ from now on, although it is not difficult to see how analogous results hold for an arbitrary phase $\theta$, which simply correspond to a choice of real and imaginary axis on $\bC$. The compactification circle length is set to $L=1$ in the interest of clarity. We set $m_{\bR} + im_{\mathbb{I}} = \frac{i w}{2}$ and denote $\mathfrak{t} = \bR^n$ with coordinate so $m_{\bR} = - \frac{w_2}{2}$. We also set $z = \beta_0 = 2m_{\mathbb{I}} + i t$ for convenience. We let $\bT \coloneqq (\bC^*)^n$ be the complexification of $T$, with coordinate $x \coloneqq e^{2 \pi z}$. In the following, to obtain the results for an arbitrary phase $\lambda = e^{i\theta}$, simply replace $(m_{\bR},z)$ by their counterparts in \eqref{eq:phase_parameter_splitting}.

\subsubsection{Geometric preliminaries}\label{sec:geometric_prelims}

We will be interested in the asymptotic form (as $|w| \rightarrow \infty$) of the central charges, which is determined by the asymptotic form of the effective twisted superpotential. In order to describe this, we will find it useful to first introduce some objects in the Higgs branch geometry of the GLSM. In this work, we will make the assumption that the Higgs branch $X$ is Goresky--Kottwitz--MacPherson (GKM) variety for the Hamiltonian $T$-action \cite{goresky1998equivariant}. This means:
\begin{itemize}
    \item If $\Phi_{\alpha}$ denotes the collection of $T$-weights in the weight decomposition of $T_{\alpha}X$, the elements of $\Phi_{\alpha}$ are pairwise linearly independent for all fixed points \textit{i.e.} vacua $\alpha$.

    \item Equivalently, for every two fixed points $\alpha,\beta$ in $X^T$, there is no more than one $T$-equivariant curve (which must be a $\bC\bP^1$, labelled by a weight $\nu$ corresponding to $T_{\alpha}\Sigma_{\nu} \subset T_{\alpha}X$) connecting them.
\end{itemize}

This assumption is satisfied for many theories, such as general abelian theories where $X$ is toric, or quiver gauge theories where $X$ is a partial flag (this includes the Grassmannian and hence supersymmetric QCD), or a cotangent bundle thereof in the eight supercharge case. We have hyperplanes in $\mathfrak{t}$
\begin{equation}\label{eq:hyperplanes}
    \mathbb{W}_{\nu} = \{\nu \cdot m_{\bR} = 0\}.
\end{equation}
Let $T_{\nu}$ be the codimension-$1$ subtorus of $T$ generated by $\ker{\nu}$. From the above description of GKM manifolds, the hyperplanes $\mathbb{W}_{\nu}$ lie where $X^{T_{\nu}}$ no longer has isolated fixed points, but some extended locus opens up (which may be regarded as an extended moduli space of supersymmetric vacua). We denote the various chambers of the complement of the above hyperplanes by $\mathfrak{C}$. We will be particularly interested in those hyperplanes for weights $\nu$ labelling $\mathbb{CP}^1$ fixed locus connecting vacua $\alpha$ and $\beta$ as described above. These label the internal edges of a GKM diagram \cite{goresky1998equivariant, guillemin20011}. 

We also have a decomposition of weights for a fixed choice of chamber $\mathfrak{C}$:
\begin{equation}
\begin{gathered}
    \Phi_{\alpha} = \Phi_{\alpha}^{+} \sqcup  \Phi_{\alpha}^{-}, \\
    \Phi_{\alpha}^+ = \{\nu \in \Phi_{\alpha}\,|\, \nu \cdot m_{\bR} > 0\}, \qquad \Phi_{\alpha}^- = \{\nu \in \Phi_{\alpha}\,|\, \nu \cdot m_{\bR} < 0\}.
\end{gathered}
\end{equation}
The decomposition changes precisely when one crosses a hyperplane as in equation \eqref{eq:hyperplanes}.

We will also be interested in the following bilinear pairing between FI parameters and masses
\begin{equation}\label{eq:momentmap_central_charge}
    \kappa_{\alpha}(m, r) = h_{m_{\bR}} (\alpha)
\end{equation}
which arises from evaluating the moment map $h_{m_{\bR}} = m_{\bR} \cdot \mu_{T}$ for the $T$-action at the various fixed points. Here, $r$ is the real FI parameter appearing in the complexified FI parameter as $\tau_0 \coloneqq \frac{\theta}{2\pi}+ir$.
The values of this moment map induces the same ordering as that by the Morse flow generated by the masses. This ordering changes precisely as one crosses the hyperplanes  \eqref{eq:hyperplanes} corresponding to equivariant curves connecting vacua, with the two vacua corresponding to $\nu$ switched when the hyperplane is crossed.

\subsubsection{Asymptotic central charges \& filtrations}

With these details in hand, we can finally state the asymptotic behaviour of the effective twisted superpotential (at energy scale $\mu$) in terms of the geometry of the Higgs branch (\textit{i.e.} from the perspective of the non--linear sigma model to which the GLSM flows)
\begin{equation}\label{eq:asymptotic_superpotential}
     \frac{\partial W_{\text{eff}}^{(\alpha)}}{\partial\vec{m}} \rightarrow -2\pi i \kappa_{\alpha}(\tau(\mu), \,\cdot\, )+ \sum_{\nu \in \Phi_{\alpha}}\nu \log \left( \frac{\nu \cdot m}{\mu}\right) 
\end{equation}
in which $ \frac{\partial W_{\text{eff}}^{(\alpha)}}{\partial\vec{m}}$ is considered an element of $\Hom(\mathfrak{t}_{\bC},\bC)$. Here, $\tau_a$ is the FI parameter, defined for each component of
\begin{equation}
    \Hom(\pi_1(G),U(1)) = U(1)^r,\qquad a = 1,\ldots,r
\end{equation}
where $G$ is the gauge group. It is given by:
\begin{equation}
    \tau_a = \frac{\theta_a}{2 \pi} +  i r_{0,a} + \frac{b_a}{2\pi i } \log(\Lambda_0/\mu)
\end{equation}
where $r_{0,a}$ and $\theta_a$ are the bare FI parameter and instanton angle. Additionally, $\Lambda_0$ is some fixed UV energy scale, $\mu$ the RG scale, and $b_a = \sum_i Q^i_a$ where $Q^i_a$ is the charge of the $i^{\text{th}}$ chiral under $U(1)_a$. 

In particular, from the asymptotic form of the Berry connection \eqref{eq:asymptotics}, the adjoint Higgs field is given by the real part of the above
\begin{equation}\label{eq:asymptotic_Higgs_field}
      i \phi^{(\alpha)} \rightarrow 2\pi \kappa_{\alpha}(r, \,\cdot\,) + \sum_{\nu \in \Phi_{\alpha}} \nu \, \log \left| \frac{\nu \cdot m}{\mu}\right|,
\end{equation}
where $ \phi^{(\alpha)}$ represents the vector $\{ \phi^{(\alpha)}_i\}_{i=1,\ldots,n}$. Here $r_a = r_{0,a} - \frac{b_a}{2\pi} \log(\Lambda_0/\mu)$ is simply the imaginary part of $\tau_a$, and is the FI parameter at scale $\mu$.

We consider the scattering problem specified by a ray $v \in \mathfrak{C}$ the real mass parameter space, in a given chamber $\mathfrak{C}$
\begin{equation}
    v^i\left( D_{m_{\bR,i}} - i \phi_i\right)\psi = 0,
\end{equation}
off to infinity. Recall that the Bogomolny equations now imply that
\begin{equation}\label{eq:bogomolnyforktheory}
    [D_{m_\bR}, \bar{D}_{\bar{z}}]  = 0,
\end{equation}
and thus the Berry connection $\bar{D}_{\bar{z}}$ determines the structure of a rank $N$ (where $N$ is the number of ground states) holomorphic vector bundle
\begin{equation}
    \mathscr{E}_{m_{\bR}} \coloneqq E|_{\{m_{\bR}\} \times \mathbb{T},  \bar{D}_{\bar{z}}}
\end{equation} on each slice $\{m_{\bR}\} \times \mathbb{T}$, varying covariantly with respect to $m_{\bR}$. 

Due to the GCK condition \eqref{eq:GCK_condition}, the adjoint Higgs field dominates the parallel transport problem. As mentioned above, it coincides asymptotically with the real part of the effective twisted superpotential. The asymptotics \eqref{eq:asymptotic_Higgs_field} then imply that in a given chamber $\mathfrak{C}$, the holomorphic vector bundle $\mathscr{E} \coloneqq \mathscr{E}_{m_{\bR}}$ admits a holomorphic filtration:
\begin{equation}\label{eq:filtration}
    F_{\mathfrak{C}}: \quad 0 \subset \mathscr{E}_{\alpha_1} \subset \mathscr{E}_{\alpha_2} \subset  \ldots \subset \mathscr{E}_{\alpha_N} = \mathscr{E},
\end{equation}
where $\mathscr{E}_{\alpha_i}$ is a rank-$i$ holomorphic subbundle labelled by a vacuum $\alpha_i$. They are generated precisely by those sections of $\mathscr{E}$ which decay at a rate fixed by $\phi^{(\alpha_i)}$ as in \eqref{eq:asymptotic_Higgs_field}. The ordering on the vacua $\alpha_i$ is the same as the one induced by the Morse flow in the chamber $\mathfrak{C}$. This follows from two facts. First, $\kappa_{\alpha}$ is given precisely by the moment map \eqref{eq:momentmap_central_charge}. Second, in the one--loop correction from $\Phi_{\alpha}$, $x \log|x|$ is a monotonic function for large $|x|$. For a fixed value of $m_{\mathbb{I}}$, the contribution to \eqref{eq:asymptotic_Higgs_field} from $m_{\bR}$ dominates asymptotically.

The set of chambers and holomorphic filtrations $\{(\mathfrak{C}, F_{\mathfrak{C}})\}$ form a set of spectral data for the supersymmetric Berry connection.


\subsubsection{Example: $\mathbb{CP}^{N-1}$ $\sigma$-model}

Recall that the $\mathbb{CP}^{N-1}$ $\sigma$-model can be engineered as the IR limit of a UV $U(1)$ gauged linear sigma model with $N$ chiral multiplets $\Phi_i$ transforming in the fundamental of a $PSU(N)$ flavour symmetry $G_F$. We will be interested in the Cartan $T$ of the flavour symmetry, for which we can introduce mass parameters $m_{\bR,i}$ and $x_i$, for $i=1,\ldots N$. These are constrained to satisfy $\sum_{i} m_{\bR,i} = 0$ and $\prod_{i} x_i=1$ as they are $PSU(N)$ fugacities. 

We denote the Higgs branch $X = \mathbb{CP}^{N-1}$, which is described as a quotient
\begin{equation}\label{eq:cpn-1}
    \left\{\sum_{i=1}^N |\Phi_i|^2 = r\right\} \Big/ U(1)
\end{equation}
in the GLSM. Here $r=r(\mu)$ is the real FI at the RG scale $\mu$. The fixed points or massive vacua are labelled by $\alpha = 1,\ldots,N$, are given by $|\Phi_{\alpha}^2|=r$ with all other chirals vanishing. The flavour moment map is given by
\begin{equation}\label{eq:sqed_moment_map}
    h_{m_{\bR}} = \sum_{i} m_{\bR,i} |\Phi_i|^2,  \qquad h_{m_{\bR}}(\alpha) = m_{\bR,\alpha} r
\end{equation}
where the second equation shows the evaluation at the fixed points.

It will be convenient to conflate weight spaces with characters and denote
\begin{equation}\label{eq:tangent}
    TX = \sum_{i=1}^{N} sx_i -1
\end{equation}
to be the equivariant K--theory class, which, upon evaluating the Chern roots at the fixed points $s x_{\alpha}=1$, yields the tangent space characters $T_{\alpha}X$
\begin{equation}
    T_{\alpha}X  = \sum_{i\neq \alpha}  \frac{x_i}{x_\alpha}.
\end{equation}
The sum in \eqref{eq:tangent} can be regarded as the representation of $G \times T$ corresponding to the $N$ chiral multiplets, and the $-1$ as arising from the gauge group quotient. Therefore
\begin{equation}\label{eq:sqed_weights}
    \Phi_{\alpha} = \{e_i -e_{\alpha}\}_{i\neq \alpha},
\end{equation}
where $e_i$ are the fundamental weights of $PSU(N)$. Note that we have used multiplicative notation, and in the sequel we will identify $x_i = e^{2\pi z_i}$.

Now, the effective twisted superpotential of SQED[$N$] is
\begin{equation}
    W_{\text{eff}}  = -2 \pi i \tau(\mu)  \sigma + \sum_{i=1}^N(\sigma + m_i)\left(  \log  \left(\frac{\sigma + m_i}{\mu}\right)  -1 \right)
\end{equation}
where $\tau(\mu)$ is the renormalised complex FI parameter
\begin{equation}\label{eq:ren_FI_param}
    \tau(\mu)= \tau_0 + \frac{N}{2\pi i }\log(\Lambda_0/\mu).
\end{equation}
The Bethe vacua correspond to solutions of
\begin{equation}
    1 = e^{\frac{\partial W_{\text{eff}}}{\partial \sigma }} = q^{-1}\prod_{i=1}^{N} (\sigma + m_i).
\end{equation}
 where
\begin{equation}\label{eq:RG_invariant_FI}
    q= \Lambda_0^N e^{-2\pi i \tau_0}
\end{equation}
is the RG--invariant FI parameter. The solutions can be labelled by $\alpha$, corresponding to the fixed points on $X$ where $|\Phi_{\alpha}|^2= r$. These can be expanded as
\begin{equation}\label{eq:sqedn_bethe_vacua}
    \sigma \sim -m_{\alpha} + \frac{q}{\prod_{i\neq \alpha}(m_i-m_{\alpha}) } 
\end{equation}
as $q \rightarrow 0$, or equivalently as $|m|\rightarrow \infty$. 

Evaluating
\begin{equation}
     \frac{\partial W_{\text{eff}}}{\partial m_i} = \log\left(\frac{\sigma+m_i}{\mu}\right)
\end{equation}
at the Bethe vacua asymptotically in $m$ gives:
\begin{equation}
    \frac{\partial W_{\text{eff}}^{(\alpha)}}{\partial m_i} \sim
    \begin{cases}
        2\pi i \tau(\mu) - \sum_{j \neq \alpha }\log\left( \frac{m_j-m_{\alpha}}{\mu}\right) \qquad &\text{if } i =\alpha \\
        \log\left(\frac{m_i-m_{\alpha}}{\mu}\right) \qquad &\text{if } i  \neq \alpha
    \end{cases}
\end{equation}
where $\sim$ indicates dropping higher order terms in $q$ and $m^{-1}$. This can be matched to the expected asymptotics of the effective twisted superpotential \eqref{eq:asymptotic_superpotential},  using the moment map \eqref{eq:sqed_moment_map} and tangent space weights \eqref{eq:sqed_weights}. We discuss the holomorphic filtrations and vacuum ordering for this example in the next subsection.

\subsection{Physical constructions: equivariant K--theory}

We now describe how the holomorphic filtrations above allow one to construct a related type of spectral data to the supersymmetric Berry connection of a sigma model, which we show coincides with the equivariant K--theory variety of the Higgs branch $X$. This construction is the analogue of the twistorial or Hitchin spectral curve \cite{hitchin1982monopoles}, but for periodic monopoles. 

\subsubsection{Equivariant K--theory of GKM varieties}\label{sec:ktheory}

Let us first briefly recap the GKM construction of the equivariant K--theory scheme. General references and further details for equivariant K--theory can be found in \textit{e.g.} \cite{chriss1997representation, rosu2003equivariant, rimányi2018elliptic}, here we simply state the main results we need. Our physical assumptions translate into supposing that $X$ is a K\"ahler quotient, by some gauge group $G$, of a $G \times T$ representation space $R$. We will assume that $T$ is an $n$-dimensional torus, and that $X$ has isolated fixed points $X^T$ under the $T$ action labelled by $\alpha = 1,\ldots,N$.

Equivariant localisation in K--theory tells us that there is an injection:
\begin{equation}\label{eq:equivariant_localisation}
    K_T(X) \rightarrow K_T(X^T) = \bigoplus_{\alpha}K_T(\alpha)  = \bigoplus_{\alpha} \bC[x_1^{\pm1},\ldots, x_n^{\pm}].
\end{equation}
For a GKM variety, the image of this injection can be described cleanly. An $N$-tuple of Laurent polynomials $\{f_{\alpha}\}$ on the right--hand side is in the image of the localisation map if, for all $\nu$ labelling a $T$-equivariant $\mathbb{CP}^1$ connecting two vacua $\alpha$ (as described in Section~\ref{sec:geometric_prelims}):
\begin{equation}
    (f_{\alpha}-f_{\beta})|_{e^{2 \pi \nu\cdot z}=1} = 0, \qquad \nu \in \Phi_{\alpha} \text{ and } -\nu \in \Phi_{\beta} \text{ or vice versa.}
\end{equation}
We have used the notation $x = e^{2\pi z}$. We will shortly identify $z$ with the coordinate defined previously $z = 2 m_{\mathbb{I}}+it$.

Recall that the $\nu $ above label internal edges of the GKM diagram.  For non--GKM varieties, the idea is similar but the identifications across components is more complicated, and we do not deal with them here.

By taking the spectrum, one can equivalently describe:
\begin{equation}
    \mathrm{Spec} K_T(X) = \Bigg(\bigsqcup_{\alpha \in X^T} \mathbb{T}_{\alpha}\Bigg) \Big/ \Delta
\end{equation}
where $\mathbb{T}_{\alpha} \cong K_T(\text{pt}) = \mathbb{T} = (\bC^*)^n $ and $\Delta$ denotes a glueing of these abelian varieties which may be described as follows. For each compact curve labelled by a weight $\nu$ connecting two fixed points $\alpha$ and $\beta$, glue the copies $\mathbb{T}_{\alpha}$ and $\mathbb{T}_{\beta}$ along the common hyperplane
\begin{equation}
    \mathbb{T}_{\alpha} \supset \{\nu\cdot z  \in i \bZ\} \subset \mathbb{T}_{\beta}.
\end{equation}

In the case where $X$ is compact \cite{kirwan1984cohomology,harada2007surjectivity}, and in certain other cases where the result has been proven (\textit{e.g.} hyper--toric varieties \cite{harada2004equivariant} and Nakajima quiver varieties  \cite{McGerty_2017}), the equivariant K--theory admits an ``off--shell'' description as follows. Suppose that $ G = \prod_{i} U(N_i)$ is the product of unitary groups. Due to the $G$-quotient, there is a natural principal $G$-bundle over $X$ and associated $T$-equivariant tautological bundles $\mathcal{V}_i$. \textit{Kirwan surjectivity} implies that $K_T(X)$ is generated by the Schur functors of the tautological bundle $\cV_i$. In other words, there is a surjection
\begin{equation}
    K_T(\text{pt}) \otimes \bC[\{s^{\pm1}\}]^{\mathfrak{G}} \rightarrow K_T(X)
\end{equation}where $\{s\}$ are collectively the equivariant Chern roots of tautological bundles $\cV_i$, and $\mathfrak{G} = \prod_i S_{N_i}$ where $S_{N_i}$ is the symmetric group. Combined with equivariant localisation \eqref{eq:equivariant_localisation}, one therefore has the explicit description:
\begin{equation}
    K_T(X) \cong \bC[x_1^{\pm1},\ldots, x_n^{\pm}] \otimes \bC[\{s^{\pm1}\}]^{\mathfrak{G}}/I,
\end{equation}
where $I$ is the ideal of polynomials vanishing when the Chern roots $\{s\}$ are evaluated at any of the fixed points (\textit{i.e.} when $s$ are fixed to the $T$-weights of the tautological bundles at $\{\alpha\}$). Thus there is an embedding
\begin{equation}
    \mathrm{Spec} K_T(X) \hookrightarrow  \mathbb{T} \times \prod_{i} S^{N_i} \bC^*.
\end{equation}

\paragraph{Example: $\mathbb{CP}^{N-1}$.}

For $\mathbb{CP}^{N-1}$ we have $R = \bC^N$ transforming with charge $+1$ under the $U(1)$ action, subject to the moment map constraint \eqref{eq:cpn-1}. Let $s$ be the equivariant Chern root corresponding to the tautological bundle $\cO(-1)$. Let $x_1,\ldots,x_{N}$ be the coordinates on $\mathbb{T}$, or equivalently the Chern roots of the trivial $T$-equivariant bundle descended from $\bC^N$. Then the Chern root evaluation at the fixed points is given by $s = x_{\alpha}^{-1}$. Therefore, we have that
\begin{equation}
    K_T(\mathbb{CP}^{N-1})  =  \bC[s^{\pm1}, x_1^{\pm1},\ldots,x_N^{\pm1}]/
    \Big\{ \prod_{i}(1-sx_i)\Big\},
\end{equation}
and so
\begin{equation}
    \mathrm{Spec} \left( K_T(X) \right) = \Big\{\prod_{i}(1-sx_i) = 0 \Big\} \subset \mathbb{T} \times \bC^*_s.
\end{equation}
Alternatively:
\begin{equation}\label{eq:cpn-1_eq_K_theory}
    \mathrm{Spec} \left( K_T(X) \right) = \Bigg(\bigsqcup_{\alpha \in X^T} \mathbb{T}_{\alpha}\Bigg) \Big/ \Delta
\end{equation}
where $\Delta$ glues pairs $\mathbb{T}_{\alpha}, \mathbb{T}_{\beta}$ along the loci $x_{\alpha}=x_{\beta}$ for all distinct pairs $\alpha < \beta$.

\subsubsection{Supersymmetric ground states}\label{sec:susy_ground_states}

We will now pass to an effective description of the theory on $\bR\times S^1$ as an infinite dimensional quantum mechanics along $\mathbb{R}$, with the 1d $\mathcal{N}=(2,2)$ supersymmetry algebra \eqref{eq:modified_algebra}. In the absence of parameters $(m_{\bR}, z)$ for $T$, the theory flows to a sigma model on the Higgs branch. It can then alternatively be described as an  $\mathcal{N}=(2,2)$ quantum mechanics with target
\begin{equation}
    \mathcal{X} = L X = \mathrm{Map}(S^1 \rightarrow X)
\end{equation}
which is the space of smooth maps from $S^1$ to $X$, and is an infinite--dimensional K\"ahler manifold. The kinetic terms involving derivatives along $S^1$ reduce to a background vector multiplet for the $U(1)$ symmetry of $\mathcal{X}$ induced by translations along $S^1$. Now, turning on $(m_{\bR}, z)$, these couple via background vector multiplets for the induced $T$ action on $\mathcal{X}$.

Since we assume the spectrum is gapped, we pass to a cohomological description of supersymmetric ground states in the usual way, by taking the supercharge
\begin{equation}
    Q= Q_{\lambda=1}:=   \frac{1}{2}(\bar{q}_{+}+q_{-} + q_{+}+\bar{q}_{-} ),
\end{equation}
or for the case when $\lambda = e^{i\theta}$, $Q_{\lambda}$ as defined in equation \eqref{eq:lambda_supercharges}. Let us reproduce the supersymmetry algebra \eqref{eq:modified_algebra} for this choice of $\lambda$:
\begin{equation}\label{eq:modified_algebra_2}
\begin{gathered}
   \{Q, Q^{\dagger}\} = 2H, \quad \{\bar{Q}, \bar{Q}^\dagger \} = 2H  \\
   \{Q,\bar{Q}\} = Z, \quad Q^2 = P.
\end{gathered}
\end{equation}
We have made some notation simplifications, and denoted
\begin{equation}
   P \coloneqq i(\partial_2 - z\cdot J_T), \qquad Z \coloneqq 2m_{\bR} \cdot J_T.
\end{equation}
Thus by restricting to the states in the supersymmetric quantum mechanics annihilated by $Z$ and $P$
\begin{equation}\label{eq:Z_P_condition}
    m_{\bR} \cdot \gamma_T = 0, \qquad n- i z \cdot \gamma_T = 0,
\end{equation}
where $\gamma_T \in \Gamma_T^{\vee}$ is the $T$-charge of the state and $n \in \bZ$ the KK momentum along $S^1$, the supercharge $Q$ becomes a differential and its cohomology gives a description of the ground states. 

In the description of the theory as a $(2,2)$ quantum mechanical $\sigma$-model to $\mathcal{X}$, the supercharge $Q$ has the form
\begin{equation}\label{eq:sigma_model_supercharge}
    Q = e^{-h_{m_{\bR}}}(d+ \iota_{V_z}) e^{h_{m_{\bR}}}
\end{equation}
where in an abuse of notation $h_{m_{\bR}}$ denotes also the moment map for the $T$ action on $\cX$ generated by $m_{\bR}$, and $V_{z}$ is the sum of the vector field $\partial_2$ generating the natural $S^1$ action on the loop space, and the vector field  corresponding to the induced $U(1) \subset T$ action on $X$ generated by the $\bC^*$ parameter $z$. Supercharges of this type were introduced in \cite{Witten:1982im}, and have been studied from the context of quiver quantum mechanics recently in \cite{Galakhov:2018lta, Galakhov:2020vyb, Galakhov:2023aev}.

From this viewpoint, the ground states can be analysed by applying the classic arguments of Witten \cite{Witten:1982im} to this infinite--dimensional model. The superpotential or moment map $h_{m_{\bR}}$ can be scaled, localising ground states around $\text{Crit}(h_{m_{\bR}}) \subset \mathcal{X}$. The supersymmetric ground states can then be obtained as
\begin{equation}
    H_{d+ \iota_{V_z}} (\text{Crit}(h_{m_{\bR}})),
\end{equation}
\textit{i.e.} the cohomology of the equivariant differential $d+ \iota_{V_z}$ on $\text{Crit}(h_{m_{\bR}})$. This is simply the equivariant cohomology of $\text{Crit}(h_{m_{\bR}})$ localised at the equivariant parameter $z$.

This description of ground states clearly depends intricately on the values of $m_{\bR}$, and we consider some cases in turn.

\paragraph{Generic mass.}

We consider first the mass $m_{\bR}$ lying in a chamber $\mathfrak{C}$, \textit{i.e.} in the complement of all the hyperplanes $\mathbb{W}_{\nu} = \{\nu \cdot m_{\bR} =0\}$. The conditions \eqref{eq:Z_P_condition} imply that the ground states have $\gamma_T=0$ (are uncharged under the flavour symmetry), and have $0$ KK momentum. 

In the $\sigma$-model to $\mathcal{X}$, the critical loci of $h_{m_{\bR}}$ are the constant maps $S^1 \rightarrow \alpha$. As usual, one can scale the coefficient in front of the moment map to infinity in the supercharge \eqref{eq:sigma_model_supercharge}, and the normalisable perturbative ground states are Gaussian wavefunctions localised at these constant maps, which may be chosen to be orthonormal. The perturbative description is exact because $h_{m_{\bR}}$ is a moment map for a Hamiltonian isometry of the K\"ahler $\cX$, see \textit{e.g.} \cite{Hori:2003ic}. From the perspective of the $2d$ path integral, one has the description of the ground states in a given chamber $\mathfrak{C}$ as
\begin{itemize}
    \item $\ket{\alpha}_{\mathfrak{C}}$ is the supersymmetric ground state obtained from performing the path integral on $S^1 \times \bR_{\geq 0}$ with vacuum $\alpha$ at $x^1 \rightarrow +\infty$.
    \item $\bra{\alpha}_{\mathfrak{C}}$ is the supersymmetric ground state obtained from performing the path integral on $S^1 \times \bR_{\leq 0}$ with vacuum $\alpha$ at $x^1 \rightarrow -\infty$.
\end{itemize}
These states are orthonormal
\begin{equation}\label{eq:orthonormality_chamber}
    \braket{\alpha|\beta}_{\mathfrak{C}} = \delta_{\alpha\beta},
\end{equation}
reflecting the values of the path integral on the infinitely long cylinder with vacua $\alpha$ and $\beta$ at $\pm \infty$. 

Additionally, these states can be generated in $Q$-cohomology by finite distance boundary conditions, which are the thimble boundary conditions discussed in Section~\ref{sec:difference_modules}. In particular, in the 2d $\sigma$-model description
\begin{itemize}
    \item $\ket{\alpha}_{\mathfrak{C}}$ is generated by a B--brane supported on the \textit{attracting} Lagrangian $\cL^{-}_{\alpha}$ for Morse flow with respect to $h_{m_{\bR}}$.
    \item $\bra{\alpha}_{\mathfrak{C}}$ is generated by a B--brane supported on the \textit{repelling} Lagrangian $\cL^{+}_{\alpha}$ for Morse flow with respect to $h_{m_{\bR}}$.
\end{itemize}
Such boundary conditions can be engineered by UV boundary conditions which flow to the B--branes in the IR, see \textit{e.g.} \cite{Bullimore:2016nji, Bullimore:2021rnr}.

\paragraph{Mass on a wall.}
We now move on to discuss the case where $m_{\bR}$ lies on a wall $\mathbb{W}_{\nu } = \{\nu \cdot m_{\bR} = 0\}$ labelled by a weight $\nu \in \Phi_{\alpha}$. As we shall see, for the purposes of our discussion we will be interested only in the case where $\nu$ is as described in Section~\ref{sec:geometric_prelims} describing hyperplanes where a $\Sigma_{\nu } \cong \mathbb{CP}^1$ of vacua open up between two fixed points $\alpha$ and $\beta$.

Suppose $\nu \in \Phi_{\alpha} \cap -\Phi_{\beta}$. Then
\begin{equation}
    \text{Crit}(h_{m_{\bR}}) = \{\gamma \neq \alpha, \beta\} \cup L\Sigma_{\nu},
\end{equation}
where $L\Sigma_{\nu}  \coloneqq \mathrm{Map}(S^1,\Sigma_{\nu})$. One still has $N-2$ vacua as described in the previous section corresponding to constant maps to fixed points $S^1 \rightarrow \gamma$. For the ground states associated to $\alpha$ and $\beta$, we must consider the $d+ \iota_{V_z}$ cohomology of $L\Sigma_{\nu}$.

Provided $\nu \cdot z \notin  i \bZ$, $V_z$ still has isolated fixed points in $L\Sigma_{\nu}$, corresponding to the constant maps. However, there is now ambiguity in the normalisation as unitarity is lost in the cohomological description of ground states. The natural choice \cite{Bullimore:2016hdc} is to define the state $\ket{\alpha}$ as the Poincaré dual of the equivariant fundamental class of the constant map $S^1 \rightarrow \alpha$ inside $L\Sigma_{\nu}$.
\begin{equation}
\begin{aligned}
    \braket{\alpha|\alpha} &= \prod_{n \in \bZ} (n - i \nu \cdot z)  \\
    &= e^{\pi \nu \cdot z}  -  e^{-\pi \nu \cdot z}\\
    &= 2 \sinh (\pi \nu \cdot z).
\end{aligned}
\end{equation}
where the first line is the equivariant Euler class of the normal bundle to the constant map inside $L\Sigma_{\nu}$. We have used the usual regularisation of the infinite product, familiar from one--loop determinants in supersymmetric quantum mechanics \cite{Hori:2014tda}, and 2d $\cN=(2,2)$ theories on a cylinder \cite{Sugiyama:2020uqh}. Similarly:
\begin{equation}
\begin{aligned}
    \braket{\beta|\beta} &= \prod_{n \in \bZ} (n + i  \nu \cdot z)  \\
    &= -2 \sinh (\pi  \nu \cdot z),
\end{aligned}
\end{equation}
and
\begin{equation}
    \braket{\alpha|\beta} = \braket{\beta|\alpha} = 0.
\end{equation}
When $\nu \cdot z \in  i \bZ$, the fixed locus of $V_z$ is no longer isolated and the above supersymmetric ground states are not linearly independent. One can pass to linear combinations of the above which extends across this locus, but we will not need them here.

\paragraph{Zero mass.}

We can continue the above process by considering ground states on the intersection of loci $\mathbb{W}_{\nu}, \mathbb{W}_{\nu'}\ldots$. Instead, we turn to the extreme case of vanishing $m_{\bR}$, or equivalently the intersection of all such hyperplanes. 

The ground states now correspond to equivariant cohomology (localised at $z$) of the entirety of $\cX$. Repeating the same arguments as before, if $z$ is generic, meaning that it lies in the complement of all hyperplanes $\mathbb{W}_{\nu}$, then the vector field $V_{z}$ has only isolated fixed point corresponding to the constant maps $S^1\rightarrow \alpha$. Thus there are $N$ supersymmetric ground states which are normalised as
\begin{equation}\label{eq:vanishing_normalisation}
    \braket{\alpha|\beta} = \delta_{\alpha\beta} \prod_{\nu \in \Phi_{\alpha}} 2\sinh(\pi  \nu \cdot z) ,
\end{equation}
which are the equivariant fundamental classes of the constant maps $S^1\rightarrow \alpha$ inside $\cX$. If $z$ is not generic but lies on a hyperplane $\mathbb{W}_{\nu}$, then if $\nu$ is an internal edge of the GKM diagram we must pass to linear combinations as mentioned briefly above, and if $\nu$ is an external (non--internal) edge the quantum mechanics is not gapped and the construction breaks down.

These states can also be generated at a finite distance by B--branes, which are simply supported solely on the fixed points $\alpha$, with all other chiral multiplets assigned Dirichlet boundary conditions (\textit{i.e.} forced to vanish). This is compatible with the normalisations \eqref{eq:vanishing_normalisation}: the overlaps of two such B--branes as a partition function on $I \times S^1$ is given by the component of the 2d chiral multiplets with Dirichlet boundary conditions corresponding to 1d Fermi multiplets, which remain after colliding the two boundaries \cite{Sugiyama:2020uqh}.

It will be important for us to discuss the relationship between ground states $\ket{\alpha}_{\mathfrak{C}}$ in a given chamber $\mathfrak{C}$ and those at the origin of the $m_{\bR}$ parameter space, $\ket{\alpha}$. In the limit as $m_{\bR} \rightarrow 0$, we know that $\ket{\alpha}_{\mathfrak{C}}$ is no longer appropriate. However, we claim that
\begin{equation}\label{eq:relative_normalisations}
\begin{aligned}
    \ket{\alpha}_{\mathfrak{C}} \prod_{\nu \in \Phi_{\alpha}^{-}} 2\sinh(\pi  \nu \cdot z) \rightarrow \ket{\alpha} \\
    \bra{\alpha}_{\mathfrak{C}} \prod_{\nu \in \Phi_{\alpha}^{+}} 2\sinh(\pi  \nu \cdot z) \rightarrow \bra{\alpha} \\
\end{aligned}
\end{equation}
where these limits are understood as holding for computations of $Q$-closed observables. This is clearly compatible with \eqref{eq:orthonormality_chamber} and \eqref{eq:vanishing_normalisation} and is consistent with our description of boundary conditions. That is, the relative normalisations in \eqref{eq:relative_normalisations} are precisely those one would obtain from computing overlaps $\braket{B|\alpha}_{\mathfrak{C}}$ and $\braket{B|\alpha}$, where $\bra{B}$ is a state generated by an arbitrary B--brane, as partition functions on $I \times S^1$. This is due to the fact that the boundary condition for $\ket{\alpha}_{\mathfrak{C}}$ assigns Neumann--type boundary conditions for fluctuations in $\cL_{\alpha}^{-}$, versus Dirichlet in $\ket{\alpha}$. See \cite{Bullimore:2021rnr} for further details in the 3d case.

We will often consider \eqref{eq:relative_normalisations} as equalities, because computations involving supersymmetric ground states on $I\times S^1$ are independent of $m_{\bR}$. More properly, one should say that the sets of ground states are related by the action of a Janus interface for $m_{\bR}$, interpolating between $\mathfrak{C}$ and the origin \cite{Dedushenko:2021mds, Bullimore:2021rnr}.

\subsubsection{Incidence variety}

We are now finally in a position to explain how the equivariant K--theory variety is related to an incidence variety for the filtrations described in Section~\ref{sec:filtrations}. Recall that in a chamber $\mathfrak{C}$, one obtains a filtration $F_{\mathfrak{C}}$ of $\mathscr{E}$ (a holomorphic vector bundle over $\mathbb{T}$), determined by the ordering on vacua induced by the Morse function $h_{m_{\bR}}$ \eqref{eq:filtration}.

One can take the associated graded bundle
\begin{equation}
    \mathcal{G}(\mathscr{E}) = \bigoplus_{\alpha \in X^T} \cL_{\alpha}
\end{equation}
which by construction splits holomorphically as the sum of holomorphic line bundles:
\begin{equation}
    \mathcal{L}_{\alpha_i} \coloneqq \mathscr{E}_{\alpha_i} / \mathscr{E}_{\alpha_{i-1}}.
\end{equation}
By construction, these line bundles are generated by the supersymmetric ground states $\ket{\alpha_i}_{\mathfrak{C}}$ described above, since the decay of sections of these line bundles is determined by the central charges in the vacua $\alpha_i$.  The associated graded and the constituent line bundles change as we go from chamber to chamber.

We can now build the incidence variety as follows. For each weight $\nu \in \Phi_{\alpha}$ labelling an equivariant curve $\Sigma_{\nu}$ connecting vacua $\alpha$ and $\beta$, let $\mathfrak{C}$ and $\mathfrak{C}'$ denote the chambers on either side of the hyperplane $\mathbb{W}_{\nu}$. The filtrations $F_{\mathfrak{C}}$ and $F_{\mathfrak{C}'}$ have the roles of $\alpha$ and $\beta$ (which are adjacent in the ordering) switched:
\begin{equation}
\begin{aligned}
    F_{\mathfrak{C}}: \quad 0 \subset \mathscr{E}_{\alpha_1} \subset \ldots \mathscr{E}_{\alpha} \subset \mathscr{E}_{\beta} \subset \ldots \subset \mathscr{E}_{\alpha_N} = \mathscr{E},\\
    F_{\mathfrak{C'}}: \quad 0 \subset \mathscr{E}'_{\alpha_1} \subset \ldots \mathscr{E}'_{\beta} \subset \mathscr{E}'_{\alpha} \subset \ldots \subset \mathscr{E}'_{\alpha_N} = \mathscr{E},
\end{aligned}
\end{equation}
where the prime and absence of a prime indicates the chamber.

We compare the holomorphic line bundles $\cL_{\alpha}$ and $\cL'_{\alpha}$ (or equivalently $\cL_{\alpha}$ and $\cL'_{\beta}$) over $\mathbb{T}$. Recall that they are generated by:
\begin{equation}
\begin{aligned}
    \ket{\alpha}_{\mathfrak{C}} = \prod_{\mu \in \Phi_{\alpha}^{\mathfrak{C},-}} (2\sinh(\pi \mu \cdot z))^{-1} \ket{\alpha}, \quad 
    \ket{\alpha}_{\mathfrak{C}'} = \prod_{\mu \in \Phi_{\alpha}^{\mathfrak{C}',-}} (2\sinh(\pi \mu \cdot z))^{-1} \ket{\alpha},\\
    \ket{\beta}_{\mathfrak{C}} = \prod_{\mu \in \Phi_{\beta}^{\mathfrak{C},-}} (2\sinh(\pi \mu \cdot z))^{-1} \ket{\beta}, \quad 
    \ket{\beta}_{\mathfrak{C}'} = \prod_{\mu \in \Phi_{\beta}^{\mathfrak{C}',-}} (2\sinh(\pi \mu \cdot z))^{-1} \ket{\beta}.
\end{aligned}
\end{equation}
We therefore have, without loss of generality
\begin{equation}
\begin{aligned}
    \ket{\alpha}_{\mathfrak{C}}  &= 2\sinh(\pi \nu \cdot z) \ket{\alpha}_{\mathfrak{C'}}, \\
    \ket{\beta}_{\mathfrak{C}'}  &= 2\sinh(-\pi \nu \cdot z)\ket{\beta}_{\mathfrak{C}}.
\end{aligned}
\end{equation}
This relative normalisation vanishes (or diverges, depending on convention), when $\nu\cdot z \in  i \bZ$, or alternatively when
\begin{equation}\label{eq:gluing_hyperplanes}
    e^{2 \pi  \nu \cdot z} = 1.
\end{equation}

The crossing of the holomorphic filtrations $F_{\mathfrak{C}}$ is therefore encoded in the following data. Take $N$ copies $\mathbb{T}_{\alpha}$ of $\mathbb{T}$. The copies  $\mathbb{T}_{\alpha}$ and $\mathbb{T}_{\beta}$ are identified along the hyperplanes \eqref{eq:gluing_hyperplanes} for $\nu \in \Phi_{\alpha} \cup -\Phi_\beta$ the tangent weight of a $T$-equivariant $\mathbb{CP}^1$ connecting $\alpha$ and $\beta$. Thus, by matching to Section~\ref{sec:ktheory}, clearly:

\begin{mdframed}[roundcorner=10pt, linewidth=0.3mm, linecolor=gray]
Consider the sets $\{\mathfrak{C}, F_{\mathfrak{C}}\}$ of the holomorphic vector bundle $\mathscr{E}$ given by restricting $E|_{m_{\bR}, \bar{D}_{\bar{z}}}$, which compromises a form of spectral data for the supersymmetric Berry connection. The incidence variety built from comparing the holomorphic filtrations in neighbouring chambers coincides precisely with $\mathrm{Spec}K_T(X)$, the equivariant K--theory scheme of the Higgs branch $X$.
\end{mdframed}

\paragraph{Example: SQED[$N$] \& $\bC\bP^{N-1}$.}

For the case of SQED[$N$] there are $N!$ chambers $\mathfrak{C}_{\sigma}$ labelled by a permutation $\sigma \in S_N$
\begin{equation}
    \mathfrak{C}_{\sigma}: \{ m_{\bR, \sigma(1)} > m_{\bR,\sigma(2)} > \ldots > m_{\bR, \sigma(N)}\},
\end{equation}
separated by $\binom{N}{2}$ hyperplanes
\begin{equation}
    W_{\alpha,\beta} = \{ (e_{\alpha} - e_{\beta})\cdot m_{\bR} = m_{\bR,\alpha}-m_{\bR,\beta} = 0 \}
\end{equation}
labelled by a pair $(\alpha,\beta)$, where without loss of generality $\alpha < \beta$. Two chambers $\mathfrak{C}_{\sigma}$ and $\mathfrak{C}_{\sigma'}$ are adjacent if $\sigma' = \tau \circ \sigma$ where $\tau$ is the transposition swapping $\sigma(i)$ and $\sigma(i+1)$ for some $i \in \{1,\ldots,N-1\}$, and are separated by the wall $\mathbb{W}_{\sigma(i),\sigma(i+1)}$.

Following the above, the crossing of the filtrations
\begin{equation}
\begin{aligned}
    F_{\mathfrak{C}_{\sigma}}&: \quad 0 \subset \mathscr{E}_{\sigma(1)} \subset \ldots \mathscr{E}_{\sigma(i)} \subset \mathscr{E}_{\sigma(i+1)} \subset \ldots \subset \mathscr{E}_{\sigma(N)} = \mathscr{E} \\
    F_{\mathfrak{C}_{\sigma;}}&: \quad 0 \subset \mathscr{E}_{\sigma(1)} \subset \ldots \mathscr{E}_{\sigma(i+1)} \subset \mathscr{E}_{\sigma(i)} \subset \ldots \subset \mathscr{E}_{\sigma(N)} = \mathscr{E} \\
\end{aligned}
\end{equation}
occurs at
\begin{equation}
    e^{2\pi  (z_{\sigma(i)}-z_{\sigma(i+1)} )} = 1.
\end{equation}
We thus build the incidence variety by repeating this for all adjacent chambers, or equivalently all hyperplanes $\mathbb{W}_{\alpha, \beta}$. Thus we take $N$ copies $\mathbb{T}_{\alpha}$ of $\mathbb{T} := (\mathbb{C}^*)^N$ and identifying all pairs $\mathbb{T}_{\alpha}, \mathbb{T}_{\beta}$ along the loci $z_{\alpha}-z_{\beta} \in i  \bZ$. This clearly matches the equivariant K--theory scheme for $\mathbb{CP}^{N-1}$ as described in equation \eqref{eq:cpn-1_eq_K_theory}.

\subsection{Example: SQED[\texorpdfstring{$2$}{}] \& \texorpdfstring{$\mathbb{CP}^{1}$}{} \texorpdfstring{$\sigma$-model}{}}

Let us review the constructions of this section in greater detail for our running example of the $\mathbb{CP}^1$ $\sigma$-model, in a way which will help elucidate the connection to the Hitchin spectral curve \cite{hitchin1982monopoles} for monopoles in $\bR^3$.

For this example, there is only one flavour symmetry and we study the scattering problem
\begin{equation}\label{eq:sqed2_scattering}
    (D_{m_{\bR}} - i \phi)\psi = 0.
\end{equation}
In this example, there are only two chambers: $m_{\bR} > 0$ and $m_{\bR} <0$. We will therefore consider the scattering problem as $m_{\bR} \rightarrow \pm\infty$. Let $\cE_{m_{\bR}} \coloneqq E|_{\{m_{\bR}\}\times \bT, \bar{D}_{\bar{z}}}$ be the rank-$2$ holomorphic vector bundle on $\bC^*$ (parameterised by $e^{2\pi  z}$) as before. Then
\begin{itemize}
    \item We first choose a solution $\psi^{-}$ to \eqref{eq:sqed2_scattering}, which decays $\psi^{-} \rightarrow 0$ as $m_{\bR} \rightarrow -\infty$, and is holomorphic $\bar{D}_{\bar{z}}\psi^{-} = 0$. This is a consistent requirement due to the Bogomolny equations \eqref{eq:bogomolnyforktheory}. By parallel transport with respect to $D_{m_{\bR}}$, $\psi^{-}$ generates a holomorphic line sub--bundle  $L^- \subset \cE_{m_{\bR}}$ for any value of $m_{\bR}$, and we denote $\cE \coloneqq \cE_{m_{\bR}}$.
    \item We may similarly define a solution $\psi^+$ and line sub--bundle $L^+ \subset \cE$ which decays $\psi^+ \rightarrow \infty$ as $m_{\bR} \rightarrow +\infty$.
\end{itemize}

There is a natural isomorphism $L^- \cong \cE/L^+$ given by the following. Let $\psi$ be some section of $\cE$ which naturally induces the generating section of $\cE/L^+$, then we must have:
\begin{equation}\label{eq:betti_scattering}
    \psi^-(m_{\bR},z,\bar{z}) = f(z) \psi(m_{\bR},z,\bar{z}) + g(z)  \psi^+ (m_{\bR},z,\bar{z})
\end{equation}
where $f(z)$ is a meromorphic function in $z$. The function $f(z)$ depends on a choice of $\psi$ and $\psi^-$, but its zeros are independent of this choice \cite{Cherkis:2012qs}. The zeros of $f(z)$ describe the spectral points or incidence variety, at which there is a normalisable solution of \eqref{eq:betti_scattering} along $m_{\bR}$.

Let us see how this coincides with the equivariant K--theory variety of $\mathbb{CP}^1$. The section $\psi^-$ corresponds to ground states which decay the fastest as they are parallel transported along $m_{\bR} \rightarrow -\infty$. As $A_w \propto 1/w$ as $|w| \rightarrow \infty$, $A_{m_{\bR}}+ i\phi$ is dominated by the adjoint Higgs field, and the fastest decaying state will simply be the largest eigenvalue of $i\phi$, which via Section~\ref{subsubsec:sqed2_asymptotics} corresponds to vacuum $(1)$, the north pole of $\mathbb{CP}^1$. 
Thus $\psi^-  = \ket{1}_{m_{\bR}<0}$ where $\ket{1}_{m_{\bR}<0}$ is the state in the sigma model corresponding to placing a supersymmetric vacuum at $\infty$ with parameter $m_{\bR}<0$ (\textit{i.e.} in this chamber). Analogously $\psi^+  = \ket{2}_{m_{\bR}>0}$.  We identify $\psi$ as the section of $\cE$ which induces $\ket{1}_{m_{\bR} > 0}$ (which is a section of $\cE/L^+$).

The incidence variety can therefore be recovered by looking at the relative normalisations of $\ket{1}_{m_{\bR}<0}$, and $\ket{1}_{m_{\bR}>0}$ (without loss of generality), which are the vacuum states corresponding to the north pole. Via the arguments in \ref{sec:susy_ground_states} (and \cite{Bullimore:2021rnr}), $\ket{1}_{m_{\bR} >0}$ is related to a reference state $\ket{1}$ (which is a vacuum state at $m_{\bR} = 0$), by the K--theory class associated to the \textit{attracting} weights in the tangent space, with respect to the action generated by $m_{\bR} > 0$, and similarly for the other state. 
Thus
\begin{equation}
    \ket{1}_{m_{\bR} < 0} = \sinh( \pi z) \ket{1}, \quad \ket{1}_{m_{\bR} > 0} = \ket{1} \quad\Rightarrow \quad \ket{1}_{m_{\bR} < 0} = \sinh( \pi z) \ket{1}_{m_{\bR} > 0} 
\end{equation}
and similarly
\begin{equation}
    \ket{2}_{m_{\bR} <0} = \ket{2}, \quad \ket{2}_{m_{\bR} > 0} = \sinh(-\pi z)\ket{2} \quad \Rightarrow 
    \quad \ket{2}_{m_{\bR} < 0} = \sinh(-\pi z) \ket{2}_{m_{\bR} <0}.
\end{equation}
Thus we can identify $f(z) = \sinh(\pi z)$, and the incidence variety is just given by the zeros of this, which can be characterised in a way which takes into account the periodicity of $z$ by
\begin{equation}
    x = e^{2\pi z} = 1.
\end{equation}
Clearly, this coincides with the loci where the two sheets $K_T(\{i\}) \cong \bC^*_{x} = \mathrm{Spec}\bC[x^{\pm1}]$ of the equivariant K--theory scheme of $\mathbb{CP}^1$:
\begin{equation}
    \mathrm{Spec}K_T(\mathbb{CP}^1) = (K_T(\{1\})\sqcup K_T(\{2\}))/\Delta
\end{equation}
are glued, as indicated by $\Delta$. This is illustrated in Figure~\ref{fig:intro-K-theory}.

\section*{Acknowledgements}

The authors are very grateful to Mathew Bullimore for several helpful discussions at an early stage of this project. It is also a pleasure to thank Christopher Beem, Samuel Crew, Mykola Dedushenko, Tudor Dimofte and Davide Gaiotto for helpful comments. The work of AEVF has been supported by the EPSRC Grant EP/T004746/1 “Supersymmetric Gauge Theory and Enumerative Geometry” and the EPSRC Grant EP/W020939/1 ``3d N=4 TQFTs". The work of DZ is supported by a Junior Research Fellowship from St John's College, Oxford. DZ is grateful to the organisers of the Theoretical Sciences Visiting Program at the Okinawa Institute of Science and Technology, where part of the work for this manuscript was conducted.

\appendix

\section{Difference Equations for SQED\texorpdfstring{$[3]$}{} \& \texorpdfstring{$\mathbb{CP}^2$ $\sigma$}{}-model}\label{appendix:sqed3}

In this appendix, we demonstrate the results of Section~\ref{sec:conf_vortex} for supersymmetric QED with $3$ chiral multiplets of charge $+1$, which flows in the IR to an NLSM to $\mathbb{CP}^2$. Namely, we show that (a basis of) hemisphere partition functions obey difference equations of the form \eqref{eq:matrix_difference_vortex}, which realise a quantisation of the Cherkis--Kapustin spectral variety as in \eqref{eq:vortex_spectral_curve} and consequently of an action on $QH_T(\mathbb{CP}^2)$.

\paragraph{Spectral variety.} 

Before proceeding, let us first write down the Cherkis--Kapustin spectral variety. We take the charges of $\Phi_{1,2,3}$ under $T=U(1)^2$ to be $(1,0),\,(0,1), (-1,-1)$ respectively, and introduce complex masses $m_{1,2}$ for $T$. The effective twisted superpotential is then
\begin{equation}
\begin{aligned}
    W_{\text{eff}}  =& \, (\sigma + m_1)\left(  \log  \left(\frac{\sigma + m_1}{\ep}\right)  -1 \right)+ (\sigma + m_2)\left(  \log  \left(\frac{\sigma + m_2}{\ep}\right)  -1 \right)\\
    &+(\sigma -m_1-m_2)\left(  \log  \left(\frac{\sigma -m_1- m_2}{\ep}\right)  -1 \right) -2 \pi i \tau(\ep)  \sigma.
\end{aligned}    
\end{equation}
Here, $\tau(\ep)$ is the renormalised complex FI parameter at scale $\ep$ \eqref{eq:ren_FI_param}. The vacua are determined by
\begin{equation}
    e^{\frac{\partial W_{\text{eff}}}{\partial \sigma}} = q^{-1}(\sigma+m_1)(\sigma+m_2)(\sigma-m_1-m_2) = 1,
\end{equation}
which is equivalently $\mathrm{Spec} \,QH_T(\mathbb{CP}^2)$. Here, $q = \ep^3 e^{2\pi i \tau(\ep)}$ is the RG--invariant combination \eqref{eq:RG_invariant_FI}. 

We may pass to the Cherkis--Kapustin spectral variety, or alternatively the momentum space representation of $\mathrm{Spec} \,QH_T(\mathbb{CP}^2)$, by appending the equations
\begin{equation}
    p_1 = e^{\frac{\partial W_{\text{eff}}}{\partial m_1}}, \qquad p_2 = e^{\frac{\partial W_{\text{eff}}}{\partial m_2}}. 
\end{equation}
Eliminating $\sigma$ gives the following two equations for the generalised (rank-$2$) Cherkis--Kapustin spectral variety
\begin{equation}\label{eq:sqed3_spectral_curve}
\begin{aligned}
    \cL_1(p,m)\, \coloneqq q(p_1-1)^3 - (2m_1+m_2)^2 p_1 (m_1-m_2+m_1 p_1 +2 m_2 p_1)=0,\\
    \cL_2(p,m)\, \coloneqq q(p_2-1)^3 - (2m_2+m_1)^2 p_2 (m_2-m_1+m_2 p_2 +2 m_1 p_2)=0.
\end{aligned}
\end{equation}
These cut out a middle--dimensional Lagrangian subvariety in $ (\bC^*)^2\times \bC^2 \ni (p,m)$.

\paragraph{The result.} We claim that there is a basis of B--branes, $\{D_1,D_2,D_3\}$, such that the corresponding hemisphere partition functions $\cZ_{D_\al}[\cO_a,m]$ (where $\cO_a$ indicates an element of a basis of twisted chiral ring elements inserted at the tip of the hemisphere) obey the difference equations
\begin{equation}\label{eq:sqed3_difference_eqns}
\begin{aligned}
    \hat{p_1} \cZ_{D_\al}[\cO_a, m_1,m_2] = \cZ_{D_\al}[\cO_a, m_1+\ep,m_2] = \widetilde{G}^{(1)}_{ab}(m,\ep) \cZ_{D_\al}[\cO_b, m_1,m_2], \\
    \hat{p_2} \cZ_{D_\al}[\cO_a, m_1,m_2] = \cZ_{D_\al}[\cO_a, m_1,m_2+\ep] = \widetilde{G}^{(2)}_{ab}(m,\ep) \cZ_{D_\al}[\cO_b, m_1,m_2], \\
\end{aligned}
\end{equation}
where:
\begin{equation}\label{eq:sqed3_diff_matrix}
\begin{aligned}
    \widetilde{G}^{(1)}(m_1,m_2, \ep) = &
    \left(
    \begin{array}{cc}
    1+\frac{m_1 m_2 \left(2 m_1+m_2+\epsilon \right)}{q} &  \ldots\\
    \left(2 m_1+m_2+\epsilon \right) \left(1+ \frac{m_1 m_2 \left(m_1+m_2+\epsilon \right)}{q}\right) & \ldots \\
    \left(m_1+m_2+\epsilon \right) \left(2 m_1+m_2+\epsilon \right) \left(1+\frac{m_1 m_2 \left(m_1+m_2+\epsilon \right)}{q}\right) & \ldots \\
\end{array}
\right. \\\\
&\quad \left.
\begin{array}{ccc} \ldots\quad  & \frac{\left(m_1+m_2\right) \left(2 m_1+m_2+\epsilon \right)}{q} & \ldots\\
 \ldots\quad  &   1+ \frac{\left(m_1+m_2\right) \left(m_1+m_2+\epsilon \right) \left(2 m_1+m_2+\epsilon \right)}{q}  & \ldots\\
 \ldots\quad  &  \left(2 m_1+m_2+\epsilon \right) \left(1+ \frac{\left(m_1+m_2\right) \left(m_1+m_2+\epsilon \right)^2}{q}\right) & \ldots \\\\
\end{array}
\right.\\
&\quad \left.
\begin{array}{cc} \ldots\quad  & \frac{2 m_1+m_2+\epsilon }{q} \\
 \ldots\quad  & \frac{\left(m_1+m_2+\epsilon \right) \left(2 m_1+m_2+\epsilon \right)}{q} \\
 \ldots\quad  & 1+ \frac{\left(2 m_1+m_2+\epsilon \right) \left(m_1+m_2+\epsilon \right){}^2}{q} \\
\end{array}
\right),
\end{aligned}
\end{equation}
\vspace{1em}
\begin{equation}\label{eq:sqed3_symmetry}
    \widetilde{G}^{(2)}(m_1,m_2, \ep) = \widetilde{G}^{(1)}(m_2,m_1, \ep).
\end{equation}
Notice that these equations are independent of the boundary condition. It will follow easily from our arguments that such a result holds for any choice of boundary condition on the chiral multiplets. 

Furthermore, we claim that:
\begin{equation}
    \cL_1(\widetilde{G}, m)|_{\ep = 0} = \cL_2(\widetilde{G}, m)|_{\ep = 0} = 0
\end{equation}
where $\cL_1$ and $\cL_2$ are given in \eqref{eq:sqed3_spectral_curve}. Thus the hemisphere partition functions provide a quantisation of the spectral variety. This can be checked by computing the characteristic polynomials of the matrices above and showing that they coincide with $\cL_1$ and $\cL_2$.

\paragraph{Proof.} We choose the B--branes given in Table \ref{tab:sqed3_branes},
\begin{table}[htbp!]
    \centering
    \begin{tabular}{c|cc}
        & $\text{Dirichlet}$ & $\,\,\,\text{Neumann}\,\,\,$ \\ \hline
        $D_1$ & & $\Phi_1, \Phi_2, \Phi_3$  \\
        $D_2$ & $\Phi_1$ & $\Phi_2, \Phi_3$  \\
        $D_3$ & $\Phi_1, \Phi_2$ & $\Phi_3$  \\
    \end{tabular}
    \caption{B--branes for SQED$[3]$.}
    \label{tab:sqed3_branes}
\end{table}
and a basis of twisted chiral ring insertions $\cO_0 = \mathbf{1}$, $\cO_1= \sigma$ and $\cO_2=\sigma^2$. The corresponding hemisphere partition functions are given by:
\begin{equation}\label{eq:sqed3_contour_integrals}
\begin{aligned}
    \cZ_{D_1}[\cO_a, m] &= \oint_{\mathcal{C}_1} \frac{d\sigma}{2\pi i \ep}
    e^{-\frac{2\pi i \sigma \tau}{\ep}}  \,\,
    {\textstyle
    \Gamma\left[\frac{\sigma+m_1}{\ep}\right]
    \Gamma\left[\frac{\sigma+m_2}{\ep}\right]\Gamma\left[\frac{\sigma-m_1-m_2}{\ep}\right] 
    }
    \cO_a,\\
    \cZ_{D_2}[\cO_a, m] &= \oint_{\mathcal{C}_2} 
    \frac{d\sigma}{2\pi i \ep}
    e^{-\frac{2\pi i \sigma \tau}{\ep}} \,\,
    {\textstyle
    \frac{(-2\pi i)e^{ \frac{\pi i (\sigma+m_1)}{\ep}}}{\Gamma\left[1-\frac{\sigma+m_1}{\ep}\right]}
    \Gamma\left[\frac{\sigma+m_2}{\ep}\right]\Gamma\left[\frac{\sigma-m_1-m_2}{\ep}\right]
    }
    \cO_a,\\
    \cZ_{D_3}[\cO_a, m] &= \oint_{\mathcal{C}_3} 
    \frac{d\sigma}{2\pi i \ep}
    e^{-\frac{2\pi i \sigma \tau}{\ep}} \,\,
    {\textstyle
    \frac{(-2\pi i)e^{ \frac{\pi i (\sigma+m_1)}{\ep}}}{\Gamma\left[1-\frac{\sigma+m_1}{\ep}\right]}
    \frac{(-2\pi i)e^{ \frac{\pi i (\sigma+m_2)}{\ep}}}{\Gamma\left[1-\frac{\sigma+m_2}{\ep}\right]}
    \Gamma\left[\frac{\sigma-m_1-m_2}{\ep}\right] 
    }
    \cO_a.
\end{aligned}
\end{equation}
In the above, the contours enclose the poles at:
\begin{equation}\label{eq:sqed3_contours}
\begin{aligned}
    \mathcal{C}_1&: \quad \sigma = -\ep k-m_1\\
    \mathcal{C}_2&: \quad \sigma = -\ep k-m_2 \qquad\quad \text{for } k \in \mathbb{N}_{0}.\\
    \mathcal{C}_3&: \quad \sigma = -\ep k+m_1+m_2
\end{aligned}
\end{equation}

It will suffice to focus on $\cZ_{D_1}$, for reasons which will become clear. Performing the contour integral yields
\begin{equation}
\begin{aligned}
    \cZ_{D_1}[\mathbf{1}] =& \, e^{\frac{2\pi i m_1 \tau}{\ep}} 
    {\textstyle \Gamma\left[1-x\right] \Gamma\left[1-y\right] {}_0F_2\left[ x, y ; -e^{2\pi i \tau}\right]
    },\\[1ex]
    \cZ_{D_1}[\sigma] =& -m_1 e^{\frac{2\pi i m_1 \tau}{\ep}} 
    \Gamma\left[1-x\right] \Gamma\left[1-y\right] {}_0F_2\left[ x, y ; -e^{2\pi i \tau}\right]
    \\
    &+\ep e^{2\pi i \tau} e^{\frac{2\pi i m_1 \tau}{\ep}} 
    \Gamma\left[-x\right] \Gamma\left[-y\right] {}_0F_2\left[x+1,y+1; -e^{2\pi i \tau}\right],
    \\[1ex]
    \cZ_{D_1}[\sigma^2] =&\, m_1^2 e^{\frac{2\pi i m_1 \tau}{\ep}} 
    \Gamma\left[1-x\right] \Gamma\left[1-y\right] {}_0F_2\left[x, y; -e^{2\pi i \tau}\right]
    \\
    &-(2m_1+\ep)\ep e^{2\pi i \tau} e^{\frac{2\pi i m_1 \tau}{\ep}} 
    \Gamma\left[-x\right] \Gamma\left[-y\right] {}_0F_2\left[x+1, y+1; -e^{2\pi i \tau}\right]
    \\
    &+\ep^2 e^{4\pi i \tau} e^{\frac{2\pi i m_1 \tau}{\ep}} 
    \Gamma\left[-x-1\right] \Gamma\left[-y-1\right] {}_0F_2\left[x+2, y+2; -e^{2\pi i \tau}\right],
\end{aligned}
\end{equation}
where:
\begin{equation}
    x \coloneqq 1-\frac{m_2-m_1}{\ep}, \qquad y\coloneqq 1+\frac{2m_1+m_2}{\ep},
\end{equation}
and ${}_0F_{2}$ is the generalised hypergeometric function
\begin{equation}
    {}_0F_{2}[a,b;z] \coloneqq \sum_{n=0}^{\infty} \frac{z^n}{(a)_n (b)_n n!}.
\end{equation}

We begin with the action of $\hat{p}_1$ on $\cZ_{D_1}[\cdot]$. We note that $\hat{p}_1: x \mapsto x+1$, $\hat{p}_1: y\mapsto~{y+2}$, and that $\cZ_{D_1}[\cdot]$ are linear combinations of ${}_0F_2\left[ x + n, y + n ; -e^{2\pi i \tau}\right]$, $n=0,1,2$. We therefore look for identities relating the above to ${}_0F_2\left[ x+1+n, y+2+n ; -e^{2\pi i \tau}\right]$, $n=0,1,2$. The required identities may be found by using the standard differentiation formula for generalised hypergeometric functions, from which it follows that any three of ${}_0F_2\left[ x, y ; \lambda \right]$, ${}_0F_2\left[ x-1, y ; \lambda \right]$, ${}_0F_2\left[ x, y-1 ; \lambda \right]$ and ${}_0F_2\left[ x+1, y+1 ; \lambda \right]$ are linearly dependent. We require:
\begin{equation}
\begin{aligned}
    {}_0F_2\left[ x-1, y ; \lambda \right] - {}_0F_2\left[ x, y ; \lambda \right] - \frac{\lambda}{x(x-1)y}{}_0F_2\left[ x+1, y+1 ; \lambda \right] = 0 ,\\
    {}_0F_2\left[ x, y-1 ; \lambda \right] - {}_0F_2\left[ x, y ; \lambda \right] - \frac{\lambda}{x y (y-1)}{}_0F_2\left[ x+1, y+1 ; \lambda \right] = 0 , \\
    (x-y) {}_0F_2\left[ x, y ; \lambda \right] + (y-1){}_0F_2\left[ x, y-1 ; \lambda \right]  -(x-1){}_0F_2\left[ x-1, y ; \lambda \right] = 0.
\end{aligned}    
\end{equation}
Applying these identities successively, it is a simple but tedious exercise to find:
\begin{equation}
    \begin{pmatrix}
        {}_0F_2\left[ x+1, y+2 ; \lambda \right]\\
        {}_0F_2\left[ x+2, y+3 ; \lambda \right]\\
        {}_0F_2\left[ x+3, y+4 ; \lambda \right]
    \end{pmatrix}
    =
     M(x,y,\lambda)
     \begin{pmatrix}
        {}_0F_2\left[ x, y ; \lambda \right]\\
        {}_0F_2\left[ x+1, y+1 ; \lambda \right]\\
        {}_0F_2\left[ x+2, y+2 ; \lambda \right]
    \end{pmatrix}
\end{equation}
where
\begin{equation}
\begin{aligned}
    M(x,y,\lambda) = &\left(
    \begin{array}{cc}
        \frac{xy(y+1)}{\lambda} & -\frac{xy(y+1)}{\lambda} \\
        -\frac{x (x+1) y (y+1)^2 (y+2)}{\lambda^2} &\frac{(x+1) (y+1) (y+2)}{\lambda} {\scriptstyle \left(1+\frac{x y (y+1)}{\lambda}\right)} 
         \\
        \frac{x (x+1) (x+2) y (y+1)^2 (y+2)^2 (y+3)}{\lambda^3}
        &-\frac{(x+1) (x+2) (y+1) (y+2)^2 (y+3)}{\lambda^2} {\scriptstyle \left(1+\frac{x y (y+1) }{\lambda}\right) } 
    \end{array}
    \right. 
    \\
    &\qquad\qquad\qquad \left.
    \begin{array}{cc}
    \ldots \quad & -\frac{y}{x+1} \\
    \ldots \quad & \frac{y (y+1) (y+2) }{\lambda}\\
    \ldots \quad & \frac{(x+2) (y+2) (y+3)}{\lambda} {\scriptstyle\left(1-\frac{y (y+1) (y+2)}{\lambda}\right) }
    \end{array}
    \right).
\end{aligned}
\end{equation}
Now, rewriting $\cZ[\cO_a, m ] = A[m, \ep;\tau]_{ab}\, {}_0F_2[x+b,y+b;-e^{2\pi i \tau}]$, then we can see immediately that
\begin{equation}
    \widetilde{G}^{(1)} = (\hat{p}_1 A) M A^{-1},
\end{equation}
and it is a simple but tedious exercise to check that $\widetilde{G}^{(1)}$ is of the form \eqref{eq:sqed3_diff_matrix}.

For $\hat{p}_2$, which shifts $\hat{p}_2: x \mapsto x-1$, $\hat{p}_2: y \mapsto y+1$, we need the identities:
\begin{equation}
    \begin{pmatrix}
        {}_0F_2\left[ x-1, y+1 ; \lambda \right]\\
        {}_0F_2\left[ x, y+2 ; \lambda \right]\\
        {}_0F_2\left[ x+1, y+3 ; \lambda \right]
    \end{pmatrix}
    =
     N(x,y,\lambda)
     \begin{pmatrix}
        {}_0F_2\left[ x, y ; \lambda \right]\\
        {}_0F_2\left[ x+1, y+1 ; \lambda \right]\\
        {}_0F_2\left[ x+2, y+2 ; \lambda \right]
    \end{pmatrix}
\end{equation}
where
\begin{equation}
\begin{aligned}
    N(x,y,\lambda) = & \left(
    \begin{array}{cc}
        \frac{y}{x-1} & \frac{x-y-1}{x-1} \\
        \frac{y (y+1) (x-y-1)}{\lambda} & \frac{y+1}{x}
        {\scriptstyle \left(1-\frac{x y  (x-y-1)}{\lambda}\right)} \\
        -\frac{x y (y+1)^2 (y+2) (x-y-1)}{\lambda^2} & 
        \frac{(y+1) (y+2) (x-y-1)}{\lambda} 
        {\scriptstyle  \left(1+ \frac{x y (y+1)}{\lambda}\right)}
        \\
    \end{array}
    \right. \\
    & \qquad\qquad\qquad\left.
    \begin{array}{cc}
        \ldots \quad & \frac{\lambda (x-y-1)}{(x-1) x (x+1) (y+1)} \\
        \ldots \quad  & -\frac{y (x-y-1)}{x (x+1)} \\
        \ldots \quad & \frac{y+2}{x+1}
        {\scriptstyle \left(1+ \frac{y(y+1)(x-y-1)}{\lambda}\right)} \\
    \end{array}
    \right).
\end{aligned}
\end{equation}
So similarly we can read off
\begin{equation}
    \widetilde{G}^{(2)} = (\hat{p}_2 A) N A^{-1}
\end{equation}
and check that indeed $\widetilde{G}^{(2)}(m_1, m_2) = \widetilde{G}^{(1)}(m_2,m_1)$.

We must now verify that the \textit{same} difference equations \eqref{eq:sqed3_difference_eqns} hold for the remaining thimble basis elements. Instead of computing the action explicitly, we can use the following arguments based on symmetry. Note that the Gamma function obeys the Euler reflection formula \eqref{eq:gamma_euler_reflection}. Applying this to the contour integral for $\cZ_{D_2}[\cO_a,m]$ in \eqref{eq:sqed3_contour_integrals}, and noting that the resulting integrand is the same as that for $\cZ_{D_1}[\cO_a,m]$ but with $m_1\leftrightarrow m_2$ in the contour, we have:
\begin{equation}
    \cZ_{D_2}[\cO_a] = C_{21}\cZ_{D_1}[\cO_a, m_2, m_1].
\end{equation}
Note that the prefactor $C_{21} \coloneqq 1-e^{\frac{2\pi i (m_1-m_2)}{\ep}}$ is invariant under $\hat{p}$. Thus:
\begin{equation}
\begin{aligned}
    \hat{p}_1 \cZ_{D_2}[\cO_a,m_1,m_2]
    &=  C_{21} \left( \widetilde{G}^{(2)}_{ab}(m_1,m_2) \cZ_{D_1}[\cO_b, m_1, m_2]  \right)\Big|_{m_1 \leftrightarrow m_2} \\ 
    &= \widetilde{G}^{(1)}_{ab}(m_1,m_2)\cZ_{D_2}[\cO_a,m_1,m_2] 
\end{aligned} 
\end{equation}
using the symmetry \eqref{eq:sqed3_symmetry}. Similarly, it is easy to show that 
\begin{equation}
    \hat{p}_2 \cZ_{D_2}[\cO_a,m_1,m_2] = \widetilde{G}^{(2)}_{ab}(m_1,m_2)\cZ_{D_2}[\cO_a,m_1,m_2].
\end{equation}

Finally, we move on to the third thimble $D_3$. Using the Euler reflection formula and the contour prescription \eqref{eq:sqed3_contours} as above, we can also show that:
\begin{equation}
    \cZ_{D_3}[\cO_a,m_1,m_2] = C_{31} \cZ_{D_1}[\cO_a,-m_1-m_2,m_2].
\end{equation}
The prefactor $C_{31} \coloneqq \Big(1-e^{\frac{2\pi i (2m_1+m_2)}{\ep}}\Big)\Big(1-e^{\frac{2\pi i (2m_2+m_1)}{\ep}}\Big)$ is also $\hat{p}$-invariant. Then:
\begin{equation}
\begin{aligned}
    \hat{p}_1  \cZ_{D_3}[\cO_a,m_1,m_2]  
    &= C_{31} \cZ_{D_1}[\cO_a,-m_1-m_2-\ep,m_2]  \\
    &= C_{31}  \left(\widetilde{G}^{(1)}(m_1-\ep, m_2)^{-1}_{ab}\cZ_{D_1}[\cO_b,m_1 ,m_2] \right)\Big|_{m_1 \rightarrow -m_1-m_2} \\
    &= \widetilde{G}^{(1)}(-m_1-m_2-\ep, m_2)^{-1}_{ab}\cZ_{D_3}[\cO_b,m_1,m_2]. 
\end{aligned}
\end{equation}
It is then easily checked that:
\begin{equation}
    \widetilde{G}^{(1)}(-m_1-m_2-\ep, m_2)^{-1}_{ab} = \widetilde{G}^{(1)}(m_1, m_2)_{ab},
\end{equation}
which is a non--trivial check on the matrix difference equation, and which implies that: 
\begin{equation}
    \hat{p}_1  \cZ_{D_3}[\cO_a,m_1,m_2]  
   = \widetilde{G}^{(1)}(m_1, m_2)_{ab}\cZ_{D_3}[\cO_b,m_1,m_2],
\end{equation}
as desired. Similarly, we have:
\begin{align}
    \hat{p}_2  \cZ_{D_3} & [\cO_a,  m_1,m_2]  = C_{31} \left( \hat{p}_2 \cZ_{D_1}[\cO_a,m_1-\ep,m_2]\right)\Big|_{m_1 \rightarrow -m_1-m_2}  \nonumber \\
    &= C_{31}  \left(\widetilde{G}^{(1)}(m_2, m_1-\ep)_{ab} \cZ_{D_1}[\cO_b,m_1-\ep,m_2]\right) \Big|_{m_1 \rightarrow -m_1-m_2} \\
    &= \left(\widetilde{G}^{(1)}(m_2, m_1-\ep)_{ab} \widetilde{G}^{(1)}(m_1-\ep, m_2)^{-1}_{bc}\right) \Big|_{m_1 \rightarrow -m_1-m_2} \cZ_{D_3}[\cO_c,m_1,m_2].\nonumber
\end{align}
Using that
\begin{equation}
     \widetilde{G}^{(1)}(m_2, -m_1-m_2-\ep) = \widetilde{G}^{(1)}(m_2, m_1)  \widetilde{G}^{(1)}(-m_1-m_2-\ep, m_2),
\end{equation}
which can be checked easily from \eqref{eq:sqed3_diff_matrix}, and \eqref{eq:sqed3_symmetry}, it follows that
\begin{equation}
    \hat{p}_2  \cZ_{D_3}[\cO_a,m_1,m_2]  
   = \widetilde{G}^{(2)}(m_1, m_1)_{ab}\cZ_{D_3}[\cO_b,m_1,m_2]
\end{equation}
as expected.

\bibliographystyle{JHEP}
\bibliography{berryconnections}

\end{document}